\documentclass[letterpaper, amsfonts, amssymb, amsmath,  showkeys, nofootinbibm,reprint, twocolumnms,prb,citeautoscript]{revtex4-2} %  twoside,aps,prb,reprint,preprint
\usepackage[english]{babel}
\usepackage[utf8]{inputenc}

\usepackage{amsthm}
\usepackage{mathtools}
\usepackage{physics}
\usepackage{xcolor}
\usepackage{graphicx}
\usepackage[left=14mm,right=14mm,top=18mm,columnsep=15pt,bottom=18mm]{geometry}

\usepackage{adjustbox}
\usepackage{placeins}
\usepackage[T1]{fontenc}
\usepackage{lipsum}
\usepackage{csquotes}
\usepackage{graphicx}
\usepackage{MnSymbol}
\usepackage{mathbbol}

\graphicspath{ {./figures/} }

\usepackage[
	hyperindex, 
	breaklinks, 
	colorlinks=true,
	hypertexnames=false]{hyperref}
\hypersetup{
    colorlinks   = true, %Colour links instead of ugly boxes
    urlcolor     = blue, %Colour for external hyperlinks
    linkcolor    = blue, %Colour of internal links
    citecolor   = blue %Colour of citations
    }
\usepackage{xcolor}    

\begin{document}
%\setcitestyle{super}
\title{Evaluation of diffuse mismatch model for phonon scattering at disordered interfaces}

\author{Qichen Song}
    \email{qcsong@mit.edu}% Your name
    \affiliation{Department of Mechanical Engineering, Massachusetts Institute of Technology, Cambridge, Massachusetts 02139, USA} 
\author{Gang Chen}
    \email{gchen2@mit.edu}% Your name
    \affiliation{Department of Mechanical Engineering, Massachusetts Institute of Technology, Cambridge, Massachusetts 02139, USA}
%\date{\today} % Leave empty to omit a date

\begin{abstract}
    Diffuse phonon scattering strongly affects the phonon transport through a disordered interface. The often-used diffuse mismatch model assumes that phonons lose memory of their origin after being scattered by the interface. Using mode-resolved atomistic Green's function simulation, we demonstrate that diffuse phonon scattering by a single disordered interface cannot make a phonon lose its memory and thus the applicability of diffusive mismatch model is limited. An analytical expression for diffuse scattering probability  based on the continuum approximation is also derived and shown to work reasonably well at low frequencies.
%\lipsum[1]
\end{abstract}


\maketitle

\section{Introduction} \label{sec:1}

 The interface between two dissimilar materials, or even same materials but different crystal orientations, can scatter phonons and gives rise to the thermal boundary resistance for heat flow across the interface\cite{RevModPhys.61.605,little1959transport,kapitza1941zh}. One model for the thermal boundary resistance is based on assuming that phonons are specularly scattered at the interface and computing the phonon transmittance and reflectance  based on acoustic wave equations, \textit{i.e.}, the acoustic mismatch model (AMM)\cite{khla,little1959transport}. However, it was found that the AMM only works at very low temperatures at which the phonon wavelengths are long.  At elevated temperatures, phonons of short wavelengths carry most of the heat, and they do not experience specular transmission/reflection due to interface imperfections, such as atomic mixing.  The diffuse mismatch model (DMM) is proposed as an extreme to describe phonon transport across such rough interfaces\cite{RevModPhys.61.605}. Two major assumptions are made in the DMM. Firstly, the transmittance is isotropic, \textit{i.e.}, transmittance is angle-independent. Secondly, phonons lose memory of their origin after being scattered by the interface such that one cannot distinguish if a phonon has just been through a transmission or reflection process.  Although the DMM has improved the agreement with experimental measurements of thermal boundary resistance at high temperatures\cite{PhysRevB.73.144301,cheng2020thermal}, the assumptions behind DMM have never been examined in detail.
 
 The thermal boundary resistance has been studied using equilibrium molecular dynamics\cite{PhysRevB.85.195302,gordiz2015formalism,gordiz2016phonon} and nonequilibrium molecular dynamics\cite{PhysRevB.80.165304,PhysRevB.90.134312,yang2015thermal}. In particular, phonon-mode-resolved transmittance had been formulated, which builds upon the atomic trajectories at steady state from molecular dynamics (MD) simulations\cite{chalopin2013microscopic}, where the anharmonicity of phonons is intrinsically included.
It has been applied to study mode-resolved transmittance through perfect interfaces, yet it has not been used to examine details of diffuse phonon transmittance across a disordered interface.
Phonon wave-packet dynamics technique has been applied to compute the transmittance and reflectance of each phonon mode, and the Kapiza resistance can be calculated using Landauer formalism\cite{PhysRevB.75.144105,deng2014kapitza}. The MD simulations and, especially the wave-packet dynamics simulations, however, require large structures in real space, including two bulk regions and the interface region, and simulating phonon transport through a rough interface with a large lateral dimension becomes computationally extensive.
 
The atomistic Green's function (AGF) has been shown as an effective method to study phonon interfacial transport\cite{zhang2007simulation,zhang2007atomistic,PhysRevB.86.235304}. The method is formulated in reciprocal space such that one does not have to deal with large-scale simulations of atomic displacements in real space. 
Recent advances in calculation of interfacial thermal resistance using the AGF have provided more insights in understanding interfacial thermal resistance with detailed information including mode-resolved transmission coefficients\cite{PhysRevB.91.174302,PhysRevB.96.174302,PhysRevB.96.104310,zhang2007simulation}. Specifically, Ong \textit{et al} studied the phonon specularity and coherence for phonon transport through a disordered grain boundary in two-dimensional graphene, and showed that incoherent phonon scatterings at interface are almost perfectly diffusive\cite{PhysRevB.101.195410}. Ong also demonstrated that the specularity parameters are different for transmittance and reflectance for graphene grain-boundaries\cite{ong2021specular}.
Using AGF combined with \textit{ab initio} inter-atomic force constants, Tian \textit{et al} found that the intermixing of atoms for Si/Ge interface can enhance interfacial thermal conductance\cite{PhysRevB.86.235304}. Sadasivam \textit{et al} demonstrated using phonon-eigenspectrum-based formulation of AGF that the enhanced interfacial thermal conductance of a Si/Ge interface with atom intermixing comes from diffuse transmission channels where the in-plane momentum is not conserved\cite{PhysRevB.96.174302}.
The scattering boundary method (SBM), a mathematically equivalent
method to AGF has been proposed by Young \textit{et al}\cite{PhysRevB.40.3685}
and generalized by Zhao \textit{et al}\cite{zhao2005lattice} to study mode-resolved phonon scattering at the interface.  Simon \textit{et al}\cite{lu2017thermal} has applied SBM to study phonon scattering at the interface between two-dimensional materials. 
Recently, Latour \textit{et al} have demonstrated the transmission spectra across a perfect interface as a function of incident angle of phonons, using mode-resolved AGF\cite{PhysRevB.96.104310}. However, in order to study diffuse phonon scattering, a supercell of a rough interface with a large lateral dimension is required. In addition, the folded lateral wavevector in the supercell must be carefully mapped back to the wavevector defined in the original unitcell. We realize that, despite these studies, none of them had critically examined the validity of DMM for diffuse phonon scattering. 

In this work, we conduct mode-resolved AGF calculation of transmittance and reflectance, and revisit the assumptions of DMM. Our study reveals that most phonons do not lose their memory of origin.  We also derive an analytical expression for the diffuse transmittance and reflectance based on a continuum model, and show that it works reasonably well at low frequencies.

\section{Methodology} \label{sec:2}
\subsection{Revisiting DMM}
To derive the DMM, Swartz and Pohl\cite{RevModPhys.61.605} have made two major assumptions. The first assumption is that phonons are diffusely scattered by the interface and the transmittance is isotropic,
    \begin{equation}
        T_{L\to R}(\omega,\mathbf{q}\nu) = T_{L\to R}(\omega)
    \end{equation} 
where $\mathbf{q}$ is the wavevector of the phonon with frequency $\omega$  and $\nu$ is the phonon branch index on the left side. The second assumption is that the transmittance from one side
must equal the reflectance from the other side, \textit{i.e.}, complete loss of memory,
\begin{equation}
    R_{R\to R}(\omega,\mathbf{q}\nu) = T_{L\to R}(\omega)
\end{equation}
Consequently, the transmittance from right side writes,
\begin{equation}
    T_{R\to L}(\omega,\mathbf{q}\nu) = 1-T_{L\to R}(\omega)
\end{equation}

At a given frequency, by invoking the principle of detailed balance, the transmittance in the elastic scattering limit writes,
\begin{equation}
    \begin{split}
    &T_{L\to R}(\omega)\\
    &=\frac{\sum^+_{\mathbf{q}\nu}\frac{v_{Rz,\mathbf{q}\nu}}{V_{\mathrm{uc},R}}\delta(\omega-\omega_{R,\mathbf{q}\nu})}{\sum^+_{\mathbf{q}\nu}\frac{v_{Lz,\mathbf{q}\nu}}{V_{\mathrm{uc},L}}\delta(\omega-\omega_{L,\mathbf{q}\nu})+\sum^+_{\mathbf{q}\nu}\frac{v_{Rz,\mathbf{q}\nu}}{V_{\mathrm{uc},R}}\delta(\omega-\omega_{R,\mathbf{q}\nu})}\\
    &=\frac{\Theta_{\mathrm{bulk},R}(\omega)}{\Theta_{\mathrm{bulk},L}(\omega)+\Theta_{\mathrm{bulk},R}(\omega)} 
    \end{split}
    \label{tdmm}
\end{equation}
where $v_{\alpha z,\mathbf{q}\nu}$ with $\alpha=L,R$ is the group velocity normal to the interface for phonons from the left or the right side, $\omega_{\alpha,\mathbf{q}\nu}$ is the phonon frequency and $V_{uc,\alpha}$ is the volume of unitcell of the left and the right side. The superscript + means that only forward-moving states with $v_{\alpha z,\mathbf{q}\nu}>0$ are included in the summation. The transmittance can also be written in terms of the ratio of transmission functions, as expressed in the second line of the equation. $\Theta_{\mathrm{bulk},L/R}(\omega)$ is the bulk transmission function for the left/right side, which is a measure of the number of heat conduction channels. 

To assess the validity of DMM,
we examine if the transmittance and reflectance are indeed isotropic and if the transmittance from one side and reflectance from the other side are the same. 

\subsection{Mode-resolved atomistic Green’s function formalism}

%We consider elastic interface scattering where the energy for phonons are conserved. 
The essential physical quantities to study diffuse phonon scattering by a rough interface are the transmission probability matrix $T_{mn}(\omega)$ and reflection probability matrix $R_{ln}(\omega)$ at a given phonon frequency $\omega$, which describe the transition probability from the initial state $n$ to the final state $l$ or $m$ via interface scattering processes. These matrices are computed from mode-resolved atomistic Green's function formalism as outlined in Ref.~\cite{PhysRevB.91.174302} and Ref.~\cite{PhysRevB.101.195410}, with details provided in the supplementary material. Specifically, we divide the system of interest into three parts, the left lead, the right lead and the device.
% We then extract the momentum and group velocity of each propagating mode of the left lead and the right lead. Finally,
A propagating state coming out of one lead can be transmitted through (or be reflected by) the device region and travel to the other lead (or the same lead).
We then compute the ratio of the heat flux along the z direction of the outgoing state $m$ (or state $l$),  to the heat flux along the z direction of the initial state $n$, which is the element of transmission probability matrix $T_{mn}(\omega)$ (or reflection probability matrix $R_{ln}(\omega)$).

For a rough interface created by atomic mixing at the interface, the transverse translational symmetry is broken by the interfacial disorders. It is impractical to compute scatterings of an infinitely large rough interface. Instead, we construct a supercell of two materials and a rough interface between them with periodic boundary conditions along the transverse directions (x-direction and y-direction).  Because of the transverse periodicity of the supercell, the phonon state of the lead region defined at a given transverse wavevector  $\mathbf{q}_{\mathrm{sc},\parallel}$ can only be scattered into phonon states of the lead region (either the left or the right lead) with the same wavevector $\mathbf{q}_{\mathrm{sc},\parallel}$. 

\begin{figure}[t]
    \includegraphics[width=0.48\textwidth]{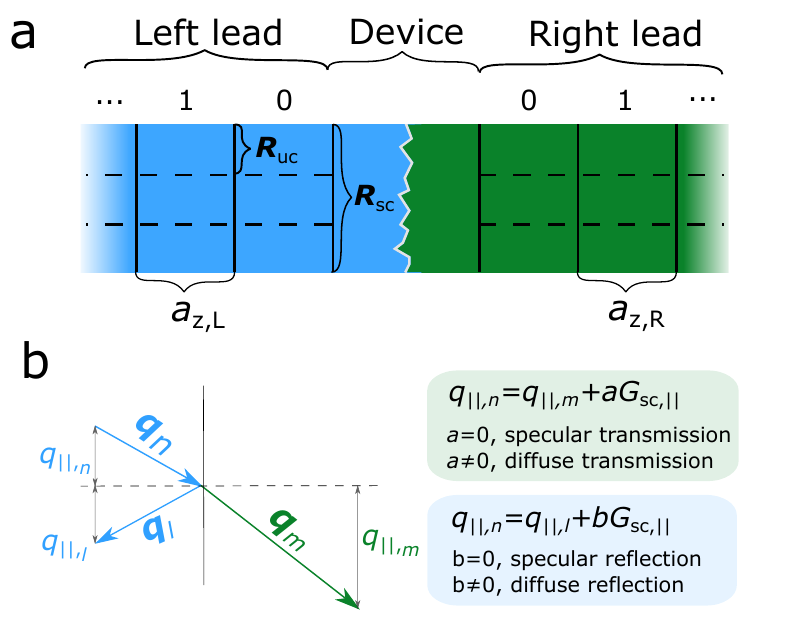}
    \caption{(a) In AGF, the system is partitioned into three parts: two semi-infinite leads and a rough interface as the device region. The lead region in the supercell, contains $N_{\mathrm{uc}}$ unitcells and transverse lattice vector $\mathbf{R}_{\mathrm{sc}} = N_{\mathrm{uc}}\mathbf{R}_{\mathrm{uc}}$. The numbers 0, 1, ... denote the index of repeated cells for left and right leads. The period lengths along the direction normal to the interface are $a_{z,L}$ and $a_{z,R}$ for the left and right lead, respectively. (b) The in-plane wavevector for incident, transmitted and reflected phonons, $\mathbf{q}_{\parallel,n}$, $\mathbf{q}_{\parallel,m}$ and $\mathbf{q}_{\parallel,l}$. 
    $a=0$ corresponds to specular transmission, while $a \neq0$ corresponds to diffuse transmission. Similarly, $b=0$ corresponds to specular reflection, while $b \neq0$ corresponds to diffuse reflection. Note the schematic is drawn for two-dimensional system for visual clarity and for three-dimensional system the partitioning and wavevector conservation laws can be analogously defined.}
    \label{fs}
\end{figure}

The lead part of the supercell contains $N_x \times N_y$ repeated unitcells, as depicted in Fig.~\ref{fs} (a). The period lengths of the lead along the direction normal to the interface are $a_{z,L}$ for the left lead and $a_{z,R}$ for the right lead. 
The phonon wavevectors parallel to the interface in the supercell and in the unitcell representations are related via $\mathbf{q}_{\mathrm{uc},\parallel} = \mathbf{q}_{\mathrm{sc},\parallel}+a\mathbf{G}_{\mathrm{sc},x}+b\mathbf{G}_{\mathrm{sc},y}$, where $a$ and $b$ are integers. $\mathbf{G}_{\mathrm{sc},x}$ and  $\mathbf{G}_{\mathrm{sc},y}$ are transverse reciprocal lattice vectors of the supercell. The phonon states at the corresponding equivalent wavevectors (with same $q_z$ and same branch index) in the two representations are equivalent\cite{parallel}. The phonon state in the unitcell representation is preferred as it is much easier to interpret than the supercell representation (we will hide subscript uc in the following for visual clarity). However, for a given supercell state $\mathbf{q}_{\mathrm{sc},\parallel}$, there are multiple possible choices of $a$ and $b$. To find out the correct pair of $a$ and $b$ for wavevector $\mathbf{q}_{\parallel}$ is known as an unfolding problem. We have adopted the unfolding scheme by Popescu \textit{et al}\cite{PhysRevLett.104.236403} and the details can be found in the supplementary material.   

%The phonon state of the lead region defined in the supercell representation at a given wavevector  $\mathbf{q}_{\mathrm{sc},\parallel}$ is equivalent to a phonon state defined in the unitcell representation with wavevector  $\mathbf{q}_{\mathrm{uc},\parallel}$ (\textit{i.e.} state $\mathbf{q}_{\mathrm{sc},\parallel}$ can be unfolded into state $\mathbf{q}_{\mathrm{uc},\parallel}$). The wavevectors in the two representations are related via  Here, .  
%The transverse reciprocal lattice vectors for supercell is smaller than the ones for unitcell,
%$\mathbf{G}_{\mathrm{sc},x}=\frac{1}{N_x}\mathbf{G}_{\mathrm{uc},x}$ and $ \mathbf{G}_{\mathrm{sc},y}=\frac{1}{N_y}\mathbf{G}_{\mathrm{uc},y}$. 
%The phonon state in the unitcell representation is preferred as it is much easier to interpret than the supercell representation  (we will hide subscript uc in the following  for visual clarity). 

For an interface scattering event, the transverse wavevectors for initial state $n$ and final state $m$ in the unitcell representation are constrained by,
\begin{equation}
\mathbf{q}_n  =  \mathbf{q}_m+a\mathbf{G}_{\mathrm{sc},x}+b\mathbf{G}_{\mathrm{sc},y},
\end{equation}
where $a$ and $b$ are unknown integers, as wavevectors of initial and final states in the supercell representation can be unfolded differently. 
This expression indicates that the interface scattering can either be a momentum conserved (specular), when $a=b=0$,  or momentum non-conserved (diffuse) process for other $a$ and $b$ values. 
%We apply the mode-resolved atomistic Green's function formalism\cite{sm} and compute the transmission matrix $\mathbb{t}_{mn}$ for phonon interfacial transport through a rough interface, where $m$ and $n$ denote propagating phonon mode of right and left lead (see schematic in Fig.~\ref{fs} (a)), respectively. The transmission probability from state $n$ to state $m$ is defined by the squared amplitude of matrix element $T_{mn}=|\mathbb{t}_{mn}|^2$. 
Depending on the transverse wavevector of state $n$ and state $m$, the transmission probability matrix can be categorized into specular and diffuse transmission parts $T_{mn}(\omega) = T_{\mathrm{s},mn}(\omega)+T_{\mathrm{d},mn}(\omega)$,
where
\begin{equation}
    \begin{cases}
    T_{\mathrm{s},mn}(\omega)  = T_{mn}(\omega), \:\mathrm{when} \:\mathbf{q}_{\parallel,n}= \mathbf{q}_{\parallel,m}\\
    T_{\mathrm{d},mn}(\omega)  = T_{mn}(\omega), \:\mathrm{when} \:\mathbf{q}_{\parallel,n}\neq \mathbf{q}_{\parallel,m} \\
    \end{cases}
\end{equation}
The reflection probability matrix can be analogously expressed by, $R_{ln}(\omega) = R_{\mathrm{s},ln}(\omega)+R_{\mathrm{d},ln}(\omega)$,
where
\begin{equation}
    \begin{cases}
    R_{\mathrm{s},ln}(\omega)  = R_{ln}(\omega), \:\mathrm{when} \:\mathbf{q}_{\parallel,n}= \mathbf{q}_{\parallel,l}\\
    R_{\mathrm{d},ln}(\omega)  = R_{ln}(\omega), \:\mathrm{when} \:\mathbf{q}_{\parallel,n}\neq \mathbf{q}_{\parallel,l} \\
    \end{cases}
\end{equation}

%In the field of acoustics and optics, it is conventional to use transmittance $T(\theta,\phi)$ to describe the elastic scattering by a disordered structure at given frequency $\omega$. Transmittance is defined by the proportion of transmitted flux in the incident flux at a given solid angle $\Omega$. The solid angle $\Omega$ of incident flux is determined by the group velocity (or momentum, due to the simple dispersion relation $\omega = c|\mathbf{q}|$) of the incident quantum state. However, in phononics, the phonon dispersion is often a complicated function. Thus, the solid angle of group velocity and momentum are generally independent. We choose the solid angle of group velocity rather than momentum to denote a incident phonon state as the momentum is not uniquely defined. The reflectance for a phonon state (either as an initial state or a final state) can be analogously defined. Note that at given frequency $\omega$, there are often multiple (or degenerate) branches such that certain solid angle corresponds to more than one quantum state. 

The diffuse transmittance for a given incident phonon $n$ from the left side is defined by summing over the scattering probabilities of all possible outgoing states,
\begin{equation} 
    \begin{split}
T_{\mathrm{d},L\to R}(\omega,\Omega_L)  = \sum_{m} T^{L\to R}_{\mathrm{d},mn}(\omega)
    \end{split}
\end{equation}
where $\Omega_L=(\theta,\phi)$ indicates transmittance is a directional quantity. The polar and azimuthal angles are defined in a coordinate system where the interface normal lies along the z-axis, $\theta = \mathrm{arccos}\frac{v_z}{|\mathbf{v}|}$, $\phi = \mathrm{arctan}\frac{v_y}{v_x}$. $\mathbf{v}=(v_x,v_y,v_z)$ is the group velocity for incident phonon $n$ from the left side and we use $L\to R$ to denote the trajectory of the phonon.
The reason  for using the angle of the group velocity rather than the phase velocity (or wavevector) is that group velocity is a uniquely defined quantity irrelevant to the choice of in-plane Brillouin zone while not for the phase velocity. 
 %We can similarly define the transmittance for a given outgoing phonon at left side by summing over the scattering probabilities of all possible incident states,
%\begin{equation} 
%    \begin{split}
%T_{R\to L}(\Omega_L) = T^{R\to L}_n = \sum_{m} T^{R\to L}_{nm}\\
%    \end{split}
%\end{equation}
%The time reversal symmetry suggests that $T^{L\to R}_{mn} = T^{R\to L}_{nm}$. Thereby, $T_{R\to L}(\Omega_L)=T_{L\to R}(\Omega_L)$, $\textit{i.e.}$ the transmittance when $n$ is an initial state is the same as the transmittance when $n$ is the final state. 
Likewise, the diffuse reflectance for a given initial state $n$ in the left side and in the right side read,
\begin{equation} 
    \begin{split}
R_{\mathrm{d},\alpha\to \alpha}(\omega,\Omega_\alpha) = \sum_{m} R^{\alpha\to \alpha}_{\mathrm{d},mn}(\omega)
    \end{split}
\end{equation}
where $\alpha=L,R$. The specular transmittance and reflectance can be similarly defined.

\begin{figure*}[t!]
    \includegraphics[width=0.9\textwidth]{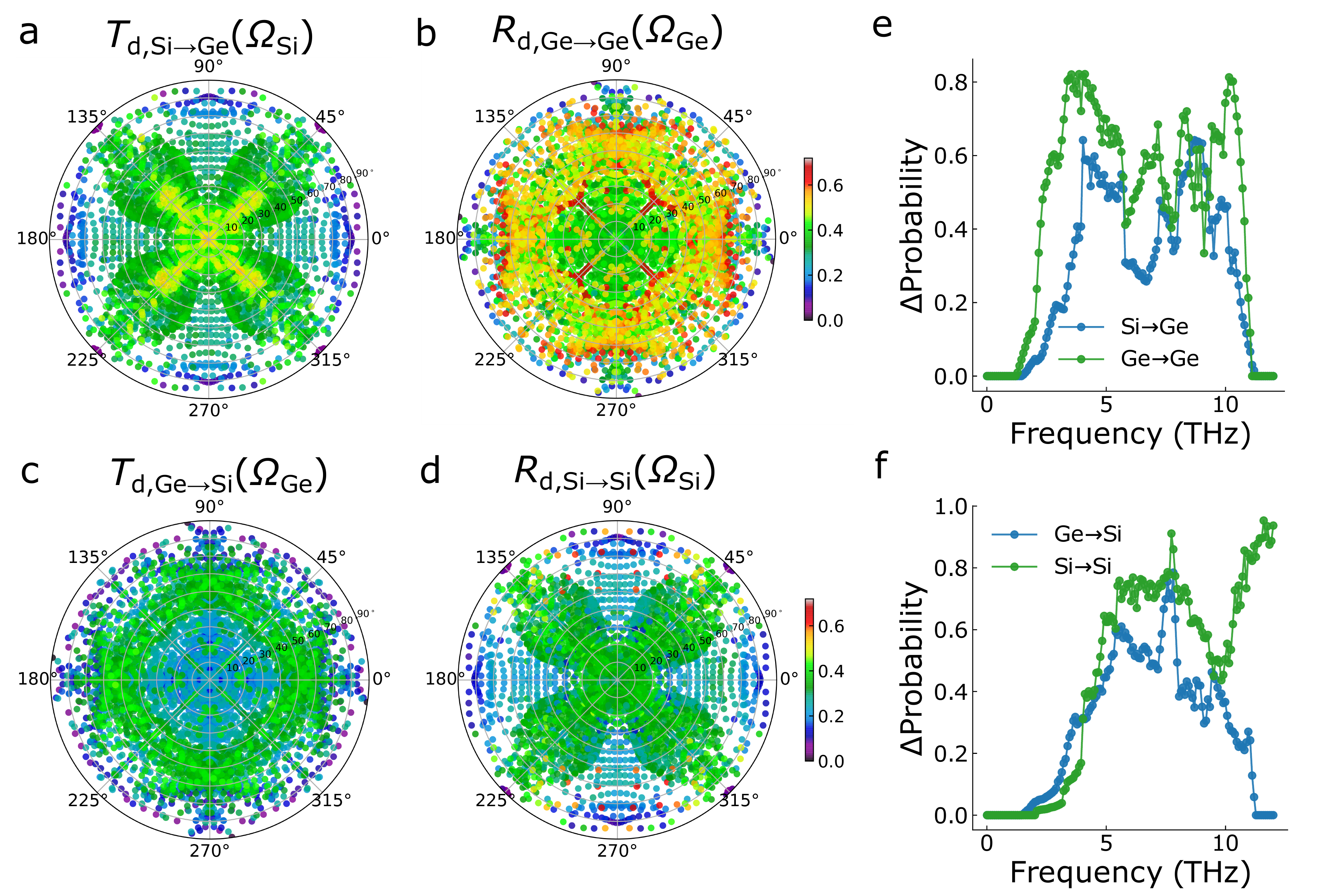}
    \caption{The angle-resolved (a) diffuse transmittance $T_{d,\mathrm{Si}\to \mathrm{Ge}}(\Omega_{\mathrm{Si}})$ from the Si side and (b) diffuse reflectance $R_{d,\mathrm{Ge}\to \mathrm{Ge}}(\Omega_{\mathrm{Ge}})$ from the Ge side at $\omega = 5 $ THz. The angle-resolved (c) diffuse transmittance $T_{d,\mathrm{Ge}\to \mathrm{Si}}(\Omega_{\mathrm{Ge}})$ from the Ge side and (d) diffuse reflectance  $R_{d,\mathrm{Si}\to \mathrm{Si}}(\Omega_{\mathrm{Si}})$ from the Si side at $\omega = 5 $ THz. $\Omega_{\mathrm{\alpha}} = (\theta_{\alpha},\phi_{\alpha})$, $\alpha=\mathrm{Si,Ge}$ is the direction of incident group velocity.  The radial coordinate corresponds to the polar angle $\theta_\alpha$ (angle of incidence) and the polar axis corresponds to the azimuthal angle $\phi_\alpha$. (e), (f) The difference between maximum  and minimum diffuse scattering probability as a 
   measure of the anisotropy of diffuse scattering probability. The diffuse scattering probability is obtained by taking the ensemble average of calculations for 21 structures of 8 ml disordered configurations with a 20$\times$20$ \mathbf{q}_{\mathrm{sc},\parallel}$-point mesh.}
    \label{f3}
\end{figure*}

Furthermore, to study the impact of diffuse phonon scattering on interfacial transport, we compute the transmission function $\Theta(\omega)$, which accounts for the total phonon conduction transmission at a given frequency $\omega$, defined by,
\begin{equation}
    \Theta(\omega) = \Theta_\mathrm{s}(\omega)+\Theta_\mathrm{d}(\omega)
\end{equation}
$\Theta_\mathrm{s}(\omega)$ and  $\Theta_\mathrm{d}(\omega)$ are specular and diffuse transmission function obtained by summing over all possible incoming and outgoing states at a given frequency,
\begin{equation}
    \begin{split}
    \Theta_\mathrm{s}(\omega) &= \sum_{mn}T^{L \to R}_{\mathrm{s},mn}(\omega)\\
    \Theta_\mathrm{d}(\omega) &= \sum_{mn}T^{L\to R}_{\mathrm{d},mn}(\omega)\\
    \end{split}
\end{equation}
The specular and diffuse reflection function is defined by,
\begin{equation}
    \begin{split}
    \Xi_{\mathrm{s},\alpha}(\omega) &= \sum_{mn}R^{ \alpha \to \alpha}_{\mathrm{s},mn}(\omega)\\
    \Xi_{\mathrm{d},\alpha}(\omega) &= \sum_{mn}R^{ \alpha\to  \alpha}_{\mathrm{d},mn}(\omega)\\
    \end{split}
\end{equation}
with $\alpha=L,R$. Note that the transmission function for two sides are the same due to the time-reversal symmetry of the transmission probability matrix but not for the reflection function.

\subsection{Continuum modeling}
In addition to the AGF simulation, we also  derived analytical formulas for diffuse the transmittance and reflectance from continuum modeling with details provided in the Appendix~\ref{sec:appendix1}. The model assumes scalar acoustic waves and random mass disorders distributed at the interface, and hence neglects mode conversion at the interface. The model is derived based on perturbation theory and it only takes the density  $\rho_L$, $\rho_R$, bulk modulus $\mu_L$, $\mu_R$ and number of pairs of swapped atoms per unit area $n$ as parameters. The model captures the specific contributions to the total transmittance/reflectance of specular and diffuse scattering processes and allows one to calculate diffuse and specular transmittance and reflectance components analytically.

We assumes a linear dispersion $\omega = c |\mathbf{q}|$, where the sound velocities for the left and right side are $c_{L} = \sqrt{\mu_{L}/\rho_{L}}$ and $c_{R} = \sqrt{\mu_{R}/\rho_{R}}$, respectively. Due to the simple dispersion relation, the transverse $\mathbf{q}_\parallel$ can uniquely define a forward-moving phonon state. Thus, we use $\mathbf{q}_\parallel$ and $\mathbf{q}^\prime_\parallel$ to denote the initial and final states, instead of using $m$ and $n$. The momentum for a phonon state in the left side is
$(\mathbf{q}_\parallel,q_L)=\frac{\omega}{c_L}(  \mathrm{sin}\theta_L\mathrm{cos}\phi, \mathrm{sin}\theta_L\mathrm{sin}\phi,\mathrm{cos}\theta_L)$, where $q_L$ is the perpendicular momentum.  For a specular transmission process from the left side to the right side, the transverse momentum is conserved. Thus, the corresponding transmitted phonon state on the right side is $(\mathbf{q}_\parallel,q_R)=\frac{\omega}{c_R}(  \mathrm{sin}\theta_R\mathrm{cos}\phi, \mathrm{sin}\theta_R\mathrm{sin}\phi,\mathrm{cos}\theta_R)$. It follows that the perpendicular velocities for the initial and final state are $v_L = c_L\mathrm{cos}\theta_L = c_L\sqrt{1-c^2_L|\mathbf{q}_\parallel|^2/\omega^2}$, $v_R=c_R\mathrm{cos}\theta_R=c_R\sqrt{1-c^2_R|\mathbf{q}_\parallel|^2/\omega^2}$, respectively. 

\begin{figure*}[t!]
\includegraphics[width=0.95\textwidth]{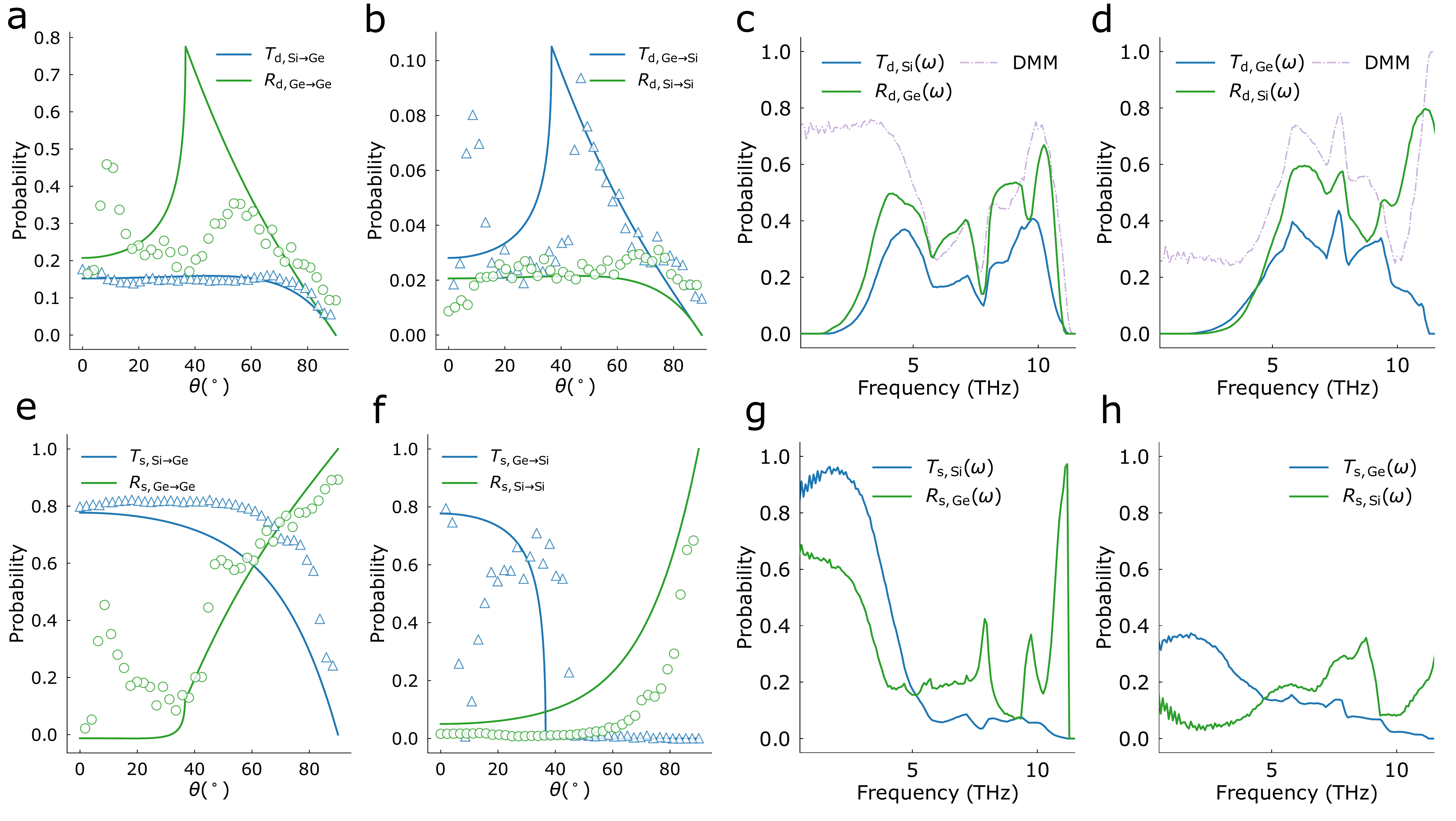}
\caption{(a), (b) The diffuse transmittance $T_{d,\alpha}$ from one side and the diffuse reflectance $R_{d,\beta}$ from the other side as a function of the polar angle of the incident phonon at $\omega = 3.3$ THz.  The markers are obtained by integrating  $T_d(\Omega)$ and $R_d(\Omega)$ from AGF calculation for 8 ml structures over the azimuthal angle $\phi$ divided by $2\pi$. The solid lines are predictions from continuum modeling with the number of pairs of swapped atoms per unit area $n = 2.78/a^2$ and $a = 5.527$ \AA.  (c), (d) The average diffuse transmittance $T_{d,\alpha}(\omega)$ from one side  and  the average diffuse reflectance $R_{d,\beta}(\omega)$  from the other side as a function of frequency from AGF calculation in solid lines, compared with DMM in dash-dot lines. (e)-(h) are the specular transmittance and specular reflectance corresponding to (a)-(d).}
\label{f4}
\end{figure*}

\begin{figure*}[t!]
    \includegraphics[width=0.97\textwidth]{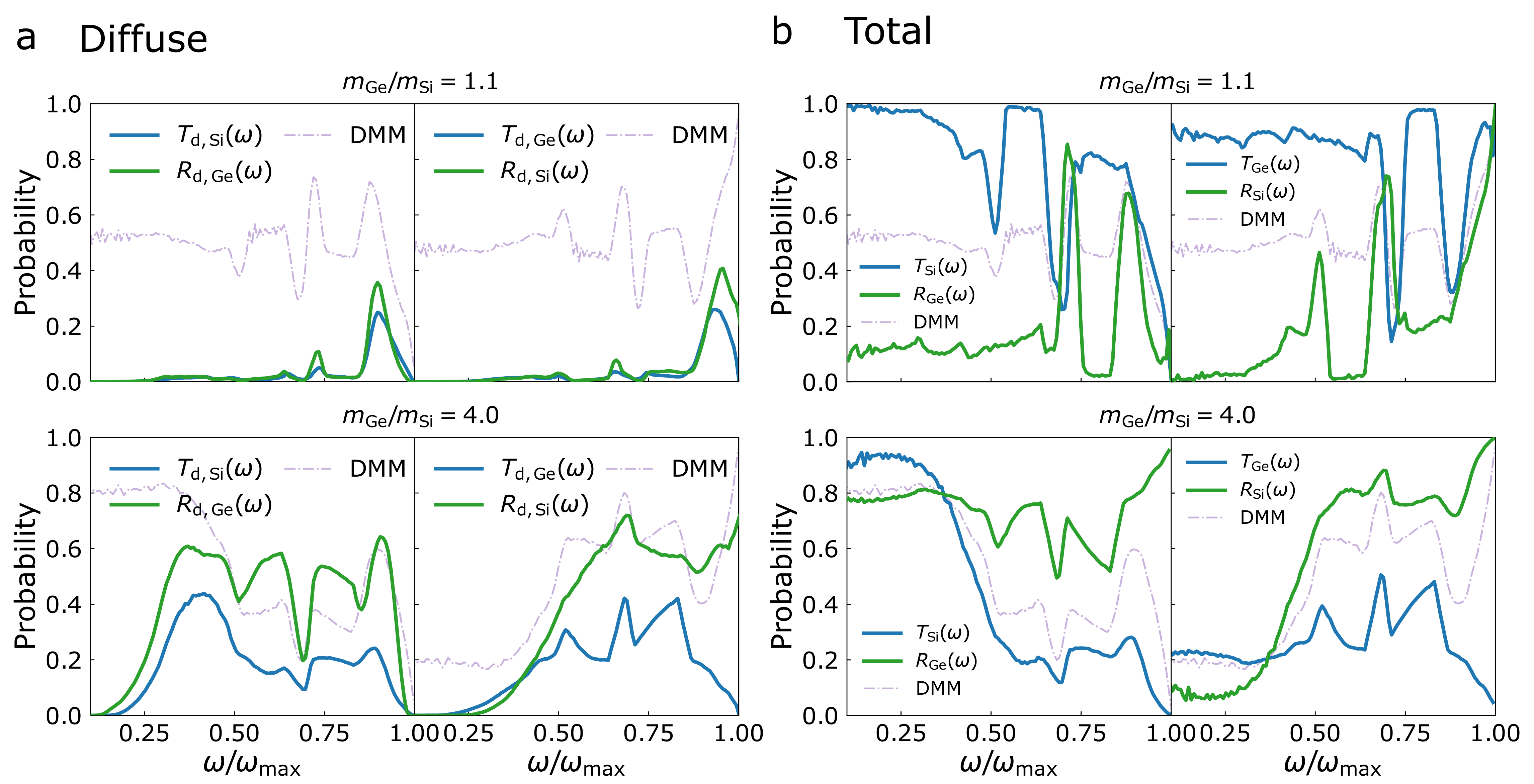}
    \caption{The frequency-resolved  (a) diffuse and (b) total transmittance/reflectance from Si and Ge side with different $m_\mathrm{Ge}/m_\mathrm{Si}$ from AGF calculation for 8 ml structures and DMM. $\omega_\mathrm{max}$ is the maximum allowed frequency for non-zero transmission function $\Theta(\omega)$.  $\omega_\mathrm{max} $ = 17.2 THz when $m_\mathrm{\mathrm{Ge}}/m_\mathrm{\mathrm{Si}} = 1.1$, and $\omega_\mathrm{max} $ = 9.1 THz when $m_\mathrm{\mathrm{Ge}}/m_\mathrm{\mathrm{Si}} = 4.0$.}
    \label{mass}
\end{figure*}

From our continuum model, the diffuse transmittance for a given incident state from the left side writes,
\begin{equation}
    \begin{split}
        T_{\mathrm{d},L\to R}(\omega,\Omega_L) = \int \frac{d^2\mathbf{q}^\prime_\parallel}{(2\pi)^2} T^{L\to R}_{\mathrm{d}}(\omega,\mathbf{q}^\prime_\parallel,\mathbf{q}_\parallel)
    \end{split}
    \label{diffuset}
\end{equation}
where
\begin{equation}
    \begin{split}
    &T^{L\to R}_{\mathrm{d}}(\omega,\mathbf{q}^\prime_\parallel,\mathbf{q}_\parallel) =\\
    & 4\omega^{-2}V_2\frac{\rho_R 
    v^\prime_R}{\left|\rho_Lv^\prime_L+\rho_Rv^\prime_R\right|^2}\frac{\rho_Lv_L}{{\left|\rho_Lv_L+\rho_Rv_R\right|^2}}
\end{split}
\end{equation}
Here, $V_2$ is a quantity related to density of mixed atoms at interface (see Appendix~\ref{sec:appendix1}).
Essentially, we have integrated over all possible final states $\mathbf{q}_\parallel^\prime$ on the right-hand side.
Note that the final transverse momentum is bounded by $|\mathbf{q}^\prime_\parallel| \le \frac{\omega}{c_R}$.

The diffuse reflectance for a phonon state from the left-hand side is,
\begin{equation}
    \begin{split}
        R_{\mathrm{d},L\to L}(\omega,\Omega_L) = \int \frac{d^2\mathbf{q}^\prime_\parallel}{(2\pi)^2} R^{L\to L}_{\mathrm{d}}(\omega,\mathbf{q}^\prime_\parallel,\mathbf{q}_\parallel)
    \end{split}
    \label{diffuser}
\end{equation}
where
\begin{equation}
    \begin{split}
       & R^{L\to L}_{\mathrm{d}}(\omega,\mathbf{q}^\prime_\parallel,\mathbf{q}_\parallel)=\\
    & 4\omega^{-2}V_2\frac{\rho_Lv^\prime_L}{\left|\rho_Lv^\prime_L+\rho_Rv^\prime_R\right|^2}\frac{\rho_L 
    v_L}{\left|\rho_Lv_L+\rho_Rv_R\right|^2}
    \end{split}
\end{equation}
and the transverse momentum of the final state is bounded by $|\mathbf{q}^\prime_\parallel| \le \frac{\omega}{c_L}$. 

If we denote $F(\omega) = \int \frac{d^2\mathbf{q}^\prime_\parallel}{(2\pi)^2} 4\omega^{-2}V_2 \frac{\rho_Lv^\prime_L}{\left|\rho_Lv^\prime_L+\rho_Rv^\prime_R\right|^2}$, 
the diffuse transmittance from one side and the diffuse reflectance from the other side can be respectively expressed by,
\begin{equation}
    T_{\mathrm{d},R\to L}(\omega,\Omega^\prime_R)=\frac{\rho_Rv^\prime_R}{\left|\rho_Lv^\prime_L+\rho_Rv^\prime_R\right|^2}F(\omega)
\end{equation}
\begin{equation}
    R_{\mathrm{d},L\to L}(\omega,\Omega_L)=\frac{\rho_Lv_L}{\left|\rho_Lv_L+\rho_Rv_R\right|^2}F(\omega)
\end{equation}
It is evident that they are both anisotropic since they depend on perpendicular incident velocity.
Their ratio writes,
\begin{equation}
    \frac{T_{\mathrm{d},R\to L}(\omega,\Omega^\prime_R)}{R_{\mathrm{d},L\to L}(\omega,\Omega_L)}=\frac{\rho_Rv^\prime_R}{\rho_Lv_L}\frac{\left|\rho_Lv_L+\rho_Rv_R\right|^2}{\left|\rho_Lv^\prime_L+\rho_Rv^\prime_R\right|^2}
\end{equation}
which is not a constant.
Thereby, our continuum model suggests that the diffuse transmittance from one side and the diffuse reflectance from the other side are generally not equal.

\section{Results and discussions} \label{sec:3}

We study phonon transport through a disordered [001] Si/Ge interface by creating a 3$\times$3 supercell (along x and y direction). We use the average of Si's and Ge's lattice constants, $a = 5.527$ \AA, as the lattice constant in generating the supercell structures and the Stillinger-Weber inter-atomic potential to compute the dynamical matrix and Green's function\cite{bi2012thermal}. The interface is constructed by randomly swapping Si and Ge atoms with the same distances to the interface and the further away from the interface the fewer atoms are swapped. At even further distances from the interface, no Si and Ge atoms are swapped. For instance, when we have 2 layers of Si and 2 layers of Ge atoms are mixed, 2 pairs of Si and Ge atoms will be swapped in the Si and the Ge layer closest to the interface and 1 pair of Si and Ge atoms will be swapped in the Si and the Ge layer secondly closest to interface. In this case, we have a 1|2|2|1 configuration, with each number denoting the number of swapped atoms within the same layer. We label such interface structure by 4 ml, in short for 4 mixing layers in total. Apparently, a larger ml number means a larger degree of disorder. We generate 21 configurations for each given total mixing layers and compute the ensemble average of the transmission and reflection probability matrix $T^{L\to R}_{mn}(\omega)$, $T^{R\to L}_{mn}(\omega)$, $R^{L\to L}_{mn}(\omega)$ and $R^{R\to R}_{mn}(\omega)$. 

We first examine the angular dependence of ensemble-averaged diffuse transmittance and reflectance at a given frequency to ascertain whether or not they are isotropic. 
In Fig.~\ref{f3}, we find strong angle dependence of diffuse transmittance and diffuse reflectance from the Si side as well as the Ge side, contradicting the assumption of isotropic reflectance and transmittance underlying the DMM. The diffuse reflectance from Ge to Ge is found to be overall higher than the diffuse transmittance from Si to Ge. The diffuse reflectance from Si to Si is in a similar range compared with the diffuse transmittance from Ge to Si, although their explicit angle dependences are drastically different.

If the DMM is valid, the diffuse transmittance from one side and the diffuse reflectance from the other side should be isotropic. We compute the difference between the maximum and minimum scattering probability at different frequencies, as a measure of anisotropy, shown in Fig.~\ref{f3} (e) and (f). We  observe that the diffuse scattering probability generally varies in a wide range. For example, the diffuse reflectance from the Si side at high frequencies ranges almost from zero to one. Hence, we conclude the diffuse transmittance and reflectance from both sides are highly anisotropic. 

Additionally, we find out that the patterns of diffuse transmittance and diffuse reflectance do not have two perfect diagonal reflection axes as the perfect Si/Ge [001] interface structure does. This is the consequence of ensemble average of  disordered structures, where each structure might have broken the reflection symmetry, and it is not guaranteed to recover the original symmetry after ensemble average. Still, we can clearly observe a clover-like pattern for the diffuse transmittance and reflectance from the Si side, which originates from the pmm symmetry of the phonon bands for a perfect [001] Si/Ge interface.

In particular, to quantitatively study how the diffuse transmittance and diffuse reflectance depend on the polar angle, we integrate the angle-dependent scattering probability over the azimuthal angle. As shown in Fig.~\ref{f4} (a), at $\omega= 3.3 $ THz, the diffuse transmittance from Si is lower than the reflectance from Ge, while in Fig.~\ref{f4} (b), the diffuse transmittance from Ge is higher than the diffuse reflectance from Si. It is interesting to note that the continuum model agrees well with the results from AGF when incident states are from Si, suggesting the model captures the physics of diffuse phonon scattering. On the contrary, the continuum model predicts a different diffuse scattering probability profile compared with the calculation from AGF when incident phonons are from Ge. From the analytical expression in the continuum model for diffuse scattering probability (Eq.~\ref{diffuset} and Eq.~\ref{diffuser}), 
we learn that the peak in diffuse transmittance and reflectance profile corresponds to the critical angle for total internal reflection (37 degrees).  In comparison, the scattering probability profile from AGF has two peaks (9 degrees and around 50 degrees). In Fig.~\ref{f4} (e) and (f), we find that the specular transmittance and reflectance have a much stronger dependence on the polar angle compared with their diffuse counterparts. Our continuum model for specular scattering probabilities again shows good agreement with the AGF calculation except incapable of capturing the multiple peaks arising from total internal reflection. Note that the continuum modeling is based on a scalar field, where the polarization vectors are not included. The difference between continuum modeling and AGF calculation suggests that without considering mode conversion among different polarizations, the scalar continuum model cannot accurately describe the actual number of available diffuse transmission and reflection pathways. 

To study the diffuse transmittance at different frequencies, we define an average diffuse transmittance for phonon modes with the same frequency by,
\begin{equation}
    T_{\mathrm{d},\alpha}(\omega) = \frac{\sum^{+}_n T^{\alpha\to \beta}_{\mathrm{d},n}(\omega)v_{z,n}}{\sum^{+}_n 1\cdot v_{z,n}}= \frac{\Theta_\mathrm{d}(\omega)}{\Theta_{\mathrm{bulk},\alpha}(\omega)}
    \label{freqavg}
\end{equation}
where $v_{z,n}$ is the group velocity normal to the interface. $\Theta_{\mathrm{bulk},\alpha}(\omega)$ is the transmission function for bulk material $\alpha$ (we use two leads and device all consisting of $\alpha$ atoms in AGF calculation).  
The average total transmittance including specular transmittance and diffuse transmittance is defined by, $T_{\alpha}(\omega)=\frac{\Theta(\omega)}{\Theta_{\mathrm{bulk},\alpha}(\omega)}$. The average diffuse and total reflectance are $R_{\mathrm{d},\alpha}(\omega)=\frac{\Xi_{\mathrm{d},\alpha}(\omega)}{\Theta_{\mathrm{bulk},\alpha}(\omega)}$ and $R_{\alpha}(\omega)=\frac{\Xi_{\alpha}(\omega)}{\Theta_{\mathrm{bulk},\alpha}(\omega)}$, respectively. Note that the sum of transmittance and reflectance is one, yet the sum of diffuse transmittance and diffuse reflectance is less than one, as not all phonons are diffusely scattered. On the other hand, the sum of transmittance and reflectance given by DMM is always unity. Therefore, when we compare the diffuse transmittance/reflectance with transmittance/reflectance by DMM, it is entirely possible their values do not match. However, what we are more interested in answering is whether or not the phonon loses its memory, \textit{i.e.}, whether the diffuse transmittance from one side equals the diffuse reflectance from the other side\cite{renorm}.

In Fig.~\ref{f4} (c), we find that the diffuse transmittance from Si is lower than the diffuse reflectance from Ge for all frequencies. They both deviate from DMM given by Eq.~\ref{tdmm} at low frequencies. At high frequencies, the DMM's prediction is close to the reflectance from Ge. In Fig.~\ref{f4} (d), the diffuse transmittance from Ge is lower than reflectance from Si except for low frequencies. And they are both different from DMM. The crossing point at 4.4 THz for transmittance and reflectance suggest that at this frequency, the average transmittance from Ge side is the same with average diffuse reflectance from Si side, although they individually have strong angle dependence. The gap between diffuse transmittance from one side and diffuse reflectance from the other side suggests that diffuse phonon scattering depends on the initial state such that phonons actually do not lose their memory of origin. In Fig.~\ref{f4} (g) and (h), we see that the specular scattering probability is generally much higher than the diffuse scattering probability at low frequencies, suggesting that at low frequencies, the interface scattering is almost all specular. This trend is also consistent with previous findings\cite{PhysRevB.96.174302,PhysRevB.97.205306}.  

An important factor that is relevant to the diffuse phonon scattering is the amount of dissimilarity between two materials.  According to Pohl and Swartz\cite{RevModPhys.61.605}, DMM predicts that diffuse scattering increases thermal boundary resistance of the interface between similar solids, suggesting that diffuse scattering plays a significant role when the mass ratio between the two materials is close to one. In Fig.~\ref{mass}, we present the frequency-resolved transmittance and reflectance with different mass ratios $m_\mathrm{Ge}/m_\mathrm{Si}$ (we fix the mass $m_\mathrm{Si}$ and vary the mass $m_\mathrm{Ge}$). We find that when the two sides are similar, the diffuse transmittance from one side is similar to the  diffuse reflectance from the other side. This partially aligns with the assumptions of DMM, although the diffuse transmittance and diffuse reflectance still have strong anisotropy in directions (see Fig. 6 in the supplementary material). When the mass ratio is large, the transmittance from one side is no longer similar to the reflectance from the other side, indicating that the scattering probability strongly depends on where the initial states are from. As for the total transmittance and reflectance, due to the inclusion of the specular scattering probability, the difference between transmittance from one side and reflectance from the other side is enlarged. It is interesting to note that when two sides are similar, although DMM cannot correctly describe either scattering probability, it is close to the average of the total transmittance from one side and the total reflectance from the other side. When the two sides are dissimilar, the DMM's prediction becomes similar to the total transmittance in Ge side at low frequencies and the total transmittance in Si side at high frequencies. This suggests that for certain cases, DMM is able to roughly describe the total transmittance from one side in a certain frequency range but not for all frequencies.

\begin{figure}[t!]
    \includegraphics[width=0.46\textwidth]{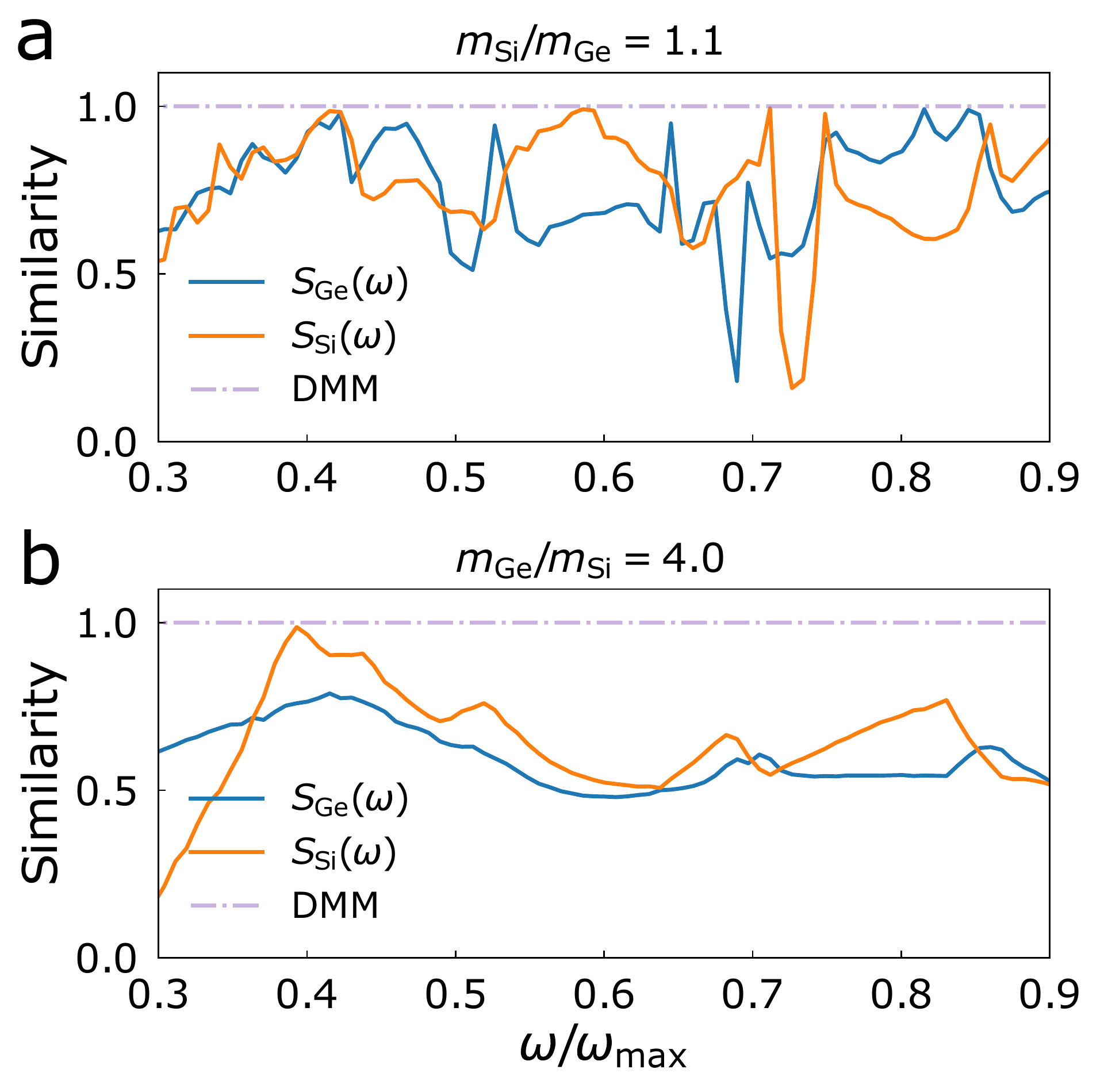}
    \caption{The similarity between the diffuse transmittance from one side and the diffuse reflectance from the other side. The blue lines correspond to the cases when the final states are on the Ge side and the orange lines correspond to the cases where the final states are on the Si side. The diffuse transmittance and reflectance for evaluating the similarity are from the data presented in Fig.~\ref{mass} (a). }
    \label{similar}
\end{figure}

\begin{figure*}[t!]
    \includegraphics[width=0.89\textwidth]{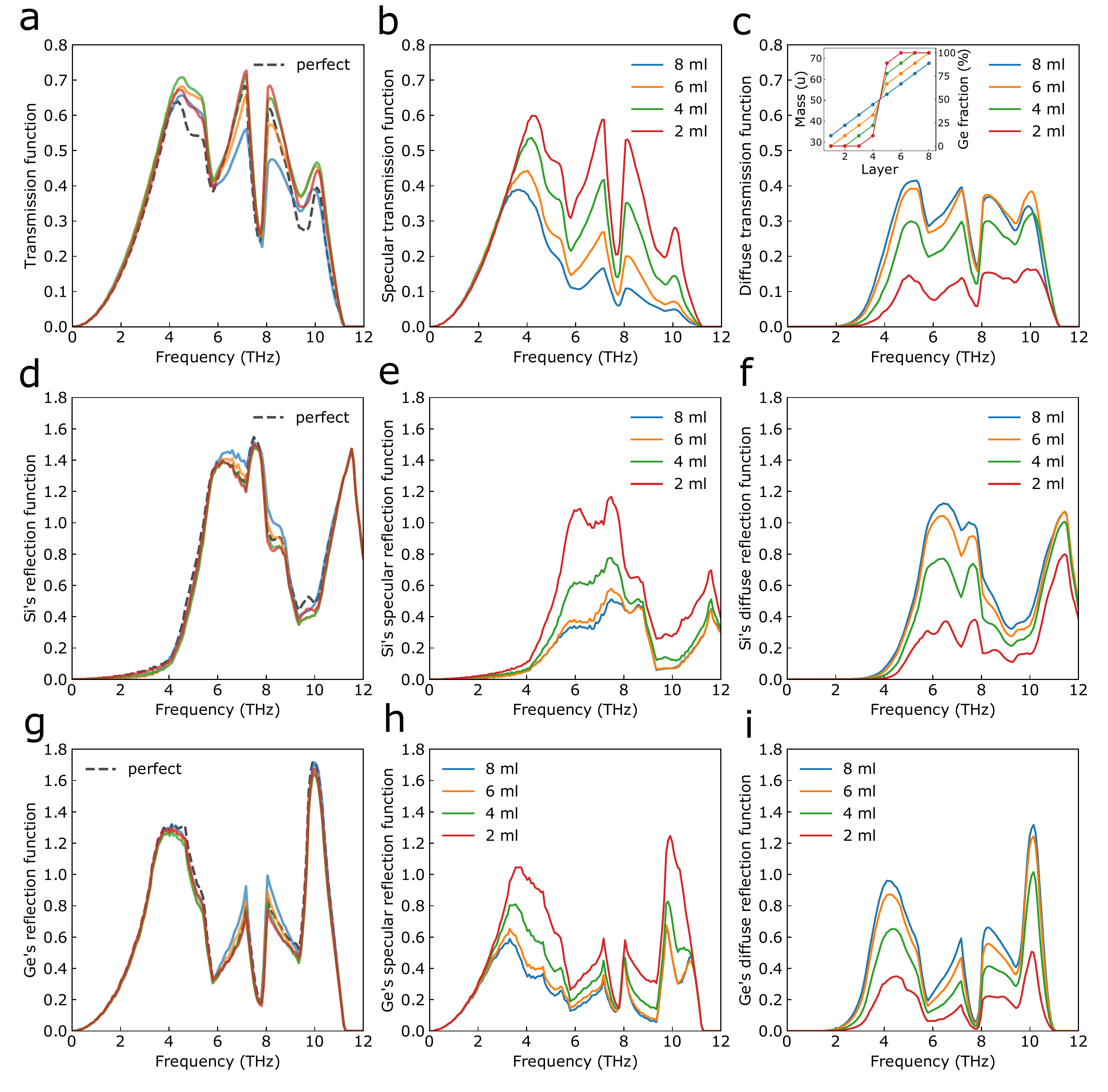}
    \caption{(a) The total transmission function, (b) the specular transmission function and (c) the diffuse transmission function versus phonon frequency for [001] Si/Ge interfaces from AGF. Inset: the ensemble-averaged mass profile as a function of atom layer number at the interface region obtained by averaging the mass of atoms within the same layer. The right axis is the Ge fraction in each atom layer. The distance between adjacent atom layers is $a/4 = 1.382$ \AA. (d)-(f) The total, specular and diffuse reflection function from the Si side. (g)-(i) The total, specular and diffuse reflection function from the Ge side.}
    \label{f2}
\end{figure*}

To study how much memory the phonon loses regarding its origin in a quantitative manner, we define a similarity measure by,
\begin{equation}
    S_\beta(\omega) = \mathrm{exp}\left(-\frac{\left|T_{\mathrm{d},\alpha}(\omega)-R_{\mathrm{d},\beta(\omega)}\right|}{R_{\mathrm{d},\beta}(\omega)}\right)
\end{equation}
where $T_\alpha(\omega)$ is the diffuse transmittance from one side and $R_\beta(\omega)$ is the diffuse reflectance from the other side, and the final states for these scattering processes are on the $\beta$ side. $S_\beta \to 1$ means high similarity between the diffuse transmittance and reflectance when the final state is on the $\beta$ side, \textit{i.e.} phonon completely loses its memory of origin. $S_\beta \to 0$ means low similarity between the diffuse
transmittance from one side and diffuse reflectance from the other side, and phonon does
not lose its memory.
A large mass ratio generally lead to a smaller similarity measure. We also find that the similarity measure depends which side the final state resides at. For example, in Fig.~\ref{similar} (b), we find that at $\omega=0.39 \omega_\mathrm{max}$, when phonon is scattered into Si side, it loses its memory. However, this is not true when phonon is scattered into Ge side. In comparison, for DMM, the similarity is always one when the final states are on either side of the interface. 

Lastly, we want to discuss the diffuse phonon scattering's role in interfacial transport.
In Fig.~\ref{f2} (a), we plot the total transmission function as a function of frequency with different numbers of mixing layers. And we find that the total transmission can either be enhanced or reduced compared with the perfect Si/Ge interface as a result of the competition between the specular transmission versus diffuse transmission. From Fig.~\ref{f2} (b) and (c), we find as the degree of disorder increases, the specular transmission decreases while the diffuse transmission increases. In other words, the disorders remove specular channels while creating new diffuse channels. As a result, we do not observe significant changes in the transmission function due to atomic mixing. The opposite trends for specular and diffuse transmission versus the degree of disorder eventually leads to the maximum thermal conductance for 4 ml structures. 
It is interesting to note that similar enhancements for transmission enabled by disorders have been discovered in electron transport in heterostructures\cite{PhysRevLett.92.106103}.
From Fig.~\ref{f2} (d)-(i), we learn that the increasing amount of disorders always reduce the specular reflection function and increases the diffuse reflection function. Thus, the disorders can both increase the diffuse transmission (enhancing the interface conductance) and the diffuse reflection (worsening the interface conductance). This competition is another reason why we cannot observe significant enhancement of thermal conductance by disorders. 
Our current analysis is based on elastic scattering, yet phonon anharmonicity can contribute to interfacial transport by enabling ``vertical coupling'' between conduction channels of different frequencies. There are several works on understanding the anharmonicity's role in thermal interface conductance for abrupt interfaces\cite{doi:10.1063/1.4859555,PhysRevB.99.045301,PhysRevB.101.041301,PhysRevB.103.174306}, but how diffuse scattering and anharmonicity affect heat conduction channels remains unclear.

\section{Conclusion}\label{sec:4}
Through AGF calculation, we have demonstrated that the diffuse phonon scattering by a single disordered interface depends both on initial incoming states as well as the final outgoing states. The transmittance and reflectance strongly depend on the polar angle of group velocity. Also, the transmittance from one side and reflectance from the other side are generally different. That is to say, phonons do not lose their memory after diffuse scattering by a single interface. When two materials are similar, the diffuse transmittance from one side and the diffuse reflectance from the other side become similar to each other. However, the total transmittance from one side and total reflectance from the other side are still different.

The number of specular transmission channels for interfacial transport is always reduced by interfacial disorders while new transmission channels are created by diffuse phonon scatterings. The competing roles of specular and diffuse transmission can lead to either enhanced or reduced transmission function and interfacial thermal conductance.

We also derived the expressions for transmittance and reflectance for diffuse scattering processes based on the continuum approximation, which works reasonably well in the low-frequency range.  The model leads to different analytical expressions for diffuse transmittance from one side and diffuse reflectance from the other side. Our model also shows that the diffuse transmission opens up new transmission channels even for those states above the critical angle for total internal reflection.
\section*{Acknowledgements}
We want to thank Dr. Jiawei Zhou and Dr. Jonathan Mendoza for helpful discussions. The funding for this work is supported by the MRSEC Program of the National Science Foundation under award number DMR-1419807.
\begin{appendix}
\section{The continuum modeling of diffuse phonon scattering} \label{sec:appendix1}
\subsection{The transmission and reflection matrix}  

%The actual interface of two materials is usually not  perfect. The roughness near 
%the interface can be regarded as random mass disorders.

We consider interfacial transport in the case of scalar phonon model in the continuum limit, where the equation of motion (EOM) for displacements
writes\cite{eason1969wave},
\begin{equation}
\rho(\mathbf{r})\frac{\partial^2 u}{\partial t^2}-\nabla\cdot\left(\mu(z)\nabla u\right) = 0
\label{eom}
\end{equation}
where $\mu(z)$ is the bulk modulus. For an interface between two dissimilar solids, we have $\mu(z) = \mu_L$ for $z<0$ and $\mu(z) = \mu_R$ for $z\ge0$.  $\rho(\mathbf{r}) = \rho_0(z)+\Delta \rho(\mathbf{r})$ is the density, where $\rho_0(z)$ is the density without mass disorder at the interface, expressed by $\rho_0(z<0)=\rho_L$ and $\rho_0(z>0)=\rho_R$. $\Delta \rho(\mathbf{r})=\sum_i \Delta m_i \delta(\mathbf{r}_{\parallel}-\mathbf{r}_{i,\parallel})\delta(z)$ is the density fluctuations due to atomic mixing at the interface, where $\Delta m_i$ is the change of mass at atom site $i$ and $\mathbf{r}_{i,\parallel}$ is the in-plane position of atom site $i$. The choice of the delta function form for density fluctuations suggests that the atomic mixing only exists exactly at the interface, thus our model does not apply to the cases where the atomic mixing exists even far away from the interface. When atomic mixing is realized by randomly swapping pairs of atoms on two sides of the interface,
the average of mass fluctuations is zero $\sum_i \Delta m_i = 0$. The mass disorders are distributed  randomly in the x-y plane.
The ensemble average of any physical quantity $P$ over many configurations of mass disorders is obtained by integrating over all possible positions of mass disorders,
\begin{equation}
  \begin{split}
  \langle P  \rangle = \int \prod_j \frac{d^2\mathbf{r}_{\parallel,j}}{A} P
   \end{split}
   \label{esmavg}
\end{equation}
where  $A$ is the cross-section area.
After ensemble average, the average of mass fluctuations is still zero $\langle\sum_i \Delta m_i = 0\rangle$.
We further assume an independent distribution of mass fluctuations such that, 
\begin{equation}
  \langle\sum_{i,j} \Delta m_i \Delta m_j\rangle = \sum_{i} \langle\left(\Delta m_i\right)^2\rangle
  \label{m2avg}
\end{equation}  

The time-harmonic solution of Eq.~\ref{eom} reads,
\begin{equation}
u=\sum_{\mathbf{q}_\parallel}u_{\mathbf{q}_\parallel}(z)\frac{e^{i\left(-\omega t +\mathbf{q}_\parallel\cdot \mathbf{r}_\parallel\right)}}{\sqrt{A}}
\end{equation}
where $\mathbf{q}_\parallel=(q_x,q_y)$ is the transverse wavevector, $\omega$ is the phonon frequency, $\mathbf{r}_\parallel=(x,y)$ is the transverse position  and $u_{\mathbf{q}_\parallel}(z)$ is the z-dependent component of the solution.
%The transverse wavevectors are the same for the left and right side of the interface due to the transverse periodicity. 
The perpendicular wavevectors $q_{\perp}(z)=q_{z,L}$, when $z< 0$, and $q_{\perp}(z)=q_{z,R}$, when $z\ge 0$, are determined by the dispersion relation $\omega^2 = c^2_{L/R}(q^2_\parallel+q_{\perp,L/R}^2)$. Here, $c_{L/R}$ is the sound velocity defined by $c_{L/R}=\sqrt{\mu_{L/R}/\rho_{L/R}}$.
Plugging in the time-harmonic solution to EOM, we have the following equation for $u_{\mathbf{q}_\parallel}(z)$,
\begin{equation}
    \sum_{\mathbf{q}_\parallel}\left[\left(\rho_0(z)+\Delta\rho(\mathbf{r})\right)\omega^2+\nabla\cdot\mu(z)\nabla\right]u_{\mathbf{q}_\parallel}(z)\frac{e^{i\mathbf{q}_\parallel\cdot\mathbf{r}_\parallel}}{\sqrt{A}}=0
    \label{EOM}
\end{equation}
Multiply both sides of Eq.~\ref{EOM} by $\int d^2\mathbf{r}_\parallel e^{-i\mathbf{q}_\parallel^\prime\cdot\mathbf{r}_\parallel}/\sqrt{A}$. The orthogonality relations for plane waves leads to,
\begin{equation}
    \left[\mu(z)q_\perp^2(z)+\frac{\partial}{\partial z}\mu(z)\frac{\partial}{\partial z}\right]u_{\mathbf{q}_\parallel^\prime}
    = \sum_{\mathbf{q}_\parallel} M_{\mathbf{q}^\prime_\parallel,\mathbf{q}_\parallel}\delta(z)u_{\mathbf{q}_\parallel}
    \label{uz}
\end{equation}
The scattering matrix   $M_{\mathbf{q}^\prime_\parallel,\mathbf{q}_\parallel}$
is defined by,
\begin{equation}
M_{\mathbf{q}^\prime_\parallel,\mathbf{q}_\parallel} = -\sum_{i} \Delta m_i\omega^2{A}^{-1}e^{i\left(\mathbf{q}_\parallel-\mathbf{q}^\prime_\parallel\right)\cdot\mathbf{r}_{\parallel,i}}.
\label{scatm}
\end{equation}
The  solution  to Eq.~\ref{uz} is expressed by,
\begin{equation}
    \begin{split}
    u_{\mathbf{q}_\parallel}(z)&= \delta_{\mathbf{q}_\parallel,\mathbf{q}^\prime_\parallel}e^{iq_L z}+r_{\mathbf{q}_\parallel,\mathbf{q}^\prime_\parallel}\frac{\sqrt{v^\prime_L}}{\sqrt{v_L}}e^{-iq_L z}, \;z\leq0    \\
    u_{\mathbf{q}_\parallel(z)}&=t_{\mathbf{q}_\parallel,\mathbf{q}^\prime_\parallel}\frac{\sqrt{\rho_L v^\prime_L}}{\sqrt{\rho_R v_R }}e^{iq_R z}, \;z \geq 0\\    
    \end{split}
    \label{uqp}
\end{equation}
where $q_{L/R}=q_{z,L/R}$ is the phonon wavevector normal to the interface. $t_{\mathbf{q}_\parallel,\mathbf{q}^\prime_\parallel}$ is the transmission matrix and 
$r_{\mathbf{q}_\parallel,\mathbf{q}^\prime_\parallel}$ is the reflection matrix. $v_{L/R} =  c_{L/R} \cos\theta_{L/R}$ is the group velocity perpendicular to the interface, where $\theta_{L/R}$ is the angle between the direction of phonon velocity and the axis normal to the interface. 

The transmission and reflection probability matrix are defined by the ratio of the transmitted flux normal to the interface of phonon $\mathbf{q}_\parallel$ and the reflected flux  normal to the interface of phonon $\mathbf{q}_\parallel$ to the incident flux normal to the interface of phonon state $\mathbf{q}^\prime_\parallel$,  $T_{\mathbf{q}_\parallel,\mathbf{q}^\prime_\parallel} =\frac{J_{\mathrm{t},\mathbf{q}_\parallel}}{J_{\mathrm{inc},\mathbf{q}^\prime_\parallel}}$ and $R_{\mathbf{q}_\parallel,\mathbf{q}^\prime_\parallel} =\frac{J_{\mathrm{r},\mathbf{q}_\parallel}}{J_{\mathrm{inc},\mathbf{q}^\prime_\parallel}}$, respectively, where the time-averaged energy flux for a phonon mode reads\cite{sgwave},
\begin{equation}
    J = \frac{\mu(z)}{-i\omega A} \int \left(u^*\frac{\partial u}{\partial z} -u\frac{\partial u^*}{\partial z} \right)d^2\mathbf{r}_\parallel
\end{equation}
The resultant expressions for transmission probability matrix and reflection probability matrix are,
\begin{subequations}
\begin{equation}
    T_{\mathbf{q}_\parallel,\mathbf{q}^\prime_\parallel} = \left|t_{\mathbf{q}_\parallel,\mathbf{q}^\prime_\parallel}\right|^2 
    \label{tandt}
\end{equation}
\begin{equation}
    R_{\mathbf{q}_\parallel,\mathbf{q}^\prime_\parallel} = \left|r_{\mathbf{q}_\parallel,\mathbf{q}^\prime_\parallel}\right|^2
    \label{randr}
\end{equation}
\end{subequations}
The boundary conditions for displacement $u_{\mathbf{q}_\parallel}(z)$ for solving the transmission and reflection matrix write,
\begin{subequations}
\begin{equation}
       u_{\mathbf{q}_\parallel} (0^-) =u_{\mathbf{q}_\parallel} (0^+)
\end{equation}
\begin{equation}
\mu(z)\frac{\partial}{\partial z}u_{\mathbf{q}_\parallel}(z)\Big\vert^{0^+}_{0^-}=\sum_{\mathbf{q}^{\prime\prime}_\parallel}M_{\mathbf{q}_\parallel,\mathbf{q}^{\prime\prime}_\parallel} u_{\mathbf{q}^{\prime\prime}_\parallel}(0),
    \label{bdc}
\end{equation}
\end{subequations}
where the second boundary condition  is obtained by integrating Eq.~\ref{uz} from $-\eta$ to $\eta$, with $\eta \to 0^+$. 
Plug in the expression in Eq.~\ref{uqp} into the boundary condition. We obtain the following expressions,
\begin{subequations}
    \begin{equation}
        \delta_{\mathbf{q}_\parallel,\mathbf{q}^\prime_{\parallel}}+r_{\mathbf{q}_\parallel,\mathbf{q}^\prime_{\parallel}} = t_{\mathbf{q}_\parallel,\mathbf{q}^\prime_{\parallel}} 
        \label{rela11}
    \end{equation}
\begin{equation}
    \sum_{\mathbf{q}^{\prime\prime}_\parallel}(\delta_{\mathbf{q}_\parallel,\mathbf{q}^{\prime\prime}_\parallel}
    +i\Gamma_{\mathbf{q}_\parallel,\mathbf{q}^{\prime\prime}_\parallel})t_{\mathbf{q}^{\prime\prime}_\parallel,\mathbf{q}^{\prime}_\parallel}=\Lambda_{\mathbf{q}_\parallel,\mathbf{q}^\prime_\parallel}
    \label{rela}
\end{equation}
\end{subequations}
where 
\begin{subequations}
\begin{equation}
    \Gamma_{\mathbf{q}_\parallel,\mathbf{q}^{\prime\prime}_\parallel}= \frac{M_{\mathbf{q}_\parallel,\mathbf{q}^{\prime\prime}_\parallel}}{2\omega\bar{\rho}\bar{v}}\sqrt{\frac{v_R}{v_R^{\prime\prime}}}
    \label{gammaqq}
  \end{equation}
  \begin{equation}
    \Lambda_{\mathbf{q}_\parallel,\mathbf{q}^\prime_\parallel} = \delta_{\mathbf{q}_\parallel,\mathbf{q}^\prime_\parallel}\frac{\sqrt{v_Lv_R}}{\bar{v}}
  \end{equation}
  \begin{equation}
    \bar{v}=\frac{\rho_L v_L + \rho_R v_R}{2\bar{\rho}}
  \end{equation}
  \begin{equation}
    \bar{\rho} = \sqrt{\rho_L\rho_R}
\end{equation}
\end{subequations}
In particular, we can reorganize Eq.~\ref{rela} and identify that the transmission matrix can be expanded in series as,
    \begin{equation}
       t=\sum_{N=0}^\infty (-i\Gamma)^N\Lambda.
       \label{series}
    \end{equation}
which is a summation of terms arising from multiple scatterings of different orders.
We can discard high-order terms to obtain the approximate expression for the transmission matrix.
\subsection{The Green's function  in the continuum limit}    
The transmission matrix can be computed from the Green's function of the whole system. We choose to compute the Green’s function because of the mathematical convenience in perturbation expansions using Dyson’s equation. In the following, we will illustrate the exact relationship between the transmission  matrix and the Green's function.

To start with, we evaluate the unperturbed  Green's function for a disorder-free interface. The unperturbed EOM writes,
\begin{equation}
\left[\rho_0(z)\omega^2+\frac{\partial}{\partial z}\mu(z)\frac{\partial}{\partial z}\right]u(z)=0
\end{equation}
which can be identified as a Sturm-Liouville equation. Two sets of solutions are given by,
\begin{equation}
u_<(z)=\begin{cases}
t_L e^{-iq_L z},&z<0\\
e^{-iq_R z}+r_L e^{iq_R z},&z>0\\
\end{cases}
\end{equation}
and
\begin{equation}
u_>(z)=\begin{cases}
t_R e^{iq_R z},&z>0\\
e^{iq_L z}+r_R e^{-iq_L z},&z<0\\
\end{cases}
\label{unp}
\end{equation}
%where $q_{L/R}$ is the momentum perpendicular to the interface.
The continuity condition at interface gives,
\begin{equation}
t_L = 1+r_L = \frac{2\mu_R q_R}{\mu_L q_L+\mu_R q_R}
\end{equation}
and
\begin{equation}
t_R = 1+r_R = \frac{2\mu_L q_L}{\mu_L q_L+\mu_R q_R}
\end{equation}
For a Sturm-Liouville equation, the
Wronskian  writes\cite{copley201412},
\begin{equation}
    \begin{split}
W & =u_<(z)\frac{du_>(z)}{z}-u_>(z)\frac{du_<(z)}{z}\\
 &=\begin{cases}\frac{4i\mu_L q_Lq_R}{\mu_Lq_L + \mu_Rq_R},&z>0\\
\frac{4i\mu_R q_Lq_R}{\mu_Lq_L + \mu_Rq_R},&z<0\end{cases}
\end{split}
\end{equation}
and the Green's function is defined by,
\begin{equation}
    \begin{split}
G_0(z,z')=&\begin{cases}\frac{u_<(z)u_>(z^\prime)}{\mu(z^\prime)W(z^\prime)}, & -\infty < z < z'\\
\frac{u_<(z')u_>(z)}{\mu(z')W(z')}, & z'<z<\infty
\end{cases}\\
=& \begin{cases}
    -\frac{i}{2\mu_L}\frac{t_R}{q_L}e^{-iq_L z}e^{iq_R z'}, &z<0, z'>0\\
    -\frac{i}{2\mu_R}\frac{t_L}{q_R}e^{-iq_L z'}e^{iq_R z}, &z>0, z'<0\\
    -\frac{i}{2\mu_L}\frac{e^{-iq_L |z'-z|}+r_Re^{-iq_L (z+z')}}{q_L}, &z<0, z'<0\\
    -\frac{i}{2\mu_R}\frac{e^{iq_R |z'-z|}+r_Le^{iq_R (z+z')}}{q_R}, &z>0, z'>0
    \end{cases}
\end{split}
\end{equation}
When $z$ and $z^\prime$ both approaches zero,  
the unperturbed Green's function at interface is
\begin{equation}
    G_0^+                                                                     = -\frac{i}{\mu_L q_L +\mu_R q_R} =-\frac{i}{2\omega\bar{\rho}\bar{v}}
\end{equation}
where the superscript +  is added  to represents the retarded Green's function.

Then, we study the Green's function for the scenario where atomic mixing is present at the interface. It is convenient to define the Green's function operator, 
\begin{equation}
    \hat{G}^\pm = \left[\rho(\mathbf{r})\omega^2-\hat{K}\pm i\eta\right]^{-1}
\end{equation}
where the operator $\hat{K} = -\nabla\cdot\mu(z)\nabla$ and $\eta$ is an infinitesimal positive real number. The Green's function in the real space representation can then be
expressed by $G^\pm(\mathbf{r},\mathbf{r}^\prime)=\bra{\mathbf{r}}\hat{G}^\pm\ket{\mathbf{r}^\prime}$.  The Green's function that describes the scattering channel between mode $a$ of left side  and mode $b$ of right side is,
\begin{equation}
    \begin{split}
    G_{b,a}^\pm(z,z^\prime) 
    = &A^{-1}\iint d \mathbf{r}_\parallel d\mathbf{r}^\prime_\parallel e^{-i\left(\mathbf{q}_\parallel\cdot\mathbf{r}_\parallel-\mathbf{q}^\prime_\parallel\cdot\mathbf{r}^\prime_\parallel\right)}\\
    &\times\bra{u_{\mathbf{q}_\parallel}}\hat{G}^\pm\ket{u^\prime_{\mathbf{q}^\prime_\parallel}}\\
    %=&G^\pm_{\mathbf{q}_\parallel,\mathbf{q}^\prime_\parallel}.
    \end{split}
\end{equation}
where the transverse wavevector for mode $a$ and mode $b$ are $\mathbf{q}^\prime_\parallel$ and $\mathbf{q}_\parallel$, respectively.

%In the following discussion, we hide subscript $a$ and $b$ for clarity.

When disorders are introduced,  the perturbed eigenvector is related to the unperturbed eigenvector via\cite{messiah1962quantum}, 
\begin{equation}
    u = u^\prime + \int G^+(\mathbf{r},\mathbf{r}^\prime)V(\mathbf{r}^\prime) u^\prime(\mathbf{r}^\prime) d\mathbf{r}^\prime,
    \label{lip}
\end{equation}
where $u^\prime$ is the eigenstate for the disorder-free case and the perturbation $V(\mathbf{r}) = -\Delta\rho(\mathbf{r}) \omega^2$. Specifically, the second term on the right-hand side of Eq.~\ref{lip} equals,
\begin{equation}
    \begin{split}
     &\int d\mathbf{r}^\prime G^+(\mathbf{r},\mathbf{r}^\prime) V(\mathbf{r}^\prime) u^\prime(\mathbf{r}^\prime)  \\
    =&\int d\mathbf{r}^\prime \bigg\{ G^+(\mathbf{r},\mathbf{r}^\prime) \hat{K}^\prime - \delta(\mathbf{r}-\mathbf{r}^\prime)-\left[\hat{K}^\prime  G^+(\mathbf{r},\mathbf{r}^\prime)\right]  \bigg\}u^\prime(\mathbf{r}^\prime)
    \end{split}
\end{equation}
Thus, Eq.~\ref{lip} is equivalent to,
\begin{equation}
    \begin{split}
    u %& = \int d\mathbf{r}^\prime  \bigg\{u^\prime(\mathbf{r}^\prime)\hat{K}^\prime  G(\mathbf{r},\mathbf{r}^\prime) - G(\mathbf{r},\mathbf{r}^\prime) \hat{K}^\prime u^\prime(\mathbf{r}^\prime)\bigg\}\\
      & = \int \mu(\mathbf{r}^\prime) \left(u^\prime \frac{\partial G^+}{\partial z^\prime}-G^+\frac{\partial u^\prime}{\partial z^\prime}\right)\hat{\mathbf{e}}_z\cdot d\mathbf{S}^\prime
    \end{split}
    \label{ugv}
\end{equation}

It is easy to show that,
\begin{equation}
    \begin{split}
    \bra{u^\prime}V\ket{u}_{b,a
    } &=\bra{\hat{K} u^\prime}\ket{u}_{b,a
    } -\bra{u^\prime}\hat{K} \ket{u}_{b,a
    }\\
    & = \int \left[ u_a \hat{K}  u^{\prime*}_b- u^{\prime*}_b \hat{K}  u_a     \right]d^3\mathbf{r}\\
    & = \int \mu(\mathbf{r})\left[ u^{\prime*}_b\frac{\partial u_a}{\partial z} - \frac{\partial u^{\prime*}_b}{\partial z}u_a\right]\hat{\mathbf{e}}_z\cdot d\mathbf{S}\\
    \end{split}
    \label{uvup}
\end{equation}
where we applied integration by part and the divergence theorem. 

Plugging in the expression of $u$ given by Eq.~\ref{ugv} into Eq.~\ref{uvup}, we have,
\begin{equation}
    \begin{split}
        &   \bra{u^\prime} V \ket{u}_{b,a} =  \int \left[ \hat{K} u^{\prime *}_b(\mathbf{r})\right]u_a(\mathbf{r}) d\mathbf{r}\\
    & -\iint u^{\prime *}_b(\mathbf{r})\mu(\mathbf{r}^\prime) \\
    & \times\left[u^\prime_a(\mathbf{r}^\prime)\frac{\partial \hat{K} G^+}{\partial z^\prime}-\frac{\partial u^\prime_a(\mathbf{r}^\prime)}{\partial z^\prime}\hat{K} G^+  \right]dS^\prime d\mathbf{r}\\
    =& \iint  \mu(\mathbf{r})\mu(\mathbf{r}^\prime)\big\{\frac{\partial u^{\prime *}_b(\mathbf{r}) }{\partial z}\\
    &\times\big[-u^{\prime}_a(\mathbf{r}^\prime  )\frac{\partial G^+}{\partial z^\prime}+ G^+\frac{\partial u^\prime_a(\mathbf{r}^\prime)}{\partial z^\prime}\big]\\ 
    &+u^{\prime *}_b(\mathbf{r})\big[u^\prime_a(\mathbf{r}^\prime)\frac{\partial^2 G^+}{\partial z\partial z^\prime}-\frac{\partial u^\prime_a(\mathbf{r}^\prime)}{\partial z^\prime}\frac{\partial G^+}{\partial z}\big] \big\}dS dS^\prime\\
    %&= \mu(\mathbf{r})\mu(\mathbf{r}^\prime)\big\{\frac{\partial u^{\prime *}_b(\mathbf{r}) }{\partial z}\\
    %\times&\big[-u^{\prime}_a(\mathbf{r}^\prime  )\frac{\partial G}{\partial z^\prime}+ G\frac{\partial u^\prime_a(\mathbf{r}^\prime)}{\partial z^\prime}\big]\\ 
    %&+u^{\prime *}_b(\mathbf{r})\big[-\frac{\partial u^\prime_a(\mathbf{r}^\prime)}{\partial z^\prime}\frac{\partial G}{\partial z}+u^\prime_a(\mathbf{r}^\prime)\frac{\partial^2 G}{\partial z\partial z^\prime}\big] \big\}\\
    %&= \iint  \mu(\mathbf{r}) \mu(\mathbf{r}^\prime)u^{\prime *}_b(\mathbf{r})u^\prime_a(\mathbf{r}^\prime)\\\
    %&\times \Big[ iq_b\left( \frac{\partial G}{\partial z^\prime} - iq_a G\right)+\left(-iq_a\frac{\partial G}{\partial z}+\frac{\partial^2 G}{\partial z \partial z^\prime}\right) dS dS^\prime\\
    %&= \iint \mu(\mathbf{r}) \mu(\mathbf{r}^\prime)u^{\prime *}(\mathbf{r})u^\prime(\mathbf{r}^\prime) \\
    %&\times\left(\frac{\partial}{\partial z}+iq_b\right)\left(\frac{\partial}{\partial z^\prime}-iq_a\right)G(\mathbf{r},\mathbf{r}^\prime)dS dS^\prime\\
    %& = 
    =& -4\mu_L\mu_R t_{b,R}^* e^{-ik_{b,R}z_2+ik_{a,L}z_1^\prime}G^+(z_2,z_1^\prime)q_{b,R}q_{a,L}\\
    &-4\mu^2_L r_{b,R}^* e^{ik_{b,L}z_1+ik_{a,L}z^\prime_1}G^+_1(z_1,z_1^\prime)q_{b,L}q_{a,L}\\
    & + 4\mu_L^2r^*_{b,R}e^{ik_{b,L}z_1-ik_{a,L}z_1^\prime} G^+_2(z_1,z_1^\prime)q_{b,L}q_{a,L}
    \end{split}
    \label{full}
\end{equation}
where $G^+(z_1,z_1^\prime) = G^+_1(z_1,z_1^\prime)+G^+_2(z_1,z_1^\prime)$ and we do not need to know the exact expression of $G^+_1$ and $G^+_2$. Note that we  have used the form of Green's function in its asymptotic limit  in deriving the above expression. Denote $L$ the length of the domain containing disorders. Then, $z_1, z_1^\prime<0$ and $z_2,z_2^\prime> L$ are the boundary for integration. Since the random mass disorders are localized at the interface at $z=0$, we have $L\to 0$, such that we can set $z_1 = z_1^\prime = 0^-$ and $z_2 = z_2^\prime = 0^+$.

Directly plugging in the general expression of $u^\prime$ for unperturbed system given by Eq.~\ref{unp} and $u$ for perturbed system given by Eq.~\ref{uqp}, we can obtain another expression for matrix element $\bra{u^\prime} V\ket{u}_{b,a}$,
\begin{equation}
    \begin{split}
        \bra{u^\prime} V\ket{u}_{b,a}   = & 2i\omega t^*_{b,R} t_{b,a}\sqrt{\rho_L v_{a,L}\rho_R v_{b,R}}\\
         +&2i\omega\left(r^*_{b,R} r_{b,a}\frac{\sqrt{v_{a,L}}}{\sqrt{v_{b,L}}}-\delta_{b,a} \right) \rho_L v_{b,L}    
    \end{split}
    \label{uvu}             
\end{equation}

By equating Eq.~\ref{full} to Eq.~\ref{uvu}, we have found the relationship between transmission matrix and Green's function,
\begin{equation}
    \begin{split}
t_{b,a}&=2i\omega G^+_{b,a}(0^+,0^-)\sqrt{\rho_L \rho_R v_{a,L} v_{b,R}}\\
 &=2i\omega \bar{\rho}\sqrt{ v_R v^\prime_L}G^+_{\mathbf{q}_{\parallel},\mathbf{q}^\prime_{\parallel}}
    \end{split}
    \label{tandG}
\end{equation}
Furthermore, using the boundary condition given by Eq.~\ref{rela11}, we find that
the reflection matrix is related to Green's function through,
\begin{equation}
    \begin{split}
    r_{b,a} &= 2i\omega G^+_{b,a}(0^-,0^-) \rho_L\sqrt{v_{a,L}v_{b,L}} - \delta_{ba}.
    %G(0^-,0^-) =\frac{i\delta_{ba}}{2\omega r^*_{b,R}\rho_L v_{L,a}}-\frac{ir_{ba}}{2\omega\rho_L\sqrt{v_{a,L}v_{b,L}}}.\\
    \\
    &=2i\omega \rho_L\sqrt{ v_Lv^\prime_L }G^+_{\mathbf{q}_{\parallel},\mathbf{q}^\prime_{\parallel}}-\delta_{\mathbf{q}_{\parallel},\mathbf{q}^\prime_{\parallel}}
    \end{split}
    \label{reflg}
\end{equation}

%The Eq.~\ref{lip} and Eq.~\ref{uvu} lead to the expression,
%\begin{equation}
    %\begin{split}
     %   &\bra{u_{b,0}}V\ket{u_{a,0}}+\bra{u_{b,0}}G^+VG^+\ket{u_{a,0}}=\\
     %= &\bra{\hat{K}  u_{0,b}}G^+V\ket{u_{0,a}}-\bra{ u_{0,b}}\hat{K}G^+V\ket{u_{0,a}}
%    \end{split}
%\end{equation}

%\begin{equation}
%t_{\mathbf{q}_{\parallel},\mathbf{q}^\prime_{\parallel}}=
%\label{TT}
%\end{equation}
%and for reflection matrix,
%\begin{equation}
%    r_{\mathbf{q}_{\parallel},\mathbf{q}^\prime_{\parallel}}=2i\omega \rho_L\sqrt{ v_Lv^\prime_L }G^\dagger_{\mathbf{q}_{\parallel},\mathbf{q}^\prime_{\parallel}}-\delta_{\mathbf{q}_{\parallel},\mathbf{q}^\prime_{\parallel}}
%\label{RR}
%\end{equation}
%which are proved in Appendix~\ref{relation}. 
%This suggests that the Green's function plays an essential role in interface scattering. We will focus on the calculation of Green's function in the following. To start with, we explicitly write down the one-particle Green's function while taking into account the all possible atomic configurations of the disorder interface,
\subsection{The ensemble averaged Green's function}

From the series expansion of the transmission matrix in Eq.~\ref{series} and relationship between transmission matrix and Green's function given by Eq.~\ref{tandG}, we can obtain the following series for the ensemble averaged Green's function,
\begin{equation}
    \begin{split}
\langle G^+_{\mathbf{q}_{\parallel},\mathbf{q}^\prime_{\parallel}} \rangle
% &=\sum_{\mathbf{q}^{\prime\prime}_{\parallel}} -i\frac{\delta_{\mathbf{k}^{\prime\prime}_{\parallel},\mathbf{q}^\prime_{\parallel}}}{2\sqrt{\mu_2q_2\mu_1q^{\prime}_1}}\frac{\sqrt{v_1^{\prime\prime} v_2^{\prime\prime}}}{\bar{v}^{\prime\prime}}\bigg\langle \sum_N^\infty (-i\Gamma)^N\bigg\rangle_{\mathbf{q}_{\parallel},\mathbf{q }^{\prime\prime}_{\parallel}} \\
      &=  -i\frac{1}{2\omega\bar{\rho}\bar{v}^{\prime}}\frac{\sqrt{ v_R^{\prime}}}{\sqrt{v_R}}\sum_N^\infty \bigg\langle (-i\Gamma)^N\bigg\rangle_{\mathbf{q}_{\parallel},\mathbf{q }^{\prime}_{\parallel}}
    \end{split}
    \label{avgi}
\end{equation}
According to Eq.~\ref{esmavg}, the ensemble average of matrix $\left(-i\Gamma\right)^N$ is obtained by integrating over
all possible impurity positions,
\begin{equation}
\begin{split}
\langle \left(-i\Gamma\right)^N \rangle = \int \prod_j \frac{d^2\mathbf{r}_{\parallel,j}}{A} \left(-i\Gamma\right)^N
\end{split}
\end{equation}
%For the case of $N = 1$, $\langle -i\Gamma\rangle  = 0 $. For the case of $N = 2$, we find that 
%\begin{equation}
%\langle (-i\mathbf{\Gamma})^2\rangle_{\mathbf{q}_{\parallel},\mathbf{q}^\prime_{\parallel}} =
%\end{equation}
In the weak perturbation limit, using  Eq.~\ref{m2avg}, Eq.~\ref{scatm} and Eq.~\ref{gammaqq}, we write down the Green's function in Eq.~\ref{avgi} up to the second order,
\begin{equation}
\langle G^+_{{\mathbf{q}_{\parallel},\mathbf{q}^\prime_{\parallel}}}\rangle = \left(G^+_0(\mathbf{q}^\prime_\parallel) +  G^{+2}_0(\mathbf{q}^\prime_\parallel)\frac{V_2}{A} \sum_{\mathbf{q}_\parallel^{\prime\prime}}G^+_0(\mathbf{q}^{\prime\prime}_\parallel)\right)\delta_{\mathbf{q}_{\parallel},\mathbf{q}^\prime_{\parallel}}
\label{g0V}
\end{equation}
where $V_2 = \langle\sum_{i} \frac{m_i^2}{A} \omega^4\rangle$.
The first-order term vanishes due to ensemble average $\langle\sum_i m_i \omega^2\rangle = 0$.
The diagonal form of Eq.~\ref{g0V} implies that the ensemble average  recovers the in-plane translational symmetry of the unperturbed Green's function.

\subsection{The transmission and reflection probability matrix}
From Eq.~\ref{tandt}, we find that the transmission probability matrix is related to the product of retarded and advanced Green's fuction, 
\begin{equation}
    \langle T_{{\mathbf{q}_{\parallel},\mathbf{q}^\prime_{\parallel}}}\rangle = 4\omega^2 \bar{\rho}^2 v^\prime_L v_R \langle  G^+_{{\mathbf{q}_{\parallel},\mathbf{q}^\prime_{\parallel}}}G^-_{{\mathbf{q}_{\parallel},\mathbf{q}^\prime_{\parallel}}}\rangle.
    \label{tg}
\end{equation}
The ensemble averaged $G^+_{{\mathbf{q}_{\parallel},\mathbf{q}^\prime_{\parallel}}}G^-_{{\mathbf{q}_{\parallel},\mathbf{q}^\prime_{\parallel}}}$ can be expressed by,
\begin{equation}
    \begin{split}
\langle &G^+_{\mathbf{q}_{\parallel},\mathbf{q}^\prime_{\parallel}}G^- _{\mathbf{q}_{\parallel},\mathbf{q}^\prime_{\parallel}}\rangle = |\langle G^+_{\mathbf{q}_{\parallel},\mathbf{q}^\prime_{\parallel}} \rangle |^2 \\
&+  \sum_{\mathbf{q}^{\prime\prime}_{\parallel},\mathbf{q}^{\prime\prime\prime}_{\parallel}} |\langle G^+(\mathbf{q}_{\parallel},\mathbf{q}^{\prime\prime}_{\parallel}) \rangle |^2 W_{\mathbf{q}^{\prime\prime}_{\parallel},\mathbf{q}^{\prime\prime\prime}_{\parallel}} |\langle G^+(\mathbf{q}^{\prime\prime\prime}_{\parallel},\mathbf{q}^\prime_{\parallel}) \rangle |^2
%\frac{1}{4\mu_1\mu_2k_1^\prime k_2}\bigg\langle \left(\sum_{N=0}^\infty(-i\Gamma)^N\right)\left(\sum_{N^\prime=0}^\infty(i\Gamma^*)^{N^\prime}\right)\bigg\rangle \frac{k^\prime_1 k^\prime_2}{\bar{k}^2}
    \end{split}
    \label{GW}
\end{equation}
where the term $W_{\mathbf{q}^{\prime\prime}_{\parallel},\mathbf{q}^{\prime\prime\prime}_{\parallel}} $  is called the reducible vertex function. To the lowest order, the vertex function reads\cite{PhysRevB.68.014433},
%\begin{equation}
%    \begin{split}
%        &\sum_{\mathbf{q}^{\prime\prime}_{\parallel},\mathbf{q}^{\prime\prime\prime}_{\parallel}} |\langle G^\dagger_{\mathbf{q}_{\parallel},\mathbf{q}^{\prime\prime}_{\parallel}} \rangle |^2 W_{\mathbf{q}^{\prime\prime}_{\parallel},\mathbf{q}^{\prime\prime\prime}_{\parallel}} |\langle G^\dagger_{\mathbf{q}^{\prime\prime\prime}_{\parallel},\mathbf{q}^\prime_{\parallel}} \rangle |^2  \\
%        &= | G^\dagger_0(\mathbf{q}_{\parallel}) |^2 \frac{V_2}{A} | G^\dagger_0(\mathbf{q}^\prime_{\parallel}) |^2
%\frac{1}{4\mu_1\mu_2k_1^\prime k_2}\bigg\langle \left(\sum_{N=0}^\infty(-i\Gamma)^N\right)\left(\sum_{N^\prime=0}^\infty(i\Gamma^*)^{N^\prime}\right)\bigg\rangle \frac{k^\prime_1 k^\prime_2}{\bar{k}^2}
%    \end{split}
%    \label{2nd}
%\end{equation}
\begin{equation}
    W_{\mathbf{q}^{\prime\prime}_{\parallel},\mathbf{q}^{\prime\prime\prime}_{\parallel}}  = \frac{V_2}{A}
    \label{2nd}
\end{equation}
From Eq.~\ref{g0V} to Eq.~\ref{2nd},
we obtain the expression for the transmission probability matrix,
\begin{equation}
    \begin{split}
        \langle &T_{{\mathbf{q}_{\parallel},\mathbf{q}^\prime_{\parallel}}}\rangle =\delta_{\mathbf{q}_{\parallel},\mathbf{q}^\prime_{\parallel}}\frac{4\rho_L\rho_Rv_Lv_R}{\left|\rho_Lv_L+\rho_Rv_R\right|^2}\\
        &\times\left[1-2\mathrm{Re}\,\mathcal{G}^+  \omega^{-1}V_2 \left(\frac{1}{\rho_Lv_L+\rho_Rv_R}\right)\right]\\ % \mathrm{Re}
     %&+ 
     %2\mathrm{Im}\mathcal{G}^+ V_2 \mathrm{Im}\left(\frac{1}{\rho_Lv_L+\rho_Rv_R}\right)\Big]\\
     &+\frac{4\omega^{-2}V_2}{A}\frac{\rho_Rv_R}{\left|\rho_Lv_L+\rho_Rv_R\right|^2}\frac{\rho_Lv^\prime_L}{{\left|\rho_Lv^\prime_L+\rho_Rv^\prime_R\right|^2}}
    \end{split}
    \label{T}
\end{equation}
where $\mathcal{G}^+  =  \left(\sum_{\mathbf{q}_\parallel^{\prime\prime}}i G^{+}_0(\mathbf{q}^{\prime\prime}_\parallel)\right)/A$ 
and the analytical expression for $\mathcal{G}^+$ can be found in the next session. We identify that the diagonal term is responsible for specular transmission, while the off-diagonal term is responsible for diffuse transmission. Similarly, from Eq.~\ref{randr}, Eq.~\ref{reflg} and Eq.~\ref{g0V} , we derive that the reflection probability matrix writes,
\begin{equation}
    \begin{split}
        &\langle R_{\mathbf{q}_{\parallel},\mathbf{q}^\prime_{\parallel}}\rangle = \delta_{\mathbf{q}_{\parallel},\mathbf{q}^\prime_{\parallel}}\frac{\left|\rho_Lv_L-\rho_Rv_R\right|^2}{\left|\rho_Lv_L+\rho_Rv_R\right|^2}\\
        &\times \left[1-4\mathrm{Re}\,\mathcal{G}^+\omega^{-1}V_2 \mathrm{Re} \left(\frac{\rho_L v_L}{\rho^2_Lv^2_L - \rho^2_R v^2_R}\right)\right]\\
    %&+4\mathrm{Im} \mathcal{G}^+V_2 \mathrm{Im} \left(\frac{\rho_L v_L}{\rho^2_Lv^2_L - \rho^2_R v^2_R}\right) \Big]\\
    &+\frac{4\omega^{-2}V_2}{A}\frac  {\rho_L 
    v_L}{\left|\rho_Lv_L+\rho_Rv_R\right|^2}\frac{\rho_Lv^\prime_L}{{\left|\rho_Lv^\prime_L+\rho_Rv^\prime_R\right|^2}}.
    \end{split}
    \label{R}
\end{equation}
\subsection{The analytical expression of \texorpdfstring{$\mathcal{G}^+$}{G}}
The term $\mathcal{G}^+$ in Eq.~\ref{T} and Eq.~\ref{R} can be written in terms of an integral over all transverse wavevectors,
\begin{equation}
\begin{split}
    &\mathcal{G}^+\\
& =\int \frac{d^2 \mathbf{q}_\parallel}{(2\pi)^2} \frac{1}{\sqrt{\mu_L(\rho_L \omega^2-\mu_Lq^2_\parallel)}+\sqrt{\mu_R(\rho_R \omega^2-\mu_Rq^2_\parallel)}}
\end{split}
\end{equation}
Introduce the ratio of bulk moduli as $a = \rho_R/\rho_L$ and $b=\mu_R/\mu_L$. 
%Also define, $a = \frac{\rho_L\omega^2}{\mu_L}$ and $c =  \frac{\rho_R\omega^2}{\mu_R}$, $d=\frac{b^2(a-c)}{1-b^2}$ and $f = \frac{c(1-b^2)}{a-c}$.
Depending on the bulk moduli and densities  of two sides, the expression for the real part of $\mathcal{G}^+$ is as follows. When $(b-a)(1-b^2)>0$, 
\begin{equation}
\begin{split}
    \mathrm{Re}\,\mathcal{G}^+
%&=\frac{\pi}{\mu_L}\frac{2}{1-b^2}\Big[\sqrt{a}-b\sqrt{c}+\sqrt{d}\\
%&\times\left(\mathrm{arctan}\sqrt{f}-\mathrm{arctan}\sqrt{\frac{a}{d}}\right)\Big]\\
& = \frac{\omega\mathcal{G}^+_0}{1-b^2}\Big[1-\sqrt{ab}+\sqrt{\frac{b(b-a)}{1-b^2}}\\
&\times\left(\mathrm{atan}\sqrt{\frac{a(1-b^2)}{b-a}}-\mathrm{atan}\sqrt{\frac{1-b^2}{b(b-a)}}\right)\Big]
\end{split}
\end{equation}
where $\mathcal{G}^+_0 = \frac{1}{2\pi\mu_L}\sqrt{\frac{\rho_L}{\mu_L}}$ and for Si, $\mathcal{G}^+_0 =2.62\times 10^{-16} \mathrm{s^3/kg}$.
When $(b-a)(1-b^2)<0$,
\begin{equation}
\begin{split}
    \mathrm{Re}\, \mathcal{G}^+
%&= \frac{A}{4\pi^2}\frac{\pi}{\mu_L}\frac{2}{1-b^2}\Big[\sqrt{a}-b\sqrt{c}+\frac{\sqrt{-d}}{2}\\
%&\times\left(\mathrm{ln}\frac{|\sqrt{a}-\sqrt{-d}|}{\sqrt{a}+\sqrt{-d}}-\mathrm{ln}\frac{|1-\sqrt{-\frac{1}{f}}|}{1+\sqrt{-\frac{1}{f}}}\right)\Big]\\
&  = \frac{\omega\mathcal{G}^+_0}{1-b^2}\Big[1-\sqrt{ab}+\frac{1}{2}\sqrt{\frac{b(a-b)}{1-b^2}}\\
&\times\left(\mathrm{ln}\frac{|1-\sqrt{\frac{b(a-b)}{1-b^2}}|}{1+\sqrt{\frac{b(a-b)}{1-b^2}}}-\mathrm{ln}\frac{|1-\sqrt{\frac{a-b}{a(1-b^2)}}|}{1+\sqrt{\frac{a-b}{a(1-b^2)}}}\right)\Big]
\end{split}
\end{equation}
When $b=1$ and $a\neq 1$,
\begin{equation}
    \mathrm{Re}\, \mathcal{G}^+= \frac{\omega\mathcal{G}^+_0}{3}\frac{1-a^{\frac{3}{2}}}{1-a} % \frac{2\pi}{3\mu_L}\frac{a^{\frac{3}{2}}-c^{\frac{3}{2}}}{a-c}=
\end{equation}
When $a= b$,
\begin{equation}
    \mathrm{Re}\, \mathcal{G}^+= \frac{\omega\mathcal{G}^+_0}{1+b} % \frac{2\pi\sqrt{a}}{\mu_L(1+b)} 
\end{equation}

\subsection{The specular and diffuse transmittance/reflectance}

The transmittance for a given initial state $\mathbf{q}_\parallel$ is defined by summing transition probabilities to different final states $\mathbf{q}^\prime_\parallel$,
\begin{equation}
    T_{L\to R}(\mathbf{q}_\parallel)=\sum_{\mathbf{q}_\parallel^\prime}\langle T_{\mathbf{q}^\prime_\parallel,\mathbf{q}_\parallel}\rangle
    \label{tltor}
\end{equation}
where $\langle T_{\mathbf{q}^\prime_\parallel,\mathbf{q}_\parallel}\rangle$ is the transmission probability matrix defined in Eq.~\ref{T}.

In the following, we will use direction $\Omega_L=(\theta_L,\phi_L)$ to denote a phonon state $\mathbf{q}_\parallel$, where $(\mathbf{q}_\parallel,q_L)=\frac{\omega}{c_L}(  \mathrm{sin}\theta_L\mathrm{cos}\phi, \mathrm{sin}\theta_L\mathrm{sin}\phi,\mathrm{cos}\theta_L)$. Note that the group velocity is parallel to the wavevector thus the angles for the group velocity and the wavevector are the same.
After integration, the transmittance in Eq.~\ref{tltor} is given by,
\begin{equation}
    \begin{split}
        T_{L \to R}(\Omega_L) 
        &=T_{\mathrm{s},L \to R}(\Omega_L)+T_{\mathrm{d},L \to R}(\Omega_L)\\
        &=T_{\mathrm{AMM}} (\Omega_L) p_T(\Omega_L) +T_{\mathrm{d},L \to R}(\Omega_L)\\ % \mathrm{Re}
     %&+ 
     %2\mathrm{Im}\mathcal{G}^+ V_2 \mathrm{Im}\left(\frac{1}{\rho_Lv_L+\rho_Rv_R}\right)\Big]\\
    \end{split}
    \label{Tana}
\end{equation}
 The first term is the specular transmittance, which is the product of transmittance from AMM\cite{khla,little1959transport},
\begin{equation}
    T_{\mathrm{AMM}}(\Omega_L) = \frac{4\rho_L\rho_Rv_Lv_R}{\left|\rho_Lv_L+\rho_Rv_R\right|^2}
\end{equation}
and the specularity parameter for transmittance,
\begin{equation}
    p_T(\Omega_L) = 1-2\mathrm{Re}\,\mathcal{G}^+  \omega^{-1}V_2 \left(\frac{1}{\rho_Lv_L+\rho_Rv_R}\right)
    \label{pt}
\end{equation}
The second term, diffuse transmittance, is given by Eq.~\ref{diffuset} in Sec.~\ref{sec:2}.

Similarly, the reflectance for a given incident state from the left side is given by,
\begin{equation}
    \begin{split}
        &R_{L\to L}(\Omega_L) =\sum_{\mathbf{q}^\prime_\parallel}\langle R_{\mathbf{q}^\prime_\parallel,\mathbf{q}_\parallel}\rangle\\
        &=R_{\mathrm{s},L\to L}(\Omega_L)+R_{\mathrm{d},L\to L}(\Omega_L)\\
        &=R_{\mathrm{AMM}} (\Omega_L) p_R(\Omega_L) +R_{\mathrm{d},L\to L}(\Omega_L)\\ % \mathrm{Re}
     %&+ 
     %2\mathrm{Im}\mathcal{G}^+ V_2 \mathrm{Im}\left(\frac{1}{\rho_Lv_L+\rho_Rv_R}\right)\Big]\\
    \end{split}
    \label{Rana}
\end{equation}
where $\langle R_{\mathbf{q}^\prime_\parallel,\mathbf{q}_\parallel}\rangle$ is the transmission probability matrix defined in Eq.~\ref{R}.
The reflectance by AMM writes,
\begin{equation}
    R_{\mathrm{AMM}}(\Omega_L) = \frac{\left|\rho_Lv_L-\rho_Rv_R\right|^2}{\left|\rho_Lv_L+\rho_Rv_R\right|^2}
\end{equation}
The specularity parameter for reflectance is given by,
\begin{equation}
    p_R(\Omega_L) = 1-4\mathrm{Re}\,\mathcal{G}^+\omega^{-1}V_2 \mathrm{Re} \left(\frac{\rho_L v_L}{\rho^2_Lv^{2}_L - \rho^2_R v^{2}_R}\right)
    \label{pr}
\end{equation}
And the diffuse reflectance $R_{d,L\to L}(\Omega_L)$ is defined by Eq.~\ref{diffuser}.
\begin{figure}[t!]
    \includegraphics[width=0.5\textwidth]{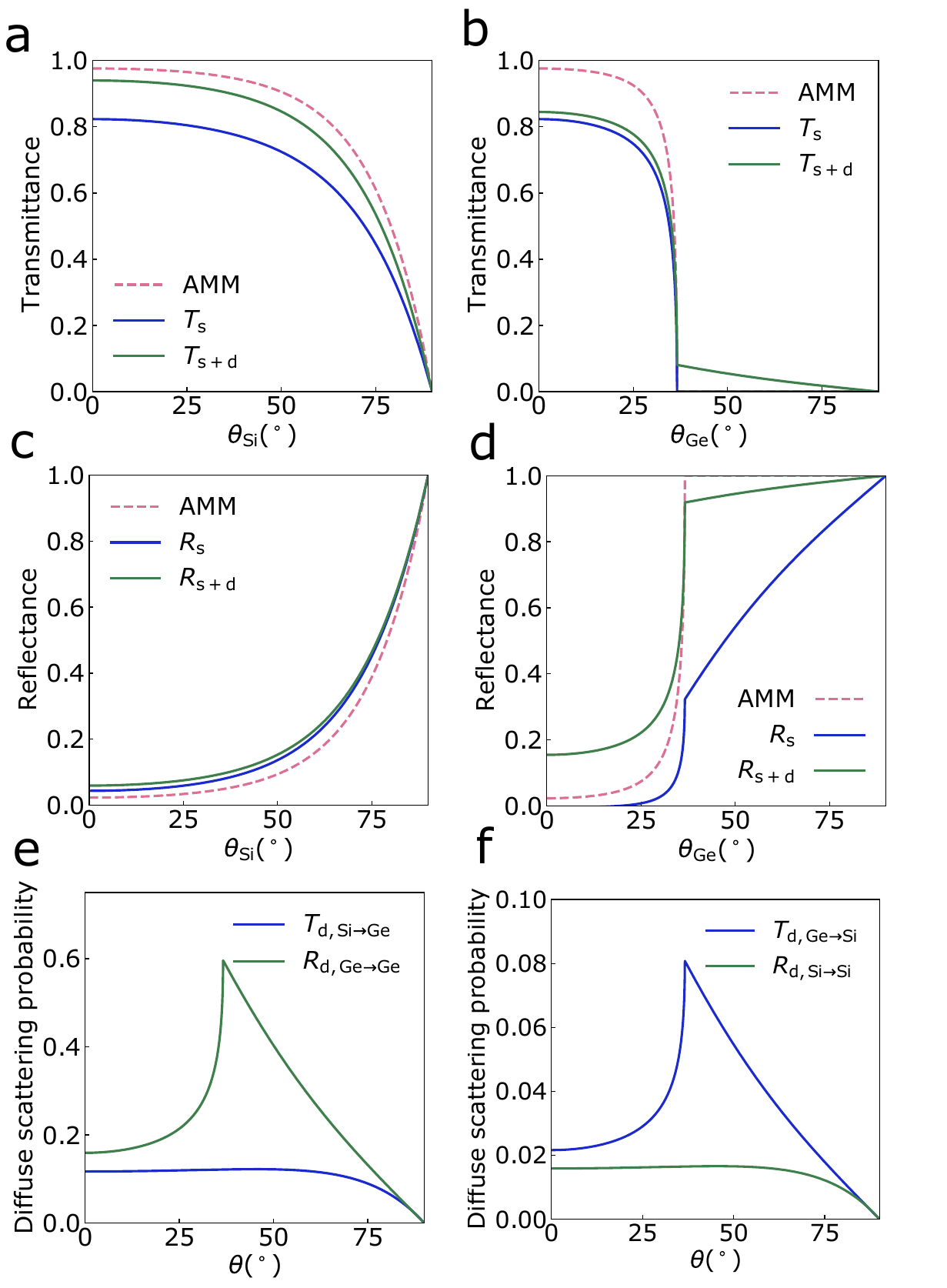}
    \caption{The transmittance and reflectance for a rough Si/Ge interface predicted from  the continuum model. (a) and (b): The specular and diffuse transmittance of acoustic phonons at 4 THz in Si and Ge compared with AMM. (c) and (d): The specular and diffuse reflectance of acoustic phonons at 4 THz in Si and Ge compared with AMM. (e) and (f): The diffuse transmittance from one side and reflectance from the other side of acoustic phonons at 4 THz. $\theta$ is the velocity angle of the incident state. }
    \label{f1}
\end{figure}
We want to stress that the expressions in Eq.~\ref{Tana} and Eq.~\ref{Rana} add up to one in the current lowest-order perturbation theory, which means our continuum model is a self-consistent theory. However, this is not a guaranteed property at higher orders.

\begin{figure*}[t]
    \includegraphics[width=0.9\textwidth]{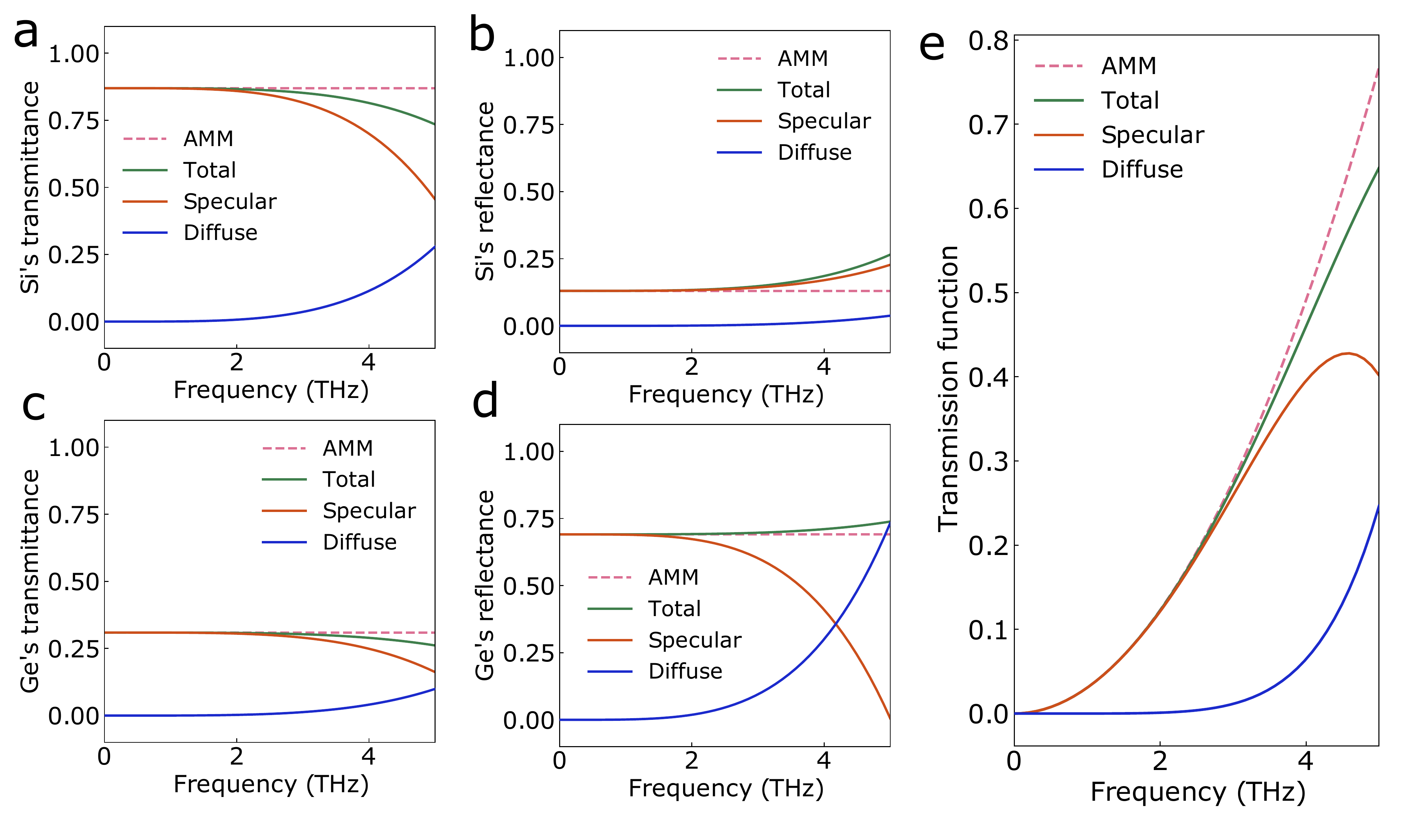}
    \caption{(a)-(d) The frequency-resolved transmittance and reflectance from Si side and Ge side from continuum modeling. (e) The transmission function $\Theta(\omega)$ for Si/Ge interface as a function of frequency. A multiplicity factor of three is multiplied in the transmission function as there are three acoustic phonon branches. When the frequency is much higher than 5 Thz, the lowest perturbation theory is no longer valid, as the perturbed part becomes large.}
\label{transmissionfunction}
\end{figure*}

In the previous study of partially specular and partially specular interface scattering by a disordered interface\cite{PhysRevB.57.14958}, the transmittance and reflectance are often phenomenologically written as,
\begin{subequations}
\begin{equation}
T(\Omega) = p(\Omega) T_\mathrm{AMM}(\Omega)+(1-p(\Omega)) T_\mathrm{DMM}(\Omega)
\label{classical1}
\end{equation}
\begin{equation}
R(\Omega) = p(\Omega) R_\mathrm{AMM}(\Omega)+(1-p(\Omega)) R_\mathrm{DMM}(\Omega)
\label{classical2}
\end{equation}
\end{subequations}
where $p$ is the specularity parameter calculated by Ziman's equation\cite{ziman2001electrons}. However, in our continuum modeling, there are two specularity parameters, one for transmittance (Eq.~\ref{pt}), one for reflectance (eq.~\ref{pr}) and they are generally not equal. The necessity of two specularity parameters has been hypothesized by Li \textit{el al}\cite{li2015phonon} and our analytical model gives direct support for the hypothesis of two different specularity parameters. What's more, it is entirely possible to have $p_R$ in our model larger than one (this is also observed in AGF calculation presented in the supplementary material), while the specularity parameter $p$ given by Ziman's equation is bounded by one. Thus, the specularity parameter is merely a correction factor and cannot be interpreted as probability of being specularly scattered.

Similar to Eq.~\ref{freqavg}, we further compute the frequency-resolved average transmittance by integrating over solid angle,
\begin{equation}
T_{L\to R}(\omega) =  2 \int_0^{\pi/2} d\theta \mathrm{sin} \theta \mathrm{cos}\theta T_{L\to R}(\Omega_L)  
\end{equation}
and the reflectance can be similarly computed.
The energy-resolved transmission function, which measures the number of conduction channels for interfacial thermal transport, is obtained by,
\begin{equation}
    \begin{split}
    \Theta(\omega) &=  A \int \frac{d^2\mathbf{q}_\parallel}{(2\pi)^2} T_{L\to R}(\omega,\mathbf{q}_\parallel)\\
    &=2\pi A\int \frac{d^2\mathbf{q}_\parallel d q_{L}}{(2\pi)^3}T_{L\to R}(\Omega_L)v_L\delta(\omega-c_L\sqrt{q^2_\parallel+q^2_L})\\
     &= \pi AD_L(\omega)c_L\int_0^{\pi/2} d\theta \mathrm{sin} \theta \mathrm{cos}\theta T_{L\to R}(\Omega_L)
    \end{split}
\end{equation}
where $D_L(\omega) = \frac{\omega^2}{2\pi^2c_L^3}$ is the density of states of left side and $\mathrm{sin}\theta = |\mathbf{q}_\parallel| c_L/\omega$.
%(omega*2*np.pi)**2/2/np.pi**2/cl**3
%The energy resolved transmission function is defined by, $T(\omega) = \frac{1}{N_{\mathbf{k}_\parallel}}\sum_{\mathbf{k}_\parallel}T(\omega,\mathbf{k}_\parallel)$.
The two-probe interfacial thermal conductance per unit area is determined by the transmission function,
\begin{equation}
    G = \frac{1}{2\pi A}\int_0^\infty \hbar\omega \Theta(\omega)\frac{\partial f(\omega,T)}{\partial T}d\omega
\end{equation}
where $f(\omega,T)$ is the Bose-Einstein distribution function.

\subsection{The interface scattering transition probability for a rough Si/Ge interface}

We apply the derived equations for transmittance and reflectance for a rough Si/Ge interface along [001] direction. If we assume the atomic mixing is realized by swapping Si and Ge atoms on two sides of interface, the variance of mass fluctuations is estimated to be $\langle m^2_i\rangle \approx  \left( m_{\mathrm{Si}}- m_{\mathrm{Ge}}\right)^2 = 1.985 \times 10^3 \mathrm{u}^2$. Thus, the parameter $V_2 = n\langle m^2_i\rangle  \omega^4$, where  $n$ is the number of pairs of swapped Si and Ge atoms per unit area. We choose $n=2/a^2$ in the following calculation, and $a = 5.527 $ \AA \, is the lattice constant, obtained by taking the average of Si's and Ge's lattice constants. The bulk moduli of Si and Ge are $\mu_L = 95$ GPa and $\mu_R = 77.2$ GPa. The densities of Si and Ge are $\rho_L = 2.329 \times10^3 \,\mathrm{kg/m^3}$ and $\rho_R = 5.323\times10^3 \,\mathrm{kg/m^3}$.

As shown in Fig.~\ref{f1}, we find that the phonon transmittance of Si is smaller compared with AMM.
Although there are more transmission channels due to diffuse scattering, the reduction of transmittance is mainly due to fewer specular transmission channels, which are removed by interfacial disorders. In addition, we note that the diffuse transmission from the Ge side opens new transmission channels above the critical angle for total reflection. Furthermore, we find that the reflectance from the Si side increases with the angle, similar to the trend of AMM. For reflectance from the Ge side, the specular part is smaller than AMM. Due to large diffuse reflectance shown in Fig.~\ref{f1} (e), the total reflection probability is eventually higher than predictions of AMM below the critical angle and lower above the critical angle. 
In fact, from the expression of Eq.~\ref{T} and Eq.~\ref{R}, we find that the specular transmittance is always reduced by disorders while the specular reflectance can either be enhanced or reduced depending on the sign of $\rho_Lv_L-\rho_Rv_R$. 

From Fig.~\ref{f1} (e) and (f), we observe that generally the diffuse transmittance from one side is different from the diffuse reflectance from the other side.Furthermore, we find that as frequency increases, the specular scattering probability decreases while the diffuse scattering probability increases, as shown in Fig.~\ref{transmissionfunction} (a)-(d). For the transmittance from both sides, the reduction in the specular part is always larger than the increment in the diffuse part, hence a reduced total transmittance. In contrast, for reflectance, the increment in the diffuse part prevails over the reduction in the specular part, causing a greater total reflectance.  When we compare the phonon transmission function for Si/Ge interface in Fig.~\ref{transmissionfunction} (e), the interface disorders lead to a smaller total transmission, thus a smaller thermal conductance. Note that when $\omega>5$ THz, the specular reflectance from Ge side will become negative, because the perturbation is no longer a small quantity. From Eq.~\ref{pr}, we see that the reduction in the specularity parameter for reflectance $p_R$ varies drastically with frequency with  $\omega^4$ scaling. When $p_R \sim 0$, we have a critical frequency $\omega\sim
\left(\frac{\mu^2 A}{\langle m^2_i \rangle}\right)^{1/4}$, and our model only works below this critical frequency.

To summarize, continuum modeling using perturbation theory to the lowest order suggests that the diffuse scattering cannot make a phonon forget its origin, opposing the picture of DMM. However, we want to point out limitations of the continuum model of scalar phonons. First of all, the phonon mode conversion is not considered. Secondly, the model is valid for low-frequency acoustic phonons thus at lifted temperatures, where high-frequency phonons are playing an important role in interfacial phonon transport, the model is no longer valid.
\end{appendix}

\bibliographystyle{apsrev4-2}
\bibliography{ref} 

\begin{thebibliography}{44}%
\makeatletter
\providecommand \@ifxundefined [1]{%
 \@ifx{#1\undefined}
}%
\providecommand \@ifnum [1]{%
 \ifnum #1\expandafter \@firstoftwo
 \else \expandafter \@secondoftwo
 \fi
}%
\providecommand \@ifx [1]{%
 \ifx #1\expandafter \@firstoftwo
 \else \expandafter \@secondoftwo
 \fi
}%
\providecommand \natexlab [1]{#1}%
\providecommand \enquote  [1]{``#1''}%
\providecommand \bibnamefont  [1]{#1}%
\providecommand \bibfnamefont [1]{#1}%
\providecommand \citenamefont [1]{#1}%
\providecommand \href@noop [0]{\@secondoftwo}%
\providecommand \href [0]{\begingroup \@sanitize@url \@href}%
\providecommand \@href[1]{\@@startlink{#1}\@@href}%
\providecommand \@@href[1]{\endgroup#1\@@endlink}%
\providecommand \@sanitize@url [0]{\catcode `\\12\catcode `\$12\catcode
  `\&12\catcode `\#12\catcode `\^12\catcode `\_12\catcode `\%12\relax}%
\providecommand \@@startlink[1]{}%
\providecommand \@@endlink[0]{}%
\providecommand \url  [0]{\begingroup\@sanitize@url \@url }%
\providecommand \@url [1]{\endgroup\@href {#1}{\urlprefix }}%
\providecommand \urlprefix  [0]{URL }%
\providecommand \Eprint [0]{\href }%
\providecommand \doibase [0]{https://doi.org/}%
\providecommand \selectlanguage [0]{\@gobble}%
\providecommand \bibinfo  [0]{\@secondoftwo}%
\providecommand \bibfield  [0]{\@secondoftwo}%
\providecommand \translation [1]{[#1]}%
\providecommand \BibitemOpen [0]{}%
\providecommand \bibitemStop [0]{}%
\providecommand \bibitemNoStop [0]{.\EOS\space}%
\providecommand \EOS [0]{\spacefactor3000\relax}%
\providecommand \BibitemShut  [1]{\csname bibitem#1\endcsname}%
\let\auto@bib@innerbib\@empty
%</preamble>
\bibitem [{\citenamefont {Swartz}\ and\ \citenamefont
  {Pohl}(1989)}]{RevModPhys.61.605}%
  \BibitemOpen
  \bibfield  {author} {\bibinfo {author} {\bibfnamefont {E.~T.}\ \bibnamefont
  {Swartz}}\ and\ \bibinfo {author} {\bibfnamefont {R.~O.}\ \bibnamefont
  {Pohl}},\ }\href {https://doi.org/10.1103/RevModPhys.61.605} {\bibfield
  {journal} {\bibinfo  {journal} {Rev. Mod. Phys.}\ }\textbf {\bibinfo {volume}
  {61}},\ \bibinfo {pages} {605} (\bibinfo {year} {1989})}\BibitemShut
  {NoStop}%
\bibitem [{\citenamefont {Little}(1959)}]{little1959transport}%
  \BibitemOpen
  \bibfield  {author} {\bibinfo {author} {\bibfnamefont {W.}~\bibnamefont
  {Little}},\ }\href {https://cdnsciencepub.com/doi/10.1139/p59-037} {\bibfield
   {journal} {\bibinfo  {journal} {Canadian Journal of Physics}\ }\textbf
  {\bibinfo {volume} {37}},\ \bibinfo {pages} {334} (\bibinfo {year}
  {1959})}\BibitemShut {NoStop}%
\bibitem [{\citenamefont {Kapitza}(1941)}]{kapitza1941zh}%
  \BibitemOpen
  \bibfield  {author} {\bibinfo {author} {\bibfnamefont {P.}~\bibnamefont
  {Kapitza}},\ }\href@noop {} {\bibfield  {journal} {\bibinfo  {journal} {J
  Phys USSR}\ }\textbf {\bibinfo {volume} {4}},\ \bibinfo {pages} {181}
  (\bibinfo {year} {1941})}\BibitemShut {NoStop}%
\bibitem [{\citenamefont {Khalatnikov}(1952)}]{khla}%
  \BibitemOpen
  \bibfield  {author} {\bibinfo {author} {\bibfnamefont {I.~M.}\ \bibnamefont
  {Khalatnikov}},\ }\href@noop {} {\bibfield  {journal} {\bibinfo  {journal}
  {Zh. Eksp. Teor. Fiz.}\ }\textbf {\bibinfo {volume} {22,}} (\bibinfo {year}
  {1952})}\BibitemShut {NoStop}%
\bibitem [{\citenamefont {Lyeo}\ and\ \citenamefont
  {Cahill}(2006)}]{PhysRevB.73.144301}%
  \BibitemOpen
  \bibfield  {author} {\bibinfo {author} {\bibfnamefont {H.-K.}\ \bibnamefont
  {Lyeo}}\ and\ \bibinfo {author} {\bibfnamefont {D.~G.}\ \bibnamefont
  {Cahill}},\ }\href {https://doi.org/10.1103/PhysRevB.73.144301} {\bibfield
  {journal} {\bibinfo  {journal} {Phys. Rev. B}\ }\textbf {\bibinfo {volume}
  {73}},\ \bibinfo {pages} {144301} (\bibinfo {year} {2006})}\BibitemShut
  {NoStop}%
\bibitem [{\citenamefont {Cheng}\ \emph {et~al.}(2020)\citenamefont {Cheng},
  \citenamefont {Koh}, \citenamefont {Ahmad}, \citenamefont {Hu}, \citenamefont
  {Shi}, \citenamefont {Liao}, \citenamefont {Wang}, \citenamefont {Bai},
  \citenamefont {Li}, \citenamefont {Lee} \emph {et~al.}}]{cheng2020thermal}%
  \BibitemOpen
  \bibfield  {author} {\bibinfo {author} {\bibfnamefont {Z.}~\bibnamefont
  {Cheng}}, \bibinfo {author} {\bibfnamefont {Y.~R.}\ \bibnamefont {Koh}},
  \bibinfo {author} {\bibfnamefont {H.}~\bibnamefont {Ahmad}}, \bibinfo
  {author} {\bibfnamefont {R.}~\bibnamefont {Hu}}, \bibinfo {author}
  {\bibfnamefont {J.}~\bibnamefont {Shi}}, \bibinfo {author} {\bibfnamefont
  {M.~E.}\ \bibnamefont {Liao}}, \bibinfo {author} {\bibfnamefont
  {Y.}~\bibnamefont {Wang}}, \bibinfo {author} {\bibfnamefont {T.}~\bibnamefont
  {Bai}}, \bibinfo {author} {\bibfnamefont {R.}~\bibnamefont {Li}}, \bibinfo
  {author} {\bibfnamefont {E.}~\bibnamefont {Lee}}, \emph {et~al.},\ }\href
  {https://www.nature.com/articles/s42005-020-0383-6} {\bibfield  {journal}
  {\bibinfo  {journal} {Communications Physics}\ }\textbf {\bibinfo {volume}
  {3}},\ \bibinfo {pages} {1} (\bibinfo {year} {2020})}\BibitemShut {NoStop}%
\bibitem [{\citenamefont {Chalopin}\ \emph {et~al.}(2012)\citenamefont
  {Chalopin}, \citenamefont {Esfarjani}, \citenamefont {Henry}, \citenamefont
  {Volz},\ and\ \citenamefont {Chen}}]{PhysRevB.85.195302}%
  \BibitemOpen
  \bibfield  {author} {\bibinfo {author} {\bibfnamefont {Y.}~\bibnamefont
  {Chalopin}}, \bibinfo {author} {\bibfnamefont {K.}~\bibnamefont {Esfarjani}},
  \bibinfo {author} {\bibfnamefont {A.}~\bibnamefont {Henry}}, \bibinfo
  {author} {\bibfnamefont {S.}~\bibnamefont {Volz}},\ and\ \bibinfo {author}
  {\bibfnamefont {G.}~\bibnamefont {Chen}},\ }\href
  {https://doi.org/10.1103/PhysRevB.85.195302} {\bibfield  {journal} {\bibinfo
  {journal} {Phys. Rev. B}\ }\textbf {\bibinfo {volume} {85}},\ \bibinfo
  {pages} {195302} (\bibinfo {year} {2012})}\BibitemShut {NoStop}%
\bibitem [{\citenamefont {Gordiz}\ and\ \citenamefont
  {Henry}(2015)}]{gordiz2015formalism}%
  \BibitemOpen
  \bibfield  {author} {\bibinfo {author} {\bibfnamefont {K.}~\bibnamefont
  {Gordiz}}\ and\ \bibinfo {author} {\bibfnamefont {A.}~\bibnamefont {Henry}},\
  }\href {https://iopscience.iop.org/article/10.1088/1367-2630/17/10/103002}
  {\bibfield  {journal} {\bibinfo  {journal} {New Journal of Physics}\ }\textbf
  {\bibinfo {volume} {17}},\ \bibinfo {pages} {103002} (\bibinfo {year}
  {2015})}\BibitemShut {NoStop}%
\bibitem [{\citenamefont {Gordiz}\ and\ \citenamefont
  {Henry}(2016)}]{gordiz2016phonon}%
  \BibitemOpen
  \bibfield  {author} {\bibinfo {author} {\bibfnamefont {K.}~\bibnamefont
  {Gordiz}}\ and\ \bibinfo {author} {\bibfnamefont {A.}~\bibnamefont {Henry}},\
  }\href {https://www.nature.com/articles/srep23139} {\bibfield  {journal}
  {\bibinfo  {journal} {Scientific Reports}\ }\textbf {\bibinfo {volume} {6}},\
  \bibinfo {pages} {1} (\bibinfo {year} {2016})}\BibitemShut {NoStop}%
\bibitem [{\citenamefont {Landry}\ and\ \citenamefont
  {McGaughey}(2009)}]{PhysRevB.80.165304}%
  \BibitemOpen
  \bibfield  {author} {\bibinfo {author} {\bibfnamefont {E.~S.}\ \bibnamefont
  {Landry}}\ and\ \bibinfo {author} {\bibfnamefont {A.~J.~H.}\ \bibnamefont
  {McGaughey}},\ }\href {https://doi.org/10.1103/PhysRevB.80.165304} {\bibfield
   {journal} {\bibinfo  {journal} {Phys. Rev. B}\ }\textbf {\bibinfo {volume}
  {80}},\ \bibinfo {pages} {165304} (\bibinfo {year} {2009})}\BibitemShut
  {NoStop}%
\bibitem [{\citenamefont {S\"a\"askilahti}\ \emph {et~al.}(2014)\citenamefont
  {S\"a\"askilahti}, \citenamefont {Oksanen}, \citenamefont {Tulkki},\ and\
  \citenamefont {Volz}}]{PhysRevB.90.134312}%
  \BibitemOpen
  \bibfield  {author} {\bibinfo {author} {\bibfnamefont {K.}~\bibnamefont
  {S\"a\"askilahti}}, \bibinfo {author} {\bibfnamefont {J.}~\bibnamefont
  {Oksanen}}, \bibinfo {author} {\bibfnamefont {J.}~\bibnamefont {Tulkki}},\
  and\ \bibinfo {author} {\bibfnamefont {S.}~\bibnamefont {Volz}},\ }\href
  {https://doi.org/10.1103/PhysRevB.90.134312} {\bibfield  {journal} {\bibinfo
  {journal} {Phys. Rev. B}\ }\textbf {\bibinfo {volume} {90}},\ \bibinfo
  {pages} {134312} (\bibinfo {year} {2014})}\BibitemShut {NoStop}%
\bibitem [{\citenamefont {Yang}\ \emph {et~al.}(2015)\citenamefont {Yang},
  \citenamefont {Luo}, \citenamefont {Esfarjani}, \citenamefont {Henry},
  \citenamefont {Tian}, \citenamefont {Shiomi}, \citenamefont {Chalopin},
  \citenamefont {Li},\ and\ \citenamefont {Chen}}]{yang2015thermal}%
  \BibitemOpen
  \bibfield  {author} {\bibinfo {author} {\bibfnamefont {N.}~\bibnamefont
  {Yang}}, \bibinfo {author} {\bibfnamefont {T.}~\bibnamefont {Luo}}, \bibinfo
  {author} {\bibfnamefont {K.}~\bibnamefont {Esfarjani}}, \bibinfo {author}
  {\bibfnamefont {A.}~\bibnamefont {Henry}}, \bibinfo {author} {\bibfnamefont
  {Z.}~\bibnamefont {Tian}}, \bibinfo {author} {\bibfnamefont {J.}~\bibnamefont
  {Shiomi}}, \bibinfo {author} {\bibfnamefont {Y.}~\bibnamefont {Chalopin}},
  \bibinfo {author} {\bibfnamefont {B.}~\bibnamefont {Li}},\ and\ \bibinfo
  {author} {\bibfnamefont {G.}~\bibnamefont {Chen}},\ }\href
  {https://doi.org/10.1166/jctn.2015.3710} {\bibfield  {journal} {\bibinfo
  {journal} {Journal of Computational and Theoretical Nanoscience}\ }\textbf
  {\bibinfo {volume} {12}},\ \bibinfo {pages} {168} (\bibinfo {year}
  {2015})}\BibitemShut {NoStop}%
\bibitem [{\citenamefont {Chalopin}\ and\ \citenamefont
  {Volz}(2013)}]{chalopin2013microscopic}%
  \BibitemOpen
  \bibfield  {author} {\bibinfo {author} {\bibfnamefont {Y.}~\bibnamefont
  {Chalopin}}\ and\ \bibinfo {author} {\bibfnamefont {S.}~\bibnamefont
  {Volz}},\ }\href {https://doi.org/10.1063/1.4816738} {\bibfield  {journal}
  {\bibinfo  {journal} {Applied Physics Letters}\ }\textbf {\bibinfo {volume}
  {103}},\ \bibinfo {pages} {051602} (\bibinfo {year} {2013})}\BibitemShut
  {NoStop}%
\bibitem [{\citenamefont {Kimmer}\ \emph {et~al.}(2007)\citenamefont {Kimmer},
  \citenamefont {Aubry}, \citenamefont {Skye},\ and\ \citenamefont
  {Schelling}}]{PhysRevB.75.144105}%
  \BibitemOpen
  \bibfield  {author} {\bibinfo {author} {\bibfnamefont {C.}~\bibnamefont
  {Kimmer}}, \bibinfo {author} {\bibfnamefont {S.}~\bibnamefont {Aubry}},
  \bibinfo {author} {\bibfnamefont {A.}~\bibnamefont {Skye}},\ and\ \bibinfo
  {author} {\bibfnamefont {P.~K.}\ \bibnamefont {Schelling}},\ }\href
  {https://doi.org/10.1103/PhysRevB.75.144105} {\bibfield  {journal} {\bibinfo
  {journal} {Phys. Rev. B}\ }\textbf {\bibinfo {volume} {75}},\ \bibinfo
  {pages} {144105} (\bibinfo {year} {2007})}\BibitemShut {NoStop}%
\bibitem [{\citenamefont {Deng}\ \emph {et~al.}(2014)\citenamefont {Deng},
  \citenamefont {Chernatynskiy}, \citenamefont {Khafizov}, \citenamefont
  {Hurley},\ and\ \citenamefont {Phillpot}}]{deng2014kapitza}%
  \BibitemOpen
  \bibfield  {author} {\bibinfo {author} {\bibfnamefont {B.}~\bibnamefont
  {Deng}}, \bibinfo {author} {\bibfnamefont {A.}~\bibnamefont {Chernatynskiy}},
  \bibinfo {author} {\bibfnamefont {M.}~\bibnamefont {Khafizov}}, \bibinfo
  {author} {\bibfnamefont {D.~H.}\ \bibnamefont {Hurley}},\ and\ \bibinfo
  {author} {\bibfnamefont {S.~R.}\ \bibnamefont {Phillpot}},\ }\href
  {https://doi.org/10.1063/1.4867047} {\bibfield  {journal} {\bibinfo
  {journal} {Journal of Applied Physics}\ }\textbf {\bibinfo {volume} {115}},\
  \bibinfo {pages} {084910} (\bibinfo {year} {2014})}\BibitemShut {NoStop}%
\bibitem [{\citenamefont {Zhang}\ \emph {et~al.}(2006)\citenamefont {Zhang},
  \citenamefont {Fisher},\ and\ \citenamefont {Mingo}}]{zhang2007simulation}%
  \BibitemOpen
  \bibfield  {author} {\bibinfo {author} {\bibfnamefont {W.}~\bibnamefont
  {Zhang}}, \bibinfo {author} {\bibfnamefont {T.~S.}\ \bibnamefont {Fisher}},\
  and\ \bibinfo {author} {\bibfnamefont {N.}~\bibnamefont {Mingo}},\ }\href
  {https://doi.org/10.1115/1.2709656} {\bibfield  {journal} {\bibinfo
  {journal} {Journal of Heat Transfer}\ }\textbf {\bibinfo {volume} {129}},\
  \bibinfo {pages} {483} (\bibinfo {year} {2006})}\BibitemShut {NoStop}%
\bibitem [{\citenamefont {Zhang}\ \emph {et~al.}(2007)\citenamefont {Zhang},
  \citenamefont {Fisher},\ and\ \citenamefont {Mingo}}]{zhang2007atomistic}%
  \BibitemOpen
  \bibfield  {author} {\bibinfo {author} {\bibfnamefont {W.}~\bibnamefont
  {Zhang}}, \bibinfo {author} {\bibfnamefont {T.}~\bibnamefont {Fisher}},\ and\
  \bibinfo {author} {\bibfnamefont {N.}~\bibnamefont {Mingo}},\ }\href
  {https://doi.org/10.1080/10407790601144755} {\bibfield  {journal} {\bibinfo
  {journal} {Numerical Heat Transfer, Part B: Fundamentals}\ }\textbf {\bibinfo
  {volume} {51}},\ \bibinfo {pages} {333} (\bibinfo {year} {2007})}\BibitemShut
  {NoStop}%
\bibitem [{\citenamefont {Tian}\ \emph {et~al.}(2012)\citenamefont {Tian},
  \citenamefont {Esfarjani},\ and\ \citenamefont {Chen}}]{PhysRevB.86.235304}%
  \BibitemOpen
  \bibfield  {author} {\bibinfo {author} {\bibfnamefont {Z.}~\bibnamefont
  {Tian}}, \bibinfo {author} {\bibfnamefont {K.}~\bibnamefont {Esfarjani}},\
  and\ \bibinfo {author} {\bibfnamefont {G.}~\bibnamefont {Chen}},\ }\href
  {https://doi.org/10.1103/PhysRevB.86.235304} {\bibfield  {journal} {\bibinfo
  {journal} {Phys. Rev. B}\ }\textbf {\bibinfo {volume} {86}},\ \bibinfo
  {pages} {235304} (\bibinfo {year} {2012})}\BibitemShut {NoStop}%
\bibitem [{\citenamefont {Ong}\ and\ \citenamefont
  {Zhang}(2015)}]{PhysRevB.91.174302}%
  \BibitemOpen
  \bibfield  {author} {\bibinfo {author} {\bibfnamefont {Z.-Y.}\ \bibnamefont
  {Ong}}\ and\ \bibinfo {author} {\bibfnamefont {G.}~\bibnamefont {Zhang}},\
  }\href {https://doi.org/10.1103/PhysRevB.91.174302} {\bibfield  {journal}
  {\bibinfo  {journal} {Phys. Rev. B}\ }\textbf {\bibinfo {volume} {91}},\
  \bibinfo {pages} {174302} (\bibinfo {year} {2015})}\BibitemShut {NoStop}%
\bibitem [{\citenamefont {Sadasivam}\ \emph {et~al.}(2017)\citenamefont
  {Sadasivam}, \citenamefont {Waghmare},\ and\ \citenamefont
  {Fisher}}]{PhysRevB.96.174302}%
  \BibitemOpen
  \bibfield  {author} {\bibinfo {author} {\bibfnamefont {S.}~\bibnamefont
  {Sadasivam}}, \bibinfo {author} {\bibfnamefont {U.~V.}\ \bibnamefont
  {Waghmare}},\ and\ \bibinfo {author} {\bibfnamefont {T.~S.}\ \bibnamefont
  {Fisher}},\ }\href {https://doi.org/10.1103/PhysRevB.96.174302} {\bibfield
  {journal} {\bibinfo  {journal} {Phys. Rev. B}\ }\textbf {\bibinfo {volume}
  {96}},\ \bibinfo {pages} {174302} (\bibinfo {year} {2017})}\BibitemShut
  {NoStop}%
\bibitem [{\citenamefont {Latour}\ \emph {et~al.}(2017)\citenamefont {Latour},
  \citenamefont {Shulumba},\ and\ \citenamefont
  {Minnich}}]{PhysRevB.96.104310}%
  \BibitemOpen
  \bibfield  {author} {\bibinfo {author} {\bibfnamefont {B.}~\bibnamefont
  {Latour}}, \bibinfo {author} {\bibfnamefont {N.}~\bibnamefont {Shulumba}},\
  and\ \bibinfo {author} {\bibfnamefont {A.~J.}\ \bibnamefont {Minnich}},\
  }\href {https://doi.org/10.1103/PhysRevB.96.104310} {\bibfield  {journal}
  {\bibinfo  {journal} {Phys. Rev. B}\ }\textbf {\bibinfo {volume} {96}},\
  \bibinfo {pages} {104310} (\bibinfo {year} {2017})}\BibitemShut {NoStop}%
\bibitem [{\citenamefont {Ong}\ \emph {et~al.}(2020)\citenamefont {Ong},
  \citenamefont {Schusteritsch},\ and\ \citenamefont
  {Pickard}}]{PhysRevB.101.195410}%
  \BibitemOpen
  \bibfield  {author} {\bibinfo {author} {\bibfnamefont {Z.-Y.}\ \bibnamefont
  {Ong}}, \bibinfo {author} {\bibfnamefont {G.}~\bibnamefont {Schusteritsch}},\
  and\ \bibinfo {author} {\bibfnamefont {C.~J.}\ \bibnamefont {Pickard}},\
  }\href {https://doi.org/10.1103/PhysRevB.101.195410} {\bibfield  {journal}
  {\bibinfo  {journal} {Phys. Rev. B}\ }\textbf {\bibinfo {volume} {101}},\
  \bibinfo {pages} {195410} (\bibinfo {year} {2020})}\BibitemShut {NoStop}%
\bibitem [{\citenamefont {Ong}(2021)}]{ong2021specular}%
  \BibitemOpen
  \bibfield  {author} {\bibinfo {author} {\bibfnamefont {Z.-Y.}\ \bibnamefont
  {Ong}},\ }\href {https://doi.org/10.1209/0295-5075/133/66002} {\bibfield
  {journal} {\bibinfo  {journal} {{EPL} (Europhysics Letters)}\ }\textbf
  {\bibinfo {volume} {133}},\ \bibinfo {pages} {66002} (\bibinfo {year}
  {2021})}\BibitemShut {NoStop}%
\bibitem [{\citenamefont {Young}\ and\ \citenamefont
  {Maris}(1989)}]{PhysRevB.40.3685}%
  \BibitemOpen
  \bibfield  {author} {\bibinfo {author} {\bibfnamefont {D.~A.}\ \bibnamefont
  {Young}}\ and\ \bibinfo {author} {\bibfnamefont {H.~J.}\ \bibnamefont
  {Maris}},\ }\href {https://doi.org/10.1103/PhysRevB.40.3685} {\bibfield
  {journal} {\bibinfo  {journal} {Phys. Rev. B}\ }\textbf {\bibinfo {volume}
  {40}},\ \bibinfo {pages} {3685} (\bibinfo {year} {1989})}\BibitemShut
  {NoStop}%
\bibitem [{\citenamefont {Zhao}\ and\ \citenamefont
  {Freund}(2005)}]{zhao2005lattice}%
  \BibitemOpen
  \bibfield  {author} {\bibinfo {author} {\bibfnamefont {H.}~\bibnamefont
  {Zhao}}\ and\ \bibinfo {author} {\bibfnamefont {J.}~\bibnamefont {Freund}},\
  }\href {https://doi.org/10.1063/1.1835565} {\bibfield  {journal} {\bibinfo
  {journal} {Journal of Applied Physics}\ }\textbf {\bibinfo {volume} {97}},\
  \bibinfo {pages} {024903} (\bibinfo {year} {2005})}\BibitemShut {NoStop}%
\bibitem [{\citenamefont {Lu}\ and\ \citenamefont
  {McGaughey}(2017)}]{lu2017thermal}%
  \BibitemOpen
  \bibfield  {author} {\bibinfo {author} {\bibfnamefont {S.}~\bibnamefont
  {Lu}}\ and\ \bibinfo {author} {\bibfnamefont {A.~J.}\ \bibnamefont
  {McGaughey}},\ }\href {https://doi.org/10.1063/1.4978362} {\bibfield
  {journal} {\bibinfo  {journal} {Journal of Applied Physics}\ }\textbf
  {\bibinfo {volume} {121}},\ \bibinfo {pages} {115103} (\bibinfo {year}
  {2017})}\BibitemShut {NoStop}%
\bibitem [{par()}]{parallel}%
  \BibitemOpen
  \href@noop {} {\bibinfo  {journal} {At a given transverse wavevector
  $\mathbf{q}_\parallel$, there are multiple phonon states with different $q_z$
  or degenerate states with the same $q_z$. $\mathrm{Here}$, we use
  $\mathbf{q}_\parallel$ to describe one of those states}\ }\BibitemShut
  {NoStop}%
\bibitem [{\citenamefont {Popescu}\ and\ \citenamefont
  {Zunger}(2010)}]{PhysRevLett.104.236403}%
  \BibitemOpen
\bibfield  {journal} {  }\bibfield  {author} {\bibinfo {author} {\bibfnamefont
  {V.}~\bibnamefont {Popescu}}\ and\ \bibinfo {author} {\bibfnamefont
  {A.}~\bibnamefont {Zunger}},\ }\href
  {https://doi.org/10.1103/PhysRevLett.104.236403} {\bibfield  {journal}
  {\bibinfo  {journal} {Phys. Rev. Lett.}\ }\textbf {\bibinfo {volume} {104}},\
  \bibinfo {pages} {236403} (\bibinfo {year} {2010})}\BibitemShut {NoStop}%
\bibitem [{\citenamefont {Bi}\ \emph {et~al.}(2012)\citenamefont {Bi},
  \citenamefont {Wang}, \citenamefont {Wang}, \citenamefont {Sha},
  \citenamefont {Wang}, \citenamefont {Chen},\ and\ \citenamefont
  {Chen}}]{bi2012thermal}%
  \BibitemOpen
  \bibfield  {author} {\bibinfo {author} {\bibfnamefont {K.}~\bibnamefont
  {Bi}}, \bibinfo {author} {\bibfnamefont {J.}~\bibnamefont {Wang}}, \bibinfo
  {author} {\bibfnamefont {Y.}~\bibnamefont {Wang}}, \bibinfo {author}
  {\bibfnamefont {J.}~\bibnamefont {Sha}}, \bibinfo {author} {\bibfnamefont
  {Z.}~\bibnamefont {Wang}}, \bibinfo {author} {\bibfnamefont {M.}~\bibnamefont
  {Chen}},\ and\ \bibinfo {author} {\bibfnamefont {Y.}~\bibnamefont {Chen}},\
  }\href {https://www.sciencedirect.com/science/article/pii/S0375960112007967}
  {\bibfield  {journal} {\bibinfo  {journal} {Physics Letters A}\ }\textbf
  {\bibinfo {volume} {376}},\ \bibinfo {pages} {2668} (\bibinfo {year}
  {2012})}\BibitemShut {NoStop}%
\bibitem [{ren()}]{renorm}%
  \BibitemOpen
  \href@noop {} {\bibinfo  {journal} {One may be concerned about the
  contribution from specularity parameter to the diffuse transmittance and
  reflectance. We have discussed the role of specularity parameter in diffuse
  scattering in the supplementary material}\ }\BibitemShut {NoStop}%
\bibitem [{\citenamefont {Yang}\ \emph {et~al.}(2018)\citenamefont {Yang},
  \citenamefont {Latour},\ and\ \citenamefont {Minnich}}]{PhysRevB.97.205306}%
  \BibitemOpen
\bibfield  {journal} {  }\bibfield  {author} {\bibinfo {author} {\bibfnamefont
  {L.}~\bibnamefont {Yang}}, \bibinfo {author} {\bibfnamefont {B.}~\bibnamefont
  {Latour}},\ and\ \bibinfo {author} {\bibfnamefont {A.~J.}\ \bibnamefont
  {Minnich}},\ }\href {https://doi.org/10.1103/PhysRevB.97.205306} {\bibfield
  {journal} {\bibinfo  {journal} {Phys. Rev. B}\ }\textbf {\bibinfo {volume}
  {97}},\ \bibinfo {pages} {205306} (\bibinfo {year} {2018})}\BibitemShut
  {NoStop}%
\bibitem [{\citenamefont {Vashaee}\ and\ \citenamefont
  {Shakouri}(2004)}]{PhysRevLett.92.106103}%
  \BibitemOpen
  \bibfield  {author} {\bibinfo {author} {\bibfnamefont {D.}~\bibnamefont
  {Vashaee}}\ and\ \bibinfo {author} {\bibfnamefont {A.}~\bibnamefont
  {Shakouri}},\ }\href {https://doi.org/10.1103/PhysRevLett.92.106103}
  {\bibfield  {journal} {\bibinfo  {journal} {Phys. Rev. Lett.}\ }\textbf
  {\bibinfo {volume} {92}},\ \bibinfo {pages} {106103} (\bibinfo {year}
  {2004})}\BibitemShut {NoStop}%
\bibitem [{\citenamefont {Wu}\ and\ \citenamefont
  {Luo}(2014)}]{doi:10.1063/1.4859555}%
  \BibitemOpen
  \bibfield  {author} {\bibinfo {author} {\bibfnamefont {X.}~\bibnamefont
  {Wu}}\ and\ \bibinfo {author} {\bibfnamefont {T.}~\bibnamefont {Luo}},\
  }\href {https://doi.org/10.1063/1.4859555} {\bibfield  {journal} {\bibinfo
  {journal} {Journal of Applied Physics}\ }\textbf {\bibinfo {volume} {115}},\
  \bibinfo {pages} {014901} (\bibinfo {year} {2014})},\ \Eprint
  {https://arxiv.org/abs/https://doi.org/10.1063/1.4859555}
  {https://doi.org/10.1063/1.4859555} \BibitemShut {NoStop}%
\bibitem [{\citenamefont {Feng}\ \emph {et~al.}(2019)\citenamefont {Feng},
  \citenamefont {Zhong}, \citenamefont {Shi},\ and\ \citenamefont
  {Ruan}}]{PhysRevB.99.045301}%
  \BibitemOpen
  \bibfield  {author} {\bibinfo {author} {\bibfnamefont {T.}~\bibnamefont
  {Feng}}, \bibinfo {author} {\bibfnamefont {Y.}~\bibnamefont {Zhong}},
  \bibinfo {author} {\bibfnamefont {J.}~\bibnamefont {Shi}},\ and\ \bibinfo
  {author} {\bibfnamefont {X.}~\bibnamefont {Ruan}},\ }\href
  {https://doi.org/10.1103/PhysRevB.99.045301} {\bibfield  {journal} {\bibinfo
  {journal} {Phys. Rev. B}\ }\textbf {\bibinfo {volume} {99}},\ \bibinfo
  {pages} {045301} (\bibinfo {year} {2019})}\BibitemShut {NoStop}%
\bibitem [{\citenamefont {Dai}\ and\ \citenamefont
  {Tian}(2020)}]{PhysRevB.101.041301}%
  \BibitemOpen
  \bibfield  {author} {\bibinfo {author} {\bibfnamefont {J.}~\bibnamefont
  {Dai}}\ and\ \bibinfo {author} {\bibfnamefont {Z.}~\bibnamefont {Tian}},\
  }\href {https://doi.org/10.1103/PhysRevB.101.041301} {\bibfield  {journal}
  {\bibinfo  {journal} {Phys. Rev. B}\ }\textbf {\bibinfo {volume} {101}},\
  \bibinfo {pages} {041301} (\bibinfo {year} {2020})}\BibitemShut {NoStop}%
\bibitem [{\citenamefont {Guo}\ \emph {et~al.}(2021)\citenamefont {Guo},
  \citenamefont {Zhang}, \citenamefont {Bescond}, \citenamefont {Xiong},
  \citenamefont {Nomura},\ and\ \citenamefont {Volz}}]{PhysRevB.103.174306}%
  \BibitemOpen
  \bibfield  {author} {\bibinfo {author} {\bibfnamefont {Y.}~\bibnamefont
  {Guo}}, \bibinfo {author} {\bibfnamefont {Z.}~\bibnamefont {Zhang}}, \bibinfo
  {author} {\bibfnamefont {M.}~\bibnamefont {Bescond}}, \bibinfo {author}
  {\bibfnamefont {S.}~\bibnamefont {Xiong}}, \bibinfo {author} {\bibfnamefont
  {M.}~\bibnamefont {Nomura}},\ and\ \bibinfo {author} {\bibfnamefont
  {S.}~\bibnamefont {Volz}},\ }\href
  {https://doi.org/10.1103/PhysRevB.103.174306} {\bibfield  {journal} {\bibinfo
   {journal} {Phys. Rev. B}\ }\textbf {\bibinfo {volume} {103}},\ \bibinfo
  {pages} {174306} (\bibinfo {year} {2021})}\BibitemShut {NoStop}%
\bibitem [{\citenamefont {Eason}(1969)}]{eason1969wave}%
  \BibitemOpen
  \bibfield  {author} {\bibinfo {author} {\bibfnamefont {G.}~\bibnamefont
  {Eason}},\ }\href {https://link.springer.com/article/10.1007/BF01176664}
  {\bibfield  {journal} {\bibinfo  {journal} {Acta Mechanica}\ }\textbf
  {\bibinfo {volume} {7}},\ \bibinfo {pages} {137} (\bibinfo {year}
  {1969})}\BibitemShut {NoStop}%
\bibitem [{\citenamefont {Johnson}(2007)}]{sgwave}%
  \BibitemOpen
  \bibfield  {author} {\bibinfo {author} {\bibfnamefont {S.~G.}\ \bibnamefont
  {Johnson}},\ }\href {https://math.mit.edu/~stevenj/18.369/wave-equations.pdf}
  {\bibfield  {journal} {\bibinfo  {journal} {Notes on the algebraic structure
  of wave equations}\ } (\bibinfo {year} {2007})}\BibitemShut {NoStop}%
\bibitem [{\citenamefont {Copley}(2014)}]{copley201412}%
  \BibitemOpen
  \bibfield  {author} {\bibinfo {author} {\bibfnamefont {L.}~\bibnamefont
  {Copley}},\ }in\ \href
  {https://degruyter.com/document/doi/10.2478/9783110409475.12/html} {\emph
  {\bibinfo {booktitle} {Mathematics for the Physical Sciences}}}\ (\bibinfo
  {publisher} {Sciendo Migration},\ \bibinfo {year} {2014})\ pp.\ \bibinfo
  {pages} {384--417}\BibitemShut {NoStop}%
\bibitem [{\citenamefont {Messiah}(1962)}]{messiah1962quantum}%
  \BibitemOpen
  \bibfield  {author} {\bibinfo {author} {\bibfnamefont {A.}~\bibnamefont
  {Messiah}},\ }\href@noop {} {\emph {\bibinfo {title} {Quantum Mechanics:
  Volume II}}}\ (\bibinfo  {publisher} {North-Holland Publishing Company
  Amsterdam},\ \bibinfo {year} {1962})\BibitemShut {NoStop}%
\bibitem [{\citenamefont {Stewart}\ \emph {et~al.}(2003)\citenamefont
  {Stewart}, \citenamefont {Butler}, \citenamefont {Zhang},\ and\ \citenamefont
  {Los}}]{PhysRevB.68.014433}%
  \BibitemOpen
  \bibfield  {author} {\bibinfo {author} {\bibfnamefont {D.~A.}\ \bibnamefont
  {Stewart}}, \bibinfo {author} {\bibfnamefont {W.~H.}\ \bibnamefont {Butler}},
  \bibinfo {author} {\bibfnamefont {X.-G.}\ \bibnamefont {Zhang}},\ and\
  \bibinfo {author} {\bibfnamefont {V.~F.}\ \bibnamefont {Los}},\ }\href
  {https://doi.org/10.1103/PhysRevB.68.014433} {\bibfield  {journal} {\bibinfo
  {journal} {Phys. Rev. B}\ }\textbf {\bibinfo {volume} {68}},\ \bibinfo
  {pages} {014433} (\bibinfo {year} {2003})}\BibitemShut {NoStop}%
\bibitem [{\citenamefont {Chen}(1998)}]{PhysRevB.57.14958}%
  \BibitemOpen
  \bibfield  {author} {\bibinfo {author} {\bibfnamefont {G.}~\bibnamefont
  {Chen}},\ }\href {https://doi.org/10.1103/PhysRevB.57.14958} {\bibfield
  {journal} {\bibinfo  {journal} {Phys. Rev. B}\ }\textbf {\bibinfo {volume}
  {57}},\ \bibinfo {pages} {14958} (\bibinfo {year} {1998})}\BibitemShut
  {NoStop}%
\bibitem [{\citenamefont {Ziman}(2001)}]{ziman2001electrons}%
  \BibitemOpen
  \bibfield  {author} {\bibinfo {author} {\bibfnamefont {J.~M.}\ \bibnamefont
  {Ziman}},\ }\href
  {https://oxford.universitypressscholarship.com/view/10.1093/acprof:oso/9780198507796.001.0001/acprof-9780198507796}
  {\emph {\bibinfo {title} {Electrons and phonons: the theory of transport
  phenomena in solids}}}\ (\bibinfo  {publisher} {Oxford university press},\
  \bibinfo {year} {2001})\BibitemShut {NoStop}%
\bibitem [{\citenamefont {Li}\ and\ \citenamefont
  {McGaughey}(2015)}]{li2015phonon}%
  \BibitemOpen
  \bibfield  {author} {\bibinfo {author} {\bibfnamefont {D.}~\bibnamefont
  {Li}}\ and\ \bibinfo {author} {\bibfnamefont {A.~J.}\ \bibnamefont
  {McGaughey}},\ }\href
  {https://www.tandfonline.com/doi/full/10.1080/15567265.2015.1035199}
  {\bibfield  {journal} {\bibinfo  {journal} {Nanoscale and Microscale
  Thermophysical Engineering}\ }\textbf {\bibinfo {volume} {19}},\ \bibinfo
  {pages} {166} (\bibinfo {year} {2015})}\BibitemShut {NoStop}%
\end{thebibliography}%


\begin{thebibliography}{9}%
\makeatletter
\providecommand \@ifxundefined [1]{%
 \@ifx{#1\undefined}
}%
\providecommand \@ifnum [1]{%
 \ifnum #1\expandafter \@firstoftwo
 \else \expandafter \@secondoftwo
 \fi
}%
\providecommand \@ifx [1]{%
 \ifx #1\expandafter \@firstoftwo
 \else \expandafter \@secondoftwo
 \fi
}%
\providecommand \natexlab [1]{#1}%
\providecommand \enquote  [1]{``#1''}%
\providecommand \bibnamefont  [1]{#1}%
\providecommand \bibfnamefont [1]{#1}%
\providecommand \citenamefont [1]{#1}%
\providecommand \href@noop [0]{\@secondoftwo}%
\providecommand \href [0]{\begingroup \@sanitize@url \@href}%
\providecommand \@href[1]{\@@startlink{#1}\@@href}%
\providecommand \@@href[1]{\endgroup#1\@@endlink}%
\providecommand \@sanitize@url [0]{\catcode `\\12\catcode `\$12\catcode
  `\&12\catcode `\#12\catcode `\^12\catcode `\_12\catcode `\%12\relax}%
\providecommand \@@startlink[1]{}%
\providecommand \@@endlink[0]{}%
\providecommand \url  [0]{\begingroup\@sanitize@url \@url }%
\providecommand \@url [1]{\endgroup\@href {#1}{\urlprefix }}%
\providecommand \urlprefix  [0]{URL }%
\providecommand \Eprint [0]{\href }%
\providecommand \doibase [0]{https://doi.org/}%
\providecommand \selectlanguage [0]{\@gobble}%
\providecommand \bibinfo  [0]{\@secondoftwo}%
\providecommand \bibfield  [0]{\@secondoftwo}%
\providecommand \translation [1]{[#1]}%
\providecommand \BibitemOpen [0]{}%
\providecommand \bibitemStop [0]{}%
\providecommand \bibitemNoStop [0]{.\EOS\space}%
\providecommand \EOS [0]{\spacefactor3000\relax}%
\providecommand \BibitemShut  [1]{\csname bibitem#1\endcsname}%
\let\auto@bib@innerbib\@empty
%</preamble>
\bibitem [{\citenamefont {Khomyakov}\ \emph {et~al.}(2005)\citenamefont
  {Khomyakov}, \citenamefont {Brocks}, \citenamefont {Karpan}, \citenamefont
  {Zwierzycki},\ and\ \citenamefont {Kelly}}]{PhysRevB.72.035450}%
  \BibitemOpen
  \bibfield  {author} {\bibinfo {author} {\bibfnamefont {P.~A.}\ \bibnamefont
  {Khomyakov}}, \bibinfo {author} {\bibfnamefont {G.}~\bibnamefont {Brocks}},
  \bibinfo {author} {\bibfnamefont {V.}~\bibnamefont {Karpan}}, \bibinfo
  {author} {\bibfnamefont {M.}~\bibnamefont {Zwierzycki}},\ and\ \bibinfo
  {author} {\bibfnamefont {P.~J.}\ \bibnamefont {Kelly}},\ }\href
  {https://doi.org/10.1103/PhysRevB.72.035450} {\bibfield  {journal} {\bibinfo
  {journal} {Phys. Rev. B}\ }\textbf {\bibinfo {volume} {72}},\ \bibinfo
  {pages} {035450} (\bibinfo {year} {2005})}\BibitemShut {NoStop}%
\bibitem [{\citenamefont {Ong}\ and\ \citenamefont
  {Zhang}(2015)}]{PhysRevB.91.174302}%
  \BibitemOpen
  \bibfield  {author} {\bibinfo {author} {\bibfnamefont {Z.-Y.}\ \bibnamefont
  {Ong}}\ and\ \bibinfo {author} {\bibfnamefont {G.}~\bibnamefont {Zhang}},\
  }\href {https://doi.org/10.1103/PhysRevB.91.174302} {\bibfield  {journal}
  {\bibinfo  {journal} {Phys. Rev. B}\ }\textbf {\bibinfo {volume} {91}},\
  \bibinfo {pages} {174302} (\bibinfo {year} {2015})}\BibitemShut {NoStop}%
\bibitem [{\citenamefont {Sancho}\ \emph {et~al.}(1985)\citenamefont {Sancho},
  \citenamefont {Sancho}, \citenamefont {Sancho},\ and\ \citenamefont
  {Rubio}}]{sancho1985highly}%
  \BibitemOpen
  \bibfield  {author} {\bibinfo {author} {\bibfnamefont {M.~L.}\ \bibnamefont
  {Sancho}}, \bibinfo {author} {\bibfnamefont {J.~L.}\ \bibnamefont {Sancho}},
  \bibinfo {author} {\bibfnamefont {J.~L.}\ \bibnamefont {Sancho}},\ and\
  \bibinfo {author} {\bibfnamefont {J.}~\bibnamefont {Rubio}},\ }\href
  {https://iopscience.iop.org/article/10.1088/0305-4608/15/4/009} {\bibfield
  {journal} {\bibinfo  {journal} {Journal of Physics F: Metal Physics}\
  }\textbf {\bibinfo {volume} {15}},\ \bibinfo {pages} {851} (\bibinfo {year}
  {1985})}\BibitemShut {NoStop}%
\bibitem [{\citenamefont {Lewenkopf}\ and\ \citenamefont
  {Mucciolo}(2013)}]{Lewenkopf2013}%
  \BibitemOpen
  \bibfield  {author} {\bibinfo {author} {\bibfnamefont {C.~H.}\ \bibnamefont
  {Lewenkopf}}\ and\ \bibinfo {author} {\bibfnamefont {E.~R.}\ \bibnamefont
  {Mucciolo}},\ }\href {https://doi.org/10.1007/s10825-013-0458-7} {\bibfield
  {journal} {\bibinfo  {journal} {Journal of Computational Electronics}\
  }\textbf {\bibinfo {volume} {12}},\ \bibinfo {pages} {203} (\bibinfo {year}
  {2013})}\BibitemShut {NoStop}%
\bibitem [{\citenamefont {Sols}\ \emph {et~al.}(1989)\citenamefont {Sols},
  \citenamefont {Macucci}, \citenamefont {Ravaioli},\ and\ \citenamefont
  {Hess}}]{Sols1989}%
  \BibitemOpen
  \bibfield  {author} {\bibinfo {author} {\bibfnamefont {F.}~\bibnamefont
  {Sols}}, \bibinfo {author} {\bibfnamefont {M.}~\bibnamefont {Macucci}},
  \bibinfo {author} {\bibfnamefont {U.}~\bibnamefont {Ravaioli}},\ and\
  \bibinfo {author} {\bibfnamefont {K.}~\bibnamefont {Hess}},\ }\href
  {https://doi.org/10.1063/1.344032} {\bibfield  {journal} {\bibinfo  {journal}
  {Journal of Applied Physics}\ }\textbf {\bibinfo {volume} {66}},\ \bibinfo
  {pages} {3892} (\bibinfo {year} {1989})}\BibitemShut {NoStop}%
\bibitem [{\citenamefont {Ong}(2018)}]{PhysRevB.98.195301}%
  \BibitemOpen
  \bibfield  {author} {\bibinfo {author} {\bibfnamefont {Z.-Y.}\ \bibnamefont
  {Ong}},\ }\href {https://doi.org/10.1103/PhysRevB.98.195301} {\bibfield
  {journal} {\bibinfo  {journal} {Phys. Rev. B}\ }\textbf {\bibinfo {volume}
  {98}},\ \bibinfo {pages} {195301} (\bibinfo {year} {2018})}\BibitemShut
  {NoStop}%
\bibitem [{\citenamefont {Caroli}\ \emph {et~al.}(1971)\citenamefont {Caroli},
  \citenamefont {Combescot}, \citenamefont {Nozieres},\ and\ \citenamefont
  {Saint-James}}]{Caroli_1971}%
  \BibitemOpen
  \bibfield  {author} {\bibinfo {author} {\bibfnamefont {C.}~\bibnamefont
  {Caroli}}, \bibinfo {author} {\bibfnamefont {R.}~\bibnamefont {Combescot}},
  \bibinfo {author} {\bibfnamefont {P.}~\bibnamefont {Nozieres}},\ and\
  \bibinfo {author} {\bibfnamefont {D.}~\bibnamefont {Saint-James}},\ }\href
  {https://doi.org/10.1088/0022-3719/4/8/018} {\bibfield  {journal} {\bibinfo
  {journal} {Journal of Physics C: Solid State Physics}\ }\textbf {\bibinfo
  {volume} {4}},\ \bibinfo {pages} {916} (\bibinfo {year} {1971})}\BibitemShut
  {NoStop}%
\bibitem [{phi()}]{phinfortran}%
  \BibitemOpen
  \href {https://github.com/KitchenSong/phinfortran} {\bibinfo  {journal}
  {https://github.com/KitchenSong/phinfortran}\ }\BibitemShut {NoStop}%
\bibitem [{\citenamefont {Bi}\ \emph {et~al.}(2012)\citenamefont {Bi},
  \citenamefont {Wang}, \citenamefont {Wang}, \citenamefont {Sha},
  \citenamefont {Wang}, \citenamefont {Chen},\ and\ \citenamefont
  {Chen}}]{bi2012thermal}%
  \BibitemOpen
\bibfield  {journal} {  }\bibfield  {author} {\bibinfo {author} {\bibfnamefont
  {K.}~\bibnamefont {Bi}}, \bibinfo {author} {\bibfnamefont {J.}~\bibnamefont
  {Wang}}, \bibinfo {author} {\bibfnamefont {Y.}~\bibnamefont {Wang}}, \bibinfo
  {author} {\bibfnamefont {J.}~\bibnamefont {Sha}}, \bibinfo {author}
  {\bibfnamefont {Z.}~\bibnamefont {Wang}}, \bibinfo {author} {\bibfnamefont
  {M.}~\bibnamefont {Chen}},\ and\ \bibinfo {author} {\bibfnamefont
  {Y.}~\bibnamefont {Chen}},\ }\href
  {https://www.sciencedirect.com/science/article/pii/S0375960112007967}
  {\bibfield  {journal} {\bibinfo  {journal} {Physics Letters A}\ }\textbf
  {\bibinfo {volume} {376}},\ \bibinfo {pages} {2668} (\bibinfo {year}
  {2012})}\BibitemShut {NoStop}%
\end{thebibliography}%
\end{document}

%\preprint{}

\title{Supplemental material to the manuscript ``Evaluation of diffuse mismatch model for phonon scattering at disordered interfaces''}
\maketitle

\section{Mode-resolved atomistic Green's function} \label{sec:3}
\subsection{Partition of the dynamical matrix}
%The continuum model is limited to acoustic phonons and the effect of different polarizations is not considered. 
In this session, we illustrate the mode-resolved atomistic Green's function (AGF) formalism to study diffuse phonon scattering. Specifically, we follow the methodology of mode-matching to obtain the mode-resolved transmission and reflection probability\cite{PhysRevB.72.035450,PhysRevB.91.174302} and take advantage of the in-plane periodicity to reduce the computational cost in numerical implementation.

We start by writing down the dynamical matrix of the system of an interface formed by two semi-infinite dissimilar materials (see Fig. 1 in main text) at given transverse wavevector $\mathbf{q}_\parallel$,
\begin{equation}
    D^{\alpha,\beta}_{i,j}(\mathbf{q}_\parallel) =\sum_\tau \frac{\Phi^{\alpha,\beta}_{0i,\tau j}}{\sqrt{m_i m_j}}e^{i\mathbf{q}_\parallel\cdot(\mathbf{r}_\tau+\mathbf{r}_j-\mathbf{r}_0-\mathbf{r}_i)},
\end{equation}
where $\Phi^{\alpha,\beta}_{0i,\tau j}$ is the harmonic force
constant between atom $i$ in zeroth supercell and atom $j$ in $\tau$th supercell and $m_i$ is the atom mass.
Along the transport direction, we partition the system into three parts: semi-infinite left lead, device region and semi-infinite right lead. 
 The dynamical matrix can be written in block-tridiagonal matrix form,
\begin{equation}
    D(\mathbf{q}_\parallel)=\begin{pmatrix}
        \ddots&&&&\\
       & D^L_{11} & D^L_{01}&&&\\
       & D^L_{10} &D^L_{00}&D_{LD}&&\\
       &&D_{DL}&D_D&D_{DR}&&\\
       &&&D_{RD}&D^{R}_{00}&D^{R}_{01}&\\
       &&&&D^R_{10}&D^R_{11}&\\
       &&&&&&\ddots
    \end{pmatrix}
\end{equation}
where $D^L_{nn}  = D^L_{00} $ and $D^R_{nn}  = D^R_{00} $ for the lead region, and $n$ denote the $n$th repeated supercell cell in the lead region as defined in Fig. 1 (a) in the main text. $D_{D}$ is the dynamical matrix corresponding to the device region. $D_{LD/DL}$ and $D_{RD/DR}$ describe interactions between the lead and the device region.

\subsection{Partition of the Green's function matrix}
The Green's function of the system is defined by,
\begin{equation}
    ((\omega^2\pm i\eta) I-D(\mathbf{q}_\parallel))G^{R/A}(\mathbf{q}_\parallel) = I
\end{equation}
where the $R$ and $A$ denote retarded and advanced Green's function, respectively. Specifically, we explicitly express the retarded Green's function matrix,
\begin{equation}
    G^R(\mathbf{q}_\parallel)=\begin{pmatrix}
        \ddots&&&&&&\\
       & G^L_{11} & G^L_{01}&&&\udots\\
       & G^L_{10} &G^L_{00}&G_{LD}&G_{0,N+1}&\\
       &&G_{DL}&G_D&G_{DR}&&\\
       &&G_{N+1,0}&G_{RD}&G^{R}_{00}&G^{R}_{01}&\\
       &\udots&&&G^R_{10}&G^R_{11}&\\
       &&&&&&\ddots
    \end{pmatrix}
    \label{fullG}
\end{equation}
Unlike the dynamical matrix in the block tridiagonal form, the Green's function $G^R(\mathbf{q}_\parallel)$ is  generally not a block  tridiagonal matrix. In particular, for matrix block $G_{0,N+1}$ the subscript $0$ denotes the cell in the left lead adjacent to the device and the subscript $N+1$ denotes the cell in the right lead adjacent to the device region. Physically, this matrix block is relevant to the phonon propagation from the $(N+1)$th cell to the $0$th cell.

\subsection{The surface Green's function}
In this session, we will discuss the procedures to obtain the surface Green's function for left and right semi-infinite lead. The surface refers to the zeroth unitcell in the semi-infinite lead region, as marked in Fig. 1(a) in the main text.
The advanced surface Green's function for left lead is  the conjugate transpose of matrix block $G^L_{00}$ in Eq.~\ref{fullG},
\begin{equation}
    g^A_{L} = \left[G^L_{00}\right]^\dagger = [(\omega^2 I - i\eta)I - D^L_{00} - \Sigma^A_{L}]^{-1}
    \label{gl} 
\end{equation}
where the self-energy of left lead is $\Sigma^A_{L} = D^L_{10} g^A_{L} D^L_{01} $. The retarded surface Green's function for left lead is $g^R_{L} = [g^A_{R}]^\dagger$. The matrix block $G^R_{00}$ is called the retarded surface Green's function for the right lead,
\begin{equation}
    g^R_{R} = [(\omega^2 I + i\eta)I - D^R_{00} - \Sigma^R_{R}]^{-1} 
\end{equation}
where the self-energy of right lead $\Sigma_{R} = D^R_{01} g^R_{R} D^R_{10} $. Another set of two surface Green's functions describing the semi-infinite leads which extends toward the opposite directions are,
\begin{equation}
    \begin{split}
    g^{R}_{L^\prime}& = [(\omega^2 I + i\eta)I - D^L_{00} - \Sigma^{R}_{L^\prime}]^{-1}\\
    g^{R}_{R^\prime}& = [(\omega^2 I + i\eta)I - D^R_{00} - \Sigma^{R}_{R^\prime}]^{-1}
    \end{split}
    \label{gl1} 
\end{equation}
where $\Sigma^{R}_{L^\prime} = D^L_{01} g^{R}_{L^\prime} D^L_{10}$ and $\Sigma^{R}_{R^\prime} = D^R_{10} g^{R}_{R^\prime} D^R_{01}$. Further we define the broadening matrix $\Gamma = i(\Sigma^R- \Sigma^{R\dagger})$ which describes the escape rate of phonons for the connection between adjacent unitcells in the lead region.

\subsection{Efficient evaluation of surface Green's function for supercells}
In numerical implantation, the surface Green's functions can be computed using decimation 
technique\cite{sancho1985highly} and the computational cost is majorly coming from the matrix inversion, which scales with $N^3$, with $N\times N$ being the dimensions of the matrix. Since we are interested in the diffuse phonon scattering by a rough interface, we create a supercell with larger transverse lattice vectors $\mathbf{R}_x$ and $\mathbf{R}_y$.
However, the dimension of the dynamical matrix becomes larger with the larger size of the in-plane unitcell, making the computational cost of calculating surface Green's function intractable. 

We can take advantage of the in-plane periodicity to reduce computational cost. Suppose the dimension for the dynamical matrix of minimal repeating unitcell is $n$, the number of unitcells in the larger supercell is $N_x \times N_y$, where $N_x$ and $N_y$ are dimensions along the first and second in-plane lattice vectors $\mathbf{R}_x$ and $\mathbf{R}_y$.
Define a matrix $P_x$,
\begin{equation}
    P_x = C_x\otimes I_x
\end{equation}
where $C_x$ is a circulant matrix  of dimension $N_x\times N_x$, the matrix element of which reads,
$C_{x,ab}= e^{i \frac{(b-1)\mathbf{G}_x}{N_x} \cdot\frac{(a-1)\mathbf{R}_x}{N_x}}$. $\mathbf{G}_x$ and $\mathbf{R}_x$ are the first in-plane lattice vectors in reciprocal space and real space, and $\mathbf{G}_x\cdot\mathbf{R}_x = 2\pi$. $I_x$ is an identity matrix with dimension $n N_y \times nN_y $.
It is easy to verify that $P_x^{-1} = \frac{1}{N_x}P_x^\dagger$

Consider the dynamical matrix for left lead of the supercell $D_L$. The Fourier transform of dynamical matrix writes,
\begin{equation}
    \tilde{D}_L=P_x^{-1}D_LP_x
    \label{block1}
\end{equation}
where $\tilde{D}_L$ is a block diagonal matrix because of the periodicity along $\mathbf{R}_x$. Let $\tilde{D}^j_L$  denotes the $j$th matrix block out of $N_x$ blocks of matrix  $\tilde{D}_L$ and $\tilde{D}^j_L$  has dimension of $nN_y\times nN_y$. For each matrix block $\tilde{D}^j_L$ , the Fourier transform along $\mathbf{R}_y$ direction gives,
\begin{equation}
    \tilde{\tilde{D}}^j_L=P_y^{-1}\tilde{D}^j_LP_y
    \label{block2}
\end{equation}
where $\tilde{\tilde{D}}^j_L$ is also a block diagonal matrix and its $k$ th diagonal block is denoted by $\tilde{\tilde{D}}^{jk}_L$. $P_y =  C_y\otimes I_y$, where the auxiliary matrix $C_{y,ab}= e^{i \frac{(b-1)\mathbf{G}_y}{N_y} \cdot\frac{(a-1)\mathbf{R}_y}{N_y}}$ and $I_y$ is identity matrix of dimension of $n\times n$. Through two consecutive Fourier transforms using Eq.~\ref{block1} and Eq.~\ref{block2}, the dynamical matrix $D_L$ is reorganized as a block diagonal matrix with $N_xN_y$ blocks and each block can be denoted by $\tilde{\tilde{D}}^{jk}_L$.  

The surface Green's function for a lead shares the same transverse periodicity as the dynamical matrix thus we can do Fourier transform for Eq.~\ref{gl} and obtain matrix block $\tilde{\tilde{g}}_L$ of dimension $n\times n$. To calculate surface Green's function of the large cell, instead of solving the matrix inversion problem for matrix of dimension $nN_xN_y \times nN_xN_y$, we solve Eq.~\ref{gl} for a matrix of dimension $n\times n$ for $N_xN_y$ times,
\begin{equation}
    \tilde{\tilde{g}}^A_{L,jk} = [(\omega^2  - i\eta)I_n - \tilde{\tilde{D}}^{L,jk}_{00} - \tilde{\tilde{D}}^{L,jk}_{10}\tilde{\tilde{g}}^A_{L,jk}\tilde{\tilde{D}}^{L,jk}_{01}]^{-1}
    \label{unit}
\end{equation}
Then, we apply the inverse Fourier transform along $\mathbf{R}_y$ and $\mathbf{R}_x$ direction and obtain the full advance surface Green's function for the larger cell $g^R_A$
\begin{equation}
    \begin{split}
    \tilde{g}^A_{L,j} &=P_y\tilde{\tilde{g}}^A_{L,j}P_y^{-1}\\
    g^A_{L}&=P_x\tilde{g}^A_{L}P_x^{-1}
    \end{split}
\end{equation}
Here,  the $k$th matrix block of matrix  $\tilde{g}^A_{L,j} $ is $\tilde{\tilde{g}}^A_{L,jk} $ obtained by Eq.~\ref{unit} and  the $j$th matrix block of matrix $\tilde{g}^A_{L}$ is $\tilde{g}^A_{L,j} $.
We can calculate other surface Green's functions following the same Fourier transform and inverse Fourier transform procedures.

%For an infinite lead, the  dynamical matrix for a unitcell writes,
%\begin{equation}
%    D^L = D^L_{00}+ D^{L}_{01}e^{ik_L a}+ D^{L}_{10}e^{-ik_L a}
%\end{equation}
%From the definition of surface Green's function, 

\subsection{The wavevector and group velocity}
The structure of an interface breaks the translational symmetry along the transport direction normal to interface, thus
 the wavevector normal to interface $k_{L/R}$ is not explicitly unknown. In addition, the evanescent waves can exist and the corresponding $k_{L/R}$ is a pure imaginary number. To resolve $k_{L,R}$ in order to identify the propagating modes that are relevant to transport, we introduce matrix\cite{PhysRevB.72.035450},
\begin{equation}
    F^A_{L} = g^A_{L}D^L_{01}.
\end{equation}
The eigenvalue of this matrix $\Lambda^A_i$ stores the phase information of the mode and the eigenvector $U^A_{L,i}$ contains the phonon eigenvectors. If $|\Lambda^A_i| \neq 1$, it corresponds to an evanescent mode. If $|\Lambda^A_i| = 1$, it corresponds to a propagating mode and we can extract the perpendicular wavevector $q_L = \frac{1}{a_L}\mathrm{log}\Lambda^A_i$. Similarly, for the right lead,
 we have 
 \begin{equation}
    F^R_{R} = g^R_{R}D^R_{10}
 \end{equation}
 and we can extract the perpendicular wavevector $q_R$ from its eigenvalues, $q_R = \frac{1}{a_R}\mathrm{log}\Lambda^R_i$. 
 
 Due to the large supercell size, the in-plane wavevector for a state $U^A_{L,i}$ in the unitcell representation, is folded into a smaller in-plane Brillouin zone of the supercell. In order to know the actual in-plane momentum for a state of left/right reservoir, we need to unfold the wavevector of the supercell state to the Brillouin zone for unitcell. Suppose the in-plane wavevector in supercell is $\mathbf{q}^{\mathrm{sc}}_\parallel$. The possible wavevectors to be unfolded differ by multiples of transverse reciprocal lattice vector compared with the wavevector of the supercell state, 
 \begin{equation}
     \mathbf{q}_\parallel = a\mathbf{G}^{\mathrm{sc}}_x+b\mathbf{G}^{\mathrm{sc}}_y + \mathbf{q}^{\mathrm{sc}}_\parallel
 \end{equation}
 where $a$ and $b$ are two integers to be determined, and $\mathbf{G}^{\mathrm{sc}}_{x/y} = \frac{1}{N_{x/y}}\mathbf{G}_{x/y}$ is the transverse reciprocal lattice vector for the supercell. For a given state $U^A_{L,i}$, the weight for a possible unfolded state with certain $a$ and $b$ is,
 \begin{equation}
    \begin{split}
     W_{i,\mathbf{q}_\parallel}  =&\frac{1}{N_xN_y} \sum^{n/3}_{j}\sum^3_{k} \Big|\sum^{N_xN_y}_{l}U^A_{L,i}(j,k,l)\times\mathrm{exp}\left(-i(-\mathbf{q}^\mathrm{sc}_{\parallel}\cdot\mathbf{r}_{jl}+\mathbf{q}_\parallel\cdot(\mathbf{r}_{jl}-\mathbf{r}^\mathrm{uc}_j))\right)\Big|^2,
    %                        exp(-i_imag*(-dot_product(matmul(ksc,recivec),pos)+&
    % dot_product(kpoint,pos-pos_uc)))
    \end{split}
    \label{weight}
 \end{equation}
 where $j$ refers to the atom index in the minimal repeating cell, $k$ denotes the Cartesian coordinate and $l$ is the index for unitcell in the larger unitcell. If the weight $W_{i,\mathbf{q}_\parallel} = 1$, the wavevector $\mathbf{q}^{\mathrm{sc}}_{\parallel}$  will be unfolded into wavevector $\mathbf{q}_{\parallel}$. When $W_{i,\mathbf{q}_\parallel} = 0$, the wavevector $\mathbf{q}^{\mathrm{uc}}_{\parallel}$  do not unfolded into wavevector $\mathbf{q}_{\parallel}$. 
 
It is worth noting that the momentum or the phase velocity is not a good measure of the angle for a propagating phonon. The phase velocity is not uniquely defined as any reciprocal lattice vector can be added to the wavevector and the choice of the reciprocal lattice vector is not uniquely defined. On the other hand, the group velocity for a state is unique, independent of the choice of reciprocal lattice vectors. Thus, we compute the group velocity $\mathbf{v}=\left(v_x,v_y,v_z\right)$ for the unfolded state using Hellman-Feynman theorem, and we use the angle of the group velocity as incident angles. The polar angle is $\theta = \mathrm{arccos}({v_z}/{|\mathbf{v}|})$ and the azimuthal angle is $\phi = \mathrm{arctan}(v_y/v_x)$.

 %Since the transmission and reflection probability is defined by the ratio of flux, we also need to know the velocity of each state along transport direction. 

 Particularly, the velocity along the transport direction (perpendicular to interface) can also be described by the velocity matrix,
 \begin{equation}
    \begin{split}
    V^A_{L}& = -\frac{ia_{L}}{2\omega}U^{A\dagger}_{L}\Gamma^A_LU^{A}_{L},\\
    V^R_{R}& = \frac{ia_{R}}{2\omega}U^{R\dagger}_{R}\Gamma^R_RU^{R}_{R}.
    \end{split}
 \end{equation}
 The diagonal elements of these matrices correspond to the group velocity along z direction.

 \subsection{Recursive Green's function}

 %To calculate transmission matrix and reflection matrix, we need surface Green's function as well as the device Green's function $G_D$. 
 The device region often contains a large amount of atoms, making the computation for device Green's function challenging. To overcome this difficulty, the device Green's function is calculated using recursive technique\cite{Lewenkopf2013,Sols1989}. 

We first write down the dynamical matrix of device region in block-tridiagonal matrix form,
\begin{equation}
    D_{D} = 
    \begin{pmatrix}
         D_{11}&D_{12}&&\\
         D_{21}&D_{22}&D_{23}&\\
         &D_{32}&D_{33}&D_{34}&\\
         &&&\ddots\\
         &&&D_{N-1,N}&D_{NN}
    \end{pmatrix}
\end{equation}
%Typically, the retarded Green's function for device is computed by,
%\begin{equation}
%G^R(\omega,\vec{k}_\parallel)=\left[\omega^2 I - D_C(\vec{k}_\parallel) - \Sigma_L(\omega,\vec{k}_\parallel) -\Sigma_R(\omega,\vec{k}_\parallel)\right]^{-1}.
%\label{GD}
%\end{equation}. 
The corresponding Green's function in Eq.~\ref{fullG} can be partitioned in a similar way. With the help of recursive Green's function algorithm\cite{Lewenkopf2013,Sols1989}, we can compute $G_{01}$, $G_{01}$, $G_{02}$, $\dots$, $G_{0N}$, recursively, using the relation,
\begin{equation}
    \begin{split}
    G_{n,n} &= \left[(\omega^2+i\eta)I-D_{n,n}- V_{n,n-1}g_{n-1,n-1}V_{n-1,n}\right]^{-1}\\
    G_{0,n} &=  g_{0,n-1}V_{n-1,n}G_{n,n}
    \end{split}
\end{equation}
As a consequence, we do not need to evaluate every matrix block of Green's function matrix presented by Eq.~\ref{fullG}. Instead, we only need to compute matrix blocks $G_{0,N+1}$,  $G_{0,0}$, and  $G_{N+1,N+1}$. The matrix block $G_{N+1,0}$ can be computed by conducting the recursive procedures starting from the right lead, towards the left lead.

\subsection{Transmission and reflection matrix}

Denote $u_L(+)$ and $u_R(+)$ the forward-propagating state in the left lead and right lead, respectively. Due to the scattering of the device region, the eigenvector of left mode becomes,
\begin{equation}
    c_L = u_L(+)+\mathcal{R}u_L(+)
\end{equation}
and the propagating modes in right lead can be expressed,
\begin{equation}
    c_R = \mathcal{T}u_R(+)
\end{equation}
where the matrix $\mathcal{R}$ and $\mathcal{T}$ are the generalized reflection and transmission matrix. The physical transmission and reflection matrix is obtained by normalizing according to the heat flux,
\begin{equation}
    t_{mn} = \sqrt{\frac{v_{z,m}a_L}{v_{z,n}a_R}}\mathcal{T}_{mn}
\end{equation}
\begin{equation}
    r_{mn} = \sqrt{\frac{v_{z,m}}{v_{z,n}}}\mathcal{R}_{mn}
\end{equation}
Note that the lattice constant along z direction is used to compute the heat flux for each phonon mode, which is defined by the velocity divided by the unitcell volume. In the following, we use the physical transmission and reflection matrix to study the transmission and reflection probability.

The transmission matrix is defined by by\cite{PhysRevB.72.035450,PhysRevB.98.195301},
 \begin{equation}
     t_{RL} = \frac{2i\omega}{\sqrt{a_R a_L}}\sqrt{V^R_R}[U^R_R]^{-1}G_{N+1,0}[U^{A\dagger}_L]^{-1}\sqrt{V^A_L}
 \end{equation}
 The reflection matrix for left and right side are,
 \begin{equation}
    \begin{split}
     r_{LL} &= \frac{i\omega}{a_L}\sqrt{V^R_L}[U^R_L]^{-1}\left(G_{0,0}-Q^{-1}_L\right)[U^{A\dagger}_L]^{-1}\sqrt{V^A_{L}}\\
     r_{RR} &= \frac{i\omega}{a_R}\sqrt{V^R_R}[U^R_R]^{-1}\left(G_{N+1,N+1}-Q^{-1}_R\right)[U^{A\dagger}_R]^{-1}\sqrt{V^A_{R}}\\
    \end{split}
 \end{equation}
where $Q^{-1}_L = (\omega^2+i\eta)I_{nNxNy} - D^L_{00}-D^{L}_{10}g^R_LD^{L}_{01}-D^{L}_{01}g^R_{L'}D^{L}_{10}$ and $Q^{-1}_R = (\omega^2+i\eta)I_{nNxNy} - D^R_{00}-D^{R}_{01}g^R_RD^{L}_{10}-D^{R}_{10}g^R_{R'}D^{R}_{01}$ are the retarded Green's functions for the corresponding semi-infinite lead. 

The transmission probability matrix element $T_{ij}$ describing the transmission probability from state $j$ to state $i$ is defined by the squared amplitude of the transmission matrix element $T_{ij}=|t_{ij}|^2$ , and the reflection probability matrix element is the squared amplitude of reflection matrix element $R_{ij}=|r_{ij}|^2$. Given that state $j$ and $i$ are both propagating modes, if the unfolded momentum calculated using Eq.~\ref{weight} for state $j$ and state $i$ are the same, \textit{i.e.} $\mathbf{q}_{j,\parallel}=\mathbf{q}_{i,\parallel}$, $T_{ij}$ refers to specular transmission probability. Similarly, $R_{ij}$ corresponds to specular reflection process if $\mathbf{q}_{j,\parallel}=\mathbf{q}_{i,\parallel}$. On the other hand, if   $\mathbf{q}_{j,\parallel}\neq\mathbf{q}_{i,\parallel}$, $T_{ij}$ and $R_{ij}$ are corresponding to the probability for diffuse transmission and reflection process, respectively.

\subsection{Transmission and transmission function}
Lastly, we want to show how to calculate the transmission $T(\omega,\mathbf{q}_\parallel)$ and transmission function $\Theta(\omega)$. We define a matrix, 
\begin{equation}
\mathbb{t}_{mn} = t_{ij},
\label{ttt1}
\end{equation}
where $i$ and $j$ are both corresponding to propagating modes (evanescent states are excluded). The transmission $ T(\omega,\mathbf{q}_\parallel) $ can be expressed by,
\begin{equation}
    T(\omega,\mathbf{q}_\parallel) = \mathrm{tr}(\mathbb{t}^\dagger\mathbb{t})
    \label{ttt2}
\end{equation}
According to Caroli formula\cite{Caroli_1971}, the transmission also reads, 
\begin{equation}  T(\omega,\mathbf{q}_\parallel) = \mathrm{tr}\left[G^R_{N+1,0}\Gamma_L G^{R\dagger}_{N+1,0}\Gamma_R\right]. 
\end{equation}
Eq.~\ref{ttt1} and Eq.~\ref{ttt2} give the same transmission. We use these two formulas as a sanity check for our calculation. We define two similar $\mathbb{t}$ matrices to account for contributions from specular transmission and diffuse transmission processes,
\begin{subequations}
  \begin{equation}
    \mathbb{t}_{s,mn} = t_{ij},\: \mathrm{when} \:\mathbf{q}_{j,\parallel}=\mathbf{q}_{i,\parallel} 
  \end{equation}
  \begin{equation}
  \mathbb{t}_{d,mn} = t_{ij},  \: \mathrm{when}\:\mathbf{q}_{j,\parallel}\neq\mathbf{q}_{i,\parallel}
\end{equation}
\end{subequations}

Then, the specular and diffuse transmission can be expressed by,
\begin{subequations}
    \begin{equation}
    T_s(\omega,\mathbf{q}_\parallel) = \mathrm{tr}(\mathbb{t}_s^\dagger\mathbb{t}_s)
    \end{equation}
    \begin{equation}
    T_d(\omega,\mathbf{q}_\parallel) = \mathrm{tr}(\mathbb{t}_d^\dagger\mathbb{t}_d)
    \end{equation}
\end{subequations}
%The specular and diffuse reflection is defined in a similar way by,
%\begin{equation}
%    \begin{split}
%    R_s(\omega,\mathbf{q}_\parallel) = \mathrm{tr}(\mathbb{r}_s^\dagger\mathbb{r}_s)\\
%    R_d(\omega,\mathbf{q}_\parallel) = \mathrm{tr}(\mathbb{r}_d^\dagger\mathbb{r}_d)
%    \end{split}
%\end{equation}
%where $\mathbb{r}_{s,mn} = r_{ij}$, $\mathbf{k}_{j,\parallel}=\mathbf{q}_{i,\parallel}$ and $\mathbb{r}_{d,mn} = t_{ij}$, $\mathbf{k}_{j,\parallel}\neq\mathbf{q}_{i,\parallel}$.
%and reflection matrix $\mathbb{r}_{mn}$. 

Finally, the total, specular and diffuse transmission function are defined by 
\begin{subequations}
  \begin{equation}
  \Theta(\omega) = \frac{1}{N_{\mathbf{q}_\parallel}}\sum_{\mathbf{q}_\parallel}T(\omega,\mathbf{q}_\parallel)
  \end{equation}
  \begin{equation}
  \Theta_s(\omega) = \frac{1}{N_{\mathbf{q}_\parallel}}\sum_{\mathbf{q}_\parallel}T_s(\omega,\mathbf{q}_\parallel)
  \end{equation}
  \begin{equation}
  \Theta_d(\omega) = \frac{1}{N_{\mathbf{q}_\parallel}}\sum_{\mathbf{q}_\parallel}T_d(\omega,\mathbf{q}_\parallel)
  \end{equation}
\end{subequations}
The total  reflection function writes,
\begin{equation}
    \Xi(\omega) = \frac{1}{N_{\mathbf{q}_\parallel}}\sum_{\mathbf{q}_\parallel}\mathrm{tr}(\mathbb{r}^\dagger\mathbb{r})
\end{equation}
where $\mathbb{r}_{mn} = r_{ij}$, and $i$, $j$ are propagating modes of the same side.
The specular and diffuse reflection function can be similarly defined. Our code implementation of the above formalism can be found in Ref.\cite{phinfortran}.

\begin{figure}[t!]
    \includegraphics[width=\textwidth]{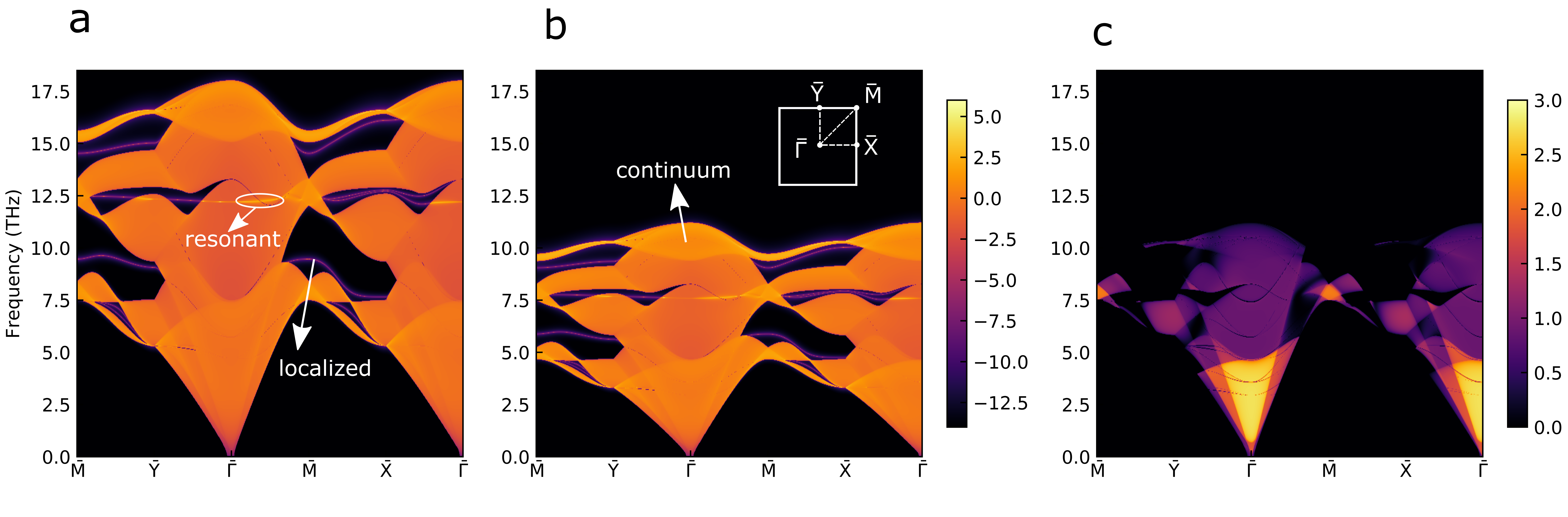}
    \caption{(a), (b) The surface density of states and (c) transmission as a function of $\omega$ and $\mathbf{q}_\parallel$ along high symmetry paths in the surface Brillouin zone for a perfect [001] Si/Ge interface.}
    \label{f1dot5}
\end{figure}

\subsection{The SDOS and transmission for a perfect Si/Ge interface}

We  study the perfect [001] Si/Ge interface without any disorder, which corresponds to the specular-only interface scattering. The lattice constant is 5.527 \AA, taken as the average of Si's and Ge's lattice constants. The force constant are computed from Stillinger-Weber potential\cite{bi2012thermal} and we assume the force constants for Ge are the same as Si.  
In Fig.~\ref{f1dot5} (a) and (b), we have plotted the surface density of states (SDOS) for semi-infinite leads along the high symmetry line in the surface Brillouin zone. The SDOS is defined by, 
\begin{equation}
 s_{L/R}(\omega,\mathbf{q}_\parallel) = -\frac{2\omega}{\pi}\mathrm{tr}\mathrm{Im}\{g^R_{L/R}(\omega,\mathbf{q}_\parallel)\} 
\end{equation}
which measures the number of states at a given frequency and in-plane wavevector. We have identified the continuum spectrum of propagating states, isolated curves of surface (localized) states between continuum spectrum and resonant states within the continuum spectrum. In particular, the continuum spectrum is corresponding to propagating phonon states. The perfect [001] Si/Ge interface structure is categorized as pmm planar symmetry group, where there are two perpendicular reflection axes (one of the axes is parallel to $\bar{\Gamma}\bar{M}$), and four rotation centers of order two (180 degree) located at the intersections of the reflection axes. The surface states and resonant states along $\bar{\Gamma}$ to $\bar{X}$ and along $\bar{\Gamma}$ to $\bar{Y}$ have distinct dispersion relation as there is no rotational symmetry of order four (90 degree). The continuum spectrum, on the other hand, has four-fold (90 degree) rotation symmetry, since it remains the same if rotating the high symmetry path by 90 degree around [001] axis. For instance, the continuum spectrum along $\bar{\Gamma}\bar{X}$ is the same with the spectrum along $\bar{\Gamma}\bar{Y}$. This is expected as the propagating states in the semi-infinite lead should be the same states in the corresponding bulk materials and in bulk Si or Ge, where the phonon dispersion has four-fold (90 degree) rotation symmetry around the [001] axis.

In Fig.~\ref{f1dot5} (c), the transmission $T(\omega,\mathbf{q}_\parallel)$ for a perfect [001] Si/Ge interface is shown. Compared with surface density of states, we find that only overlapped surface density of states can lead to non-zero transmission. Also, the acoustic branches have relatively higher transmission compared with other branches, implying the important role acoustic phonons play in interfacial phonon transport. On the contrary, the mismatch of energies  due to large mass contrast of Si and Ge causes smaller transmission  of optical branches.

\begin{figure}[t!]
    \includegraphics[width=0.8\textwidth]{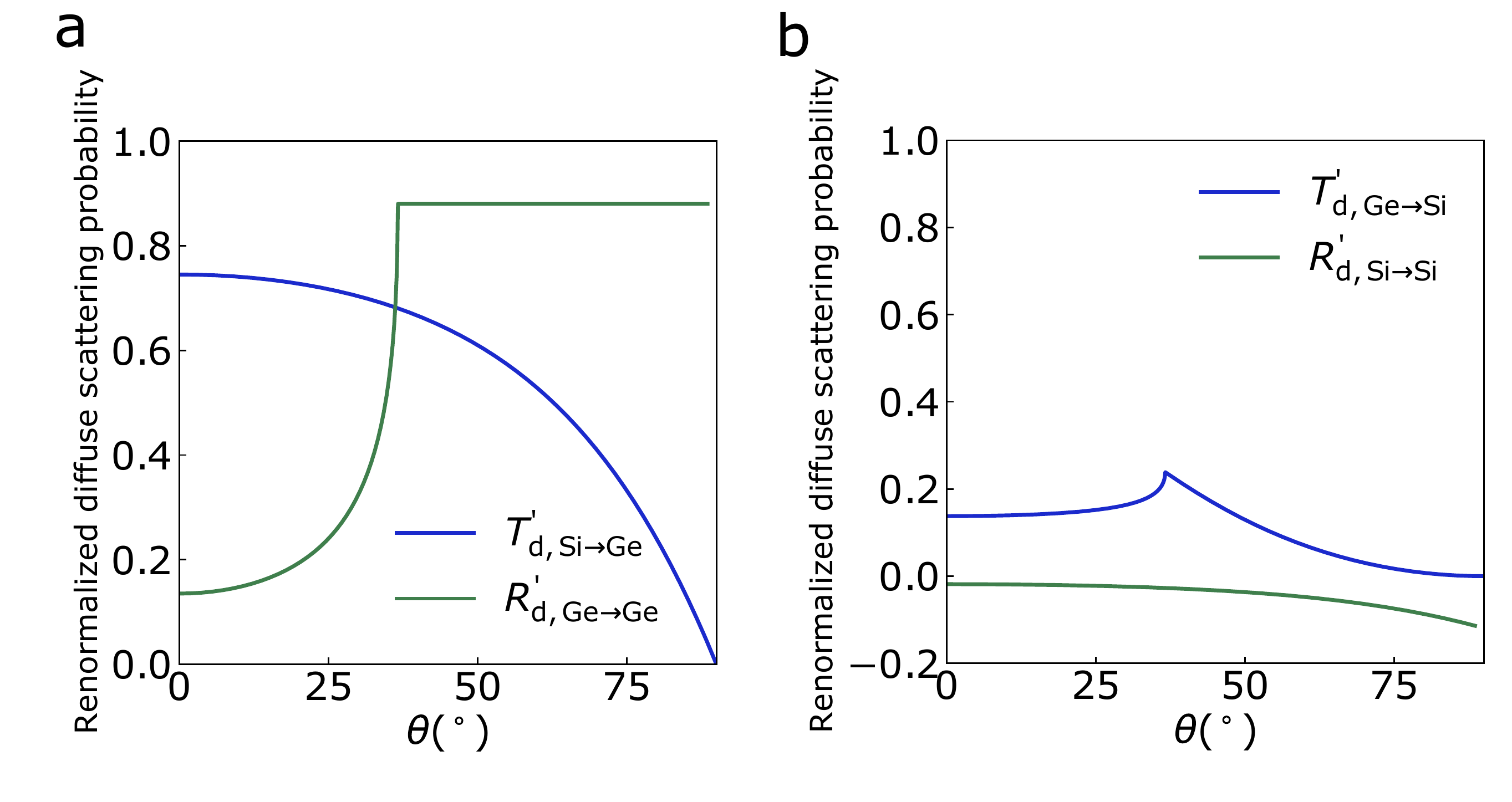}
    \caption{(a) The renormalized diffuse transmittance from Si and reflectance from Ge at $\omega=  4$  THz from the continuum model. (b) The renormalized diffuse transmittance from Ge and reflectance from Si at $\omega=  4$  THz from the continuum model. Note that $R^\prime_{\mathrm{Si} \to \mathrm{Si}}$ is negative because the corresponding specularity parameter is larger than one.}
    \label{renorm}
\end{figure}

\begin{figure}[t]
    \includegraphics[width=0.65\textwidth]{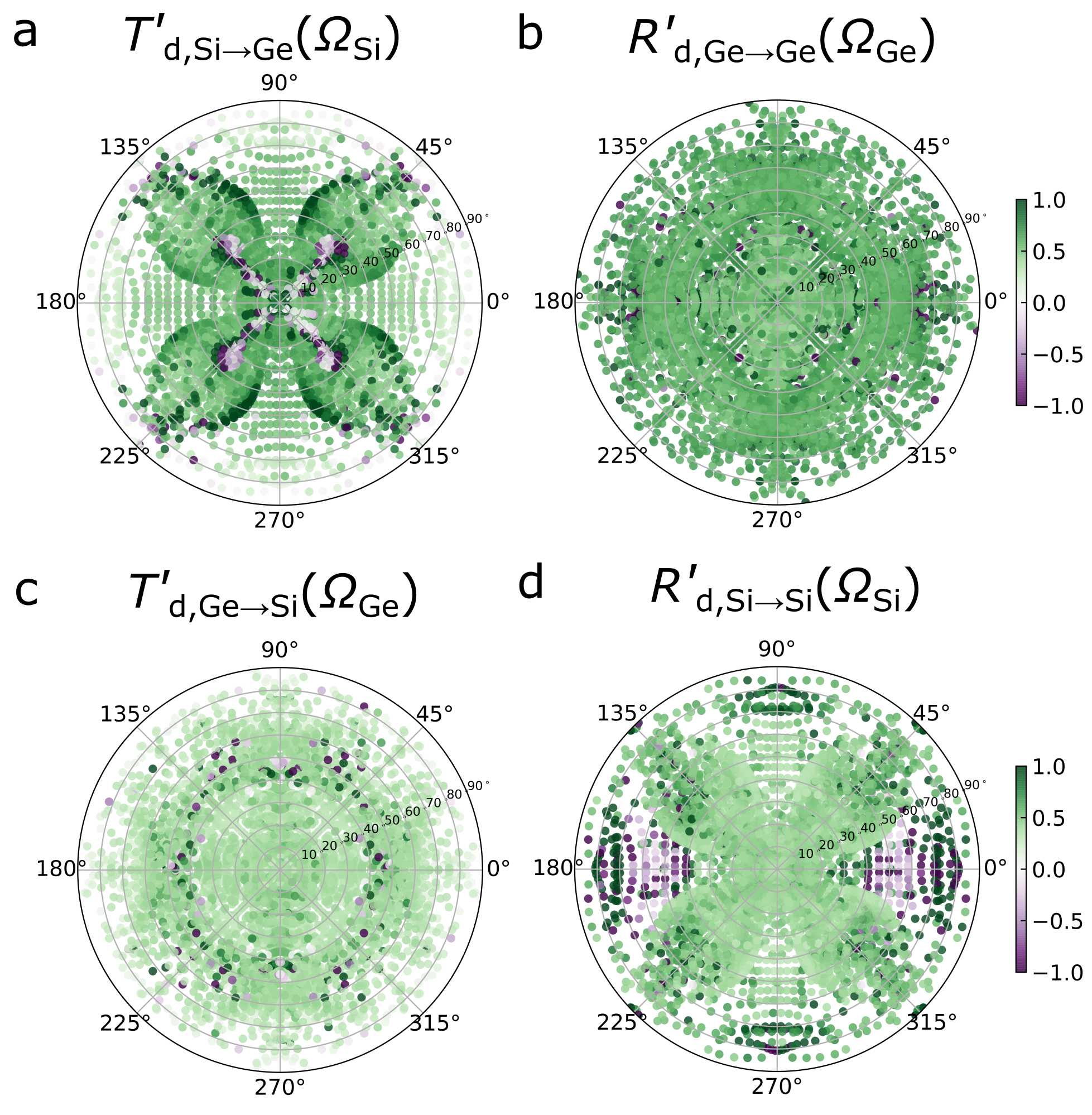}
    \caption{The directional (a) renormalized diffuse transmittance from Si and (b) renormalized diffuse reflectance from Ge at $\omega=  5$  THz. The directional (c) renormalized diffuse transmittance from Ge and (d) renormalized diffuse reflectance from Si at $\omega=  5$  THz. The scattering probabilities are acquired by taking the ensemble average of 21 AGF calculations on 8 ml structures. Note the negative value originates from the specularity parameter larger than one. }
    \label{renormagf}
\end{figure}

\begin{figure}[t]
    \includegraphics[width=0.95\textwidth]{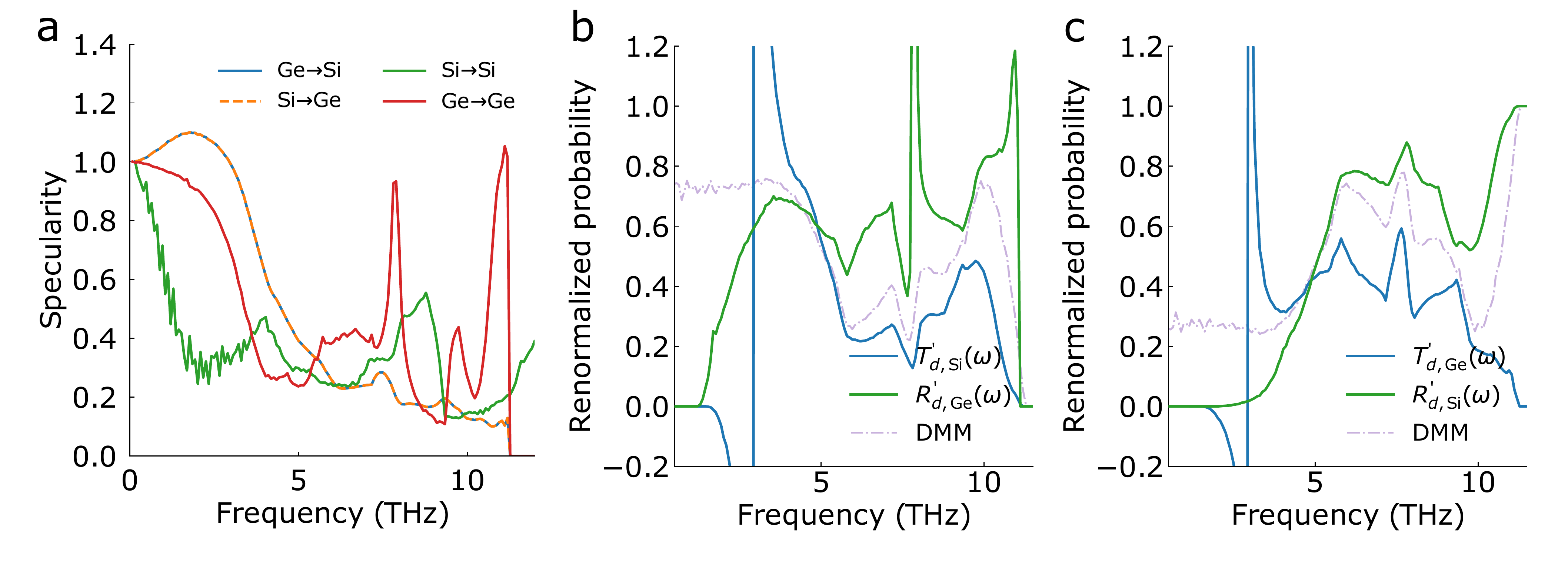}
    \caption{(a) The frequency-resolved specularity parameter from AGF calculation for 8 ml structures. (b) The frequency-resolved renormalized diffuse transmittance from the Si side compared with the renormalized diffuse reflectance from the Ge side. (c) The frequency-resolved renormalized diffuse transmittance from Ge side compared with the renormalized diffuse reflectance from Si side. }
    \label{renormomega}
\end{figure}

\begin{figure}[b]
    \includegraphics[width=0.9\textwidth]{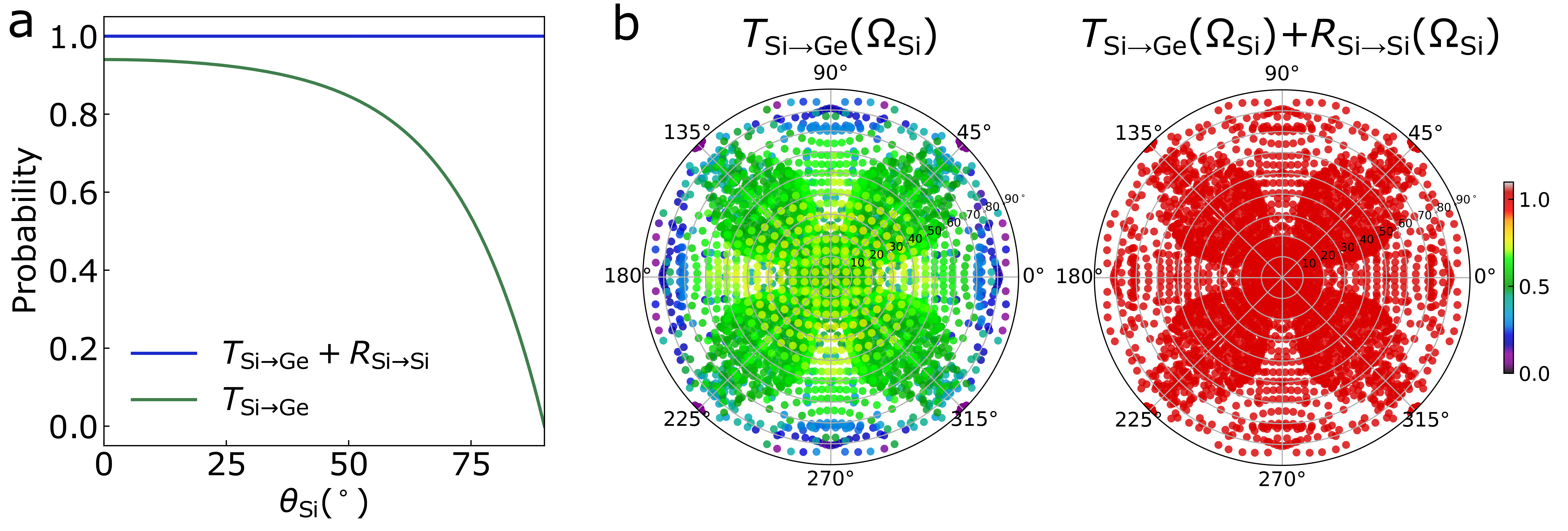}
\caption{(a) The transmittance as a function of grazing angle at $\omega = 4 $ THz from the Si side calculated using the continuum model. The sum of transmittance and reflectance is one, suggesting our continuum model up to second-order perturbation is a consistent theory such that the summation of specular channels and diffuse channels is conserved. The parameters used in calculation are the same with those in Appendix. (b) The transmittance as a function of polar angle and azimuthal angle at $\omega = 5 $ THz from Si from AGF. The summation of transmittance and reflectance is found to be unity, serving as a sanity check for our calculation. The ensemble average is done over 21 8 ml structures.}
\label{check}
\end{figure}

\section{The role of specularity parameter in diffuse scattering} \label{sec:pheno}

The conventional way to study interface scattering probability is to use the specularity parameter $p$ by Ziman's equation, as expressed in Eq. A53a and Eq. A53b in the main text. In this case, for a given incident state, the diffuse transmittance is $T_d = (1-p)T_{\mathrm{DMM}}$ and the diffuse reflectance from the other side is $R_d = (1-p)T_{\mathrm{DMM}}$. 

On the other hand, using AGF and continuum modeling, we have directly compared  diffuse transmittance $T_d$ from one side and diffuse reflectance  $R_d$ from the other side to assess the original argument by Swartz and Pohl that phonon loses its memory of origin. One might argue that $T_d$ and $R_d$ also contain the contribution from the specularity parameter and  $\frac{T_d}{1-p_T}$ and $\frac{R_d}{1-p_R}$ should be the actual diffuse scattering probability. 

To examine the role of specularity parameter in diffuse scattering, we define,
\begin{equation}
    \begin{split}
    T(\Omega) = p_T(\Omega) T_{\mathrm{perfect}}(\Omega)  + (1-p_T(\Omega)) T^\prime_{d}(\Omega)        
    \end{split}
\end{equation}
where 
$T_{\mathrm{perfect}}(\Omega)$ is the transmittance for a perfect interface without atomic mixing. The specularity parameter for transmittance is, 
\begin{equation}
    p_T(\Omega) = \frac{T_{s}(\Omega)}{ T_{\mathrm{perfect}}(\Omega)}
\end{equation}
and the renormalized diffuse transmittance is defined by,
\begin{equation}
    T^\prime_{d}(\Omega) = \frac{T_d(\Omega)}{1-p_T(\Omega)}
    \label{renomralized1}
\end{equation}
For reflectance for a given incident state, we have,
\begin{equation}
    \begin{split}
    R(\Omega) = p_R(\Omega) R_{\mathrm{perfect}}(\Omega)  + (1-p_R(\Omega)) R^\prime_{d}(\Omega)        
    \end{split}
\end{equation}
where
$R_{\mathrm{perfect}}(\Omega)$ is the reflectance for a perfect interface without atomic mixing. The specularity parameter for reflectance is, 
\begin{equation}
    p_R(\Omega) = \frac{R_{s}(\Omega)}{ R_{\mathrm{perfect}}(\Omega)}
\end{equation}
and the renormalized reflectance is,
\begin{equation}
    R^\prime_{d}(\Omega) = \frac{R_d(\Omega)}{1-p_R(\Omega)}
    \label{renomralized2}
\end{equation}
In this case, we compare if the renormalized transmittance from one side using Eq.~\ref{renomralized1}  and reflectance from the other side using Eq.~\ref{renomralized2} are equal.

\begin{figure}
    \includegraphics[width=0.7\textwidth]{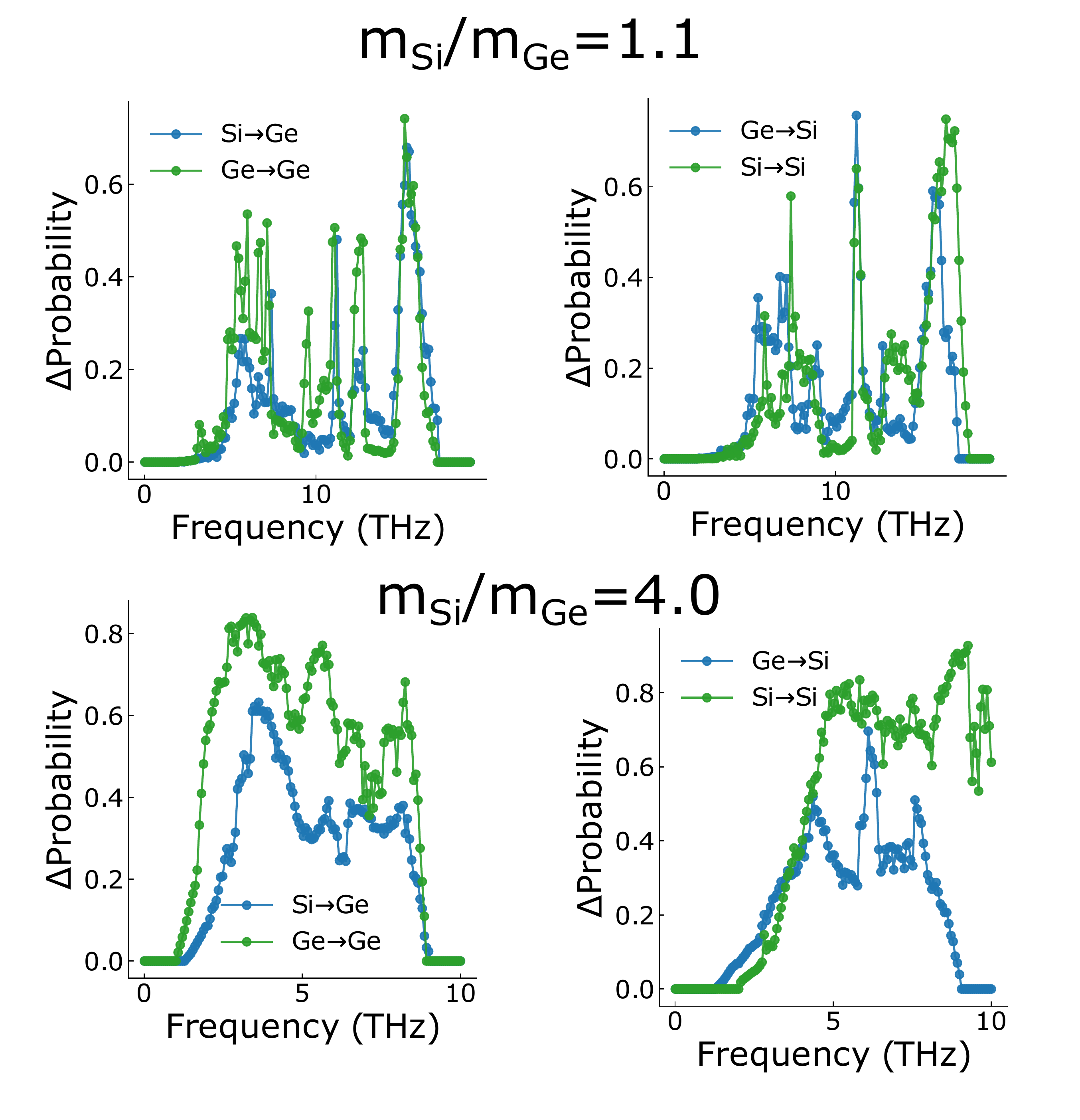}
\caption{The difference between maximum  and minimum diffuse scattering probability as an indicator of the anisotropy of scattering probability with different mass ratios $m_\mathrm{Ge}/m_\mathrm{Si}$. The scattering probability is obtained by taking ensemble average of 21 structures of 8 ml disordered configurations and a 20$\times$20 $
\mathbf{q}_{\mathrm{sc},\parallel}$-point mesh is used in each calculation.}
\label{variation}
\end{figure}

In Fig.~\ref{renorm}, we present the renormalized scattering probability from continuum modeling at $\omega = 4$ THz. We find that, the renormalized transmittance from the Si side and reflectance from the Ge side are not the same. Also, the renormalized reflectance from the Si side is negative, since the specularity parameter for reflectance is larger than one. Notice that the renormalized transmittance and reflectance are the same, when the initial state from the Ge side is normal to the interface ($\theta=0$). Above the critical angle, the renormalized reflectance from the Ge side is independent with angles, while the renormalized transmittance from the Si side decreases with the angle. 

In Fig.~\ref{renormagf}, we show that the renormalized scattering probability from AGF at $\omega = 5$ THz. We find that the renormalized transmittance and reflectance from the Si side  both contain negative values, as a result of larger-than-one corresponding specularity parameter. A relatively weak angle dependence of renormalized transmittance and reflectance from Ge side has been found. However, we can still clearly observe that the renormalized diffuse transmittance from one side and the renormalized diffuse reflectance from the other side are not equal.

In Fig.~\ref{renormomega} (a), we present the specularity parameters for transmittance and reflectance. The specularity parameter for the transmittance at low frequencies is larger than one, suggesting that more specular transmission channels are created by the disorders. In Fig.~\ref{renormomega} (b) and (c), we compare the renormalized diffuse transmittance from one side and renormalized diffuse reflectance from the other side at different frequencies. The divergence near certain frequencies comes from close-to-one specularity parameter. Nevertheless, the renormalized diffuse transmittance and reflectance are largely different and we conclude that DMM is still invalid.

In Fig.~\ref{check}, we present a sanity check for our mode-resolved AGF calculation. In Fig.~\ref{variation}, we show the anisotropy of diffuse scattering probabilities with different mass ratios. With small mass ratio, the anisotropy is reduced mainly due to smaller scattering probabilities, as shown in Fig. 4 (a) in the main text. With a larger mass ratio, the anisotropy of diffuse scattering probabilities is enlarged. 

\begin{figure}
    \includegraphics[width=0.9\textwidth]{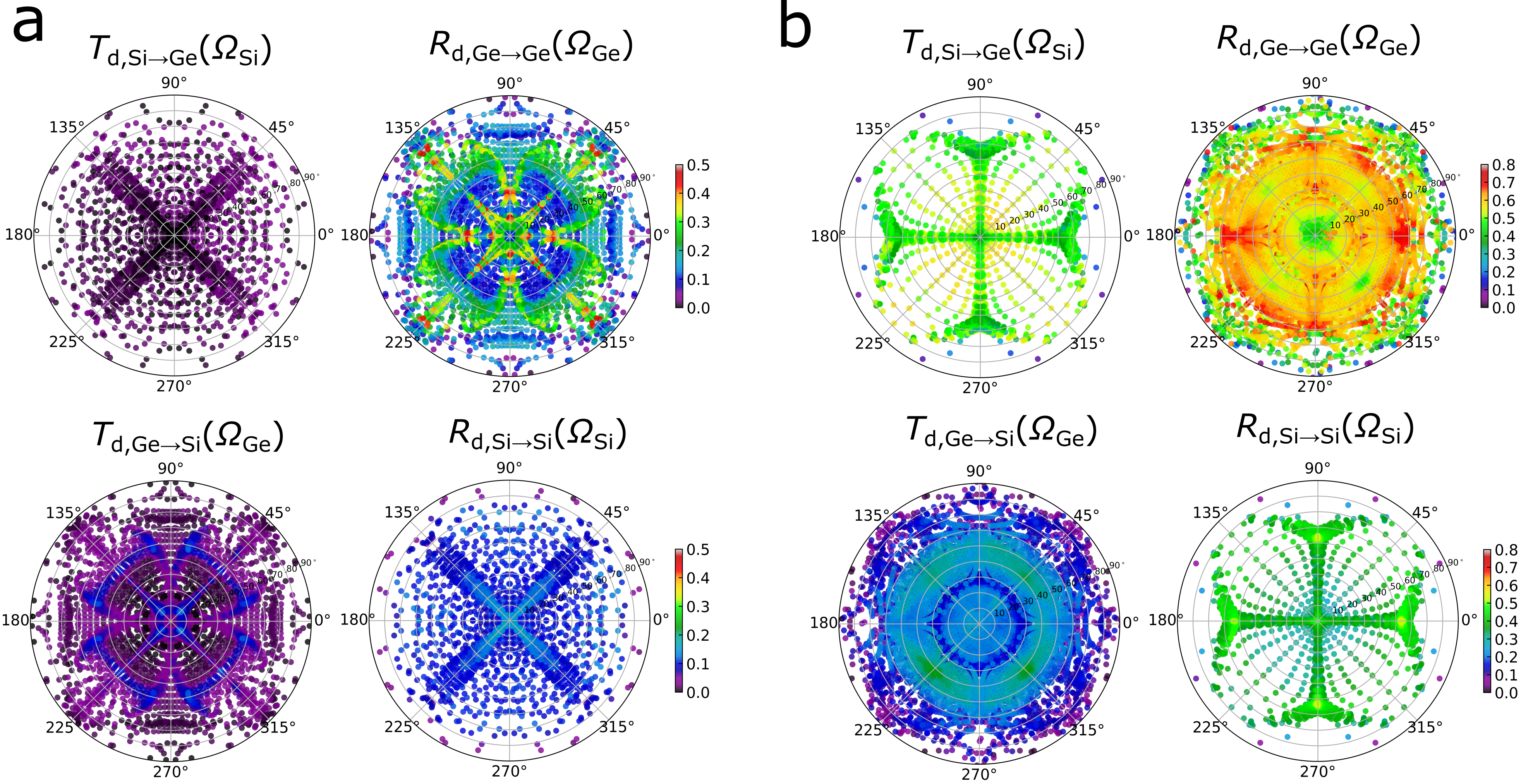}
\caption{The directional diffuse transmittance and diffuse reflectance for 8 ml structures at (a) $\omega$ = 3 THz and (b) $\omega$ = 10 THz.}
\label{low_high}
\end{figure}

\begin{figure}
    \includegraphics[width=0.9\textwidth]{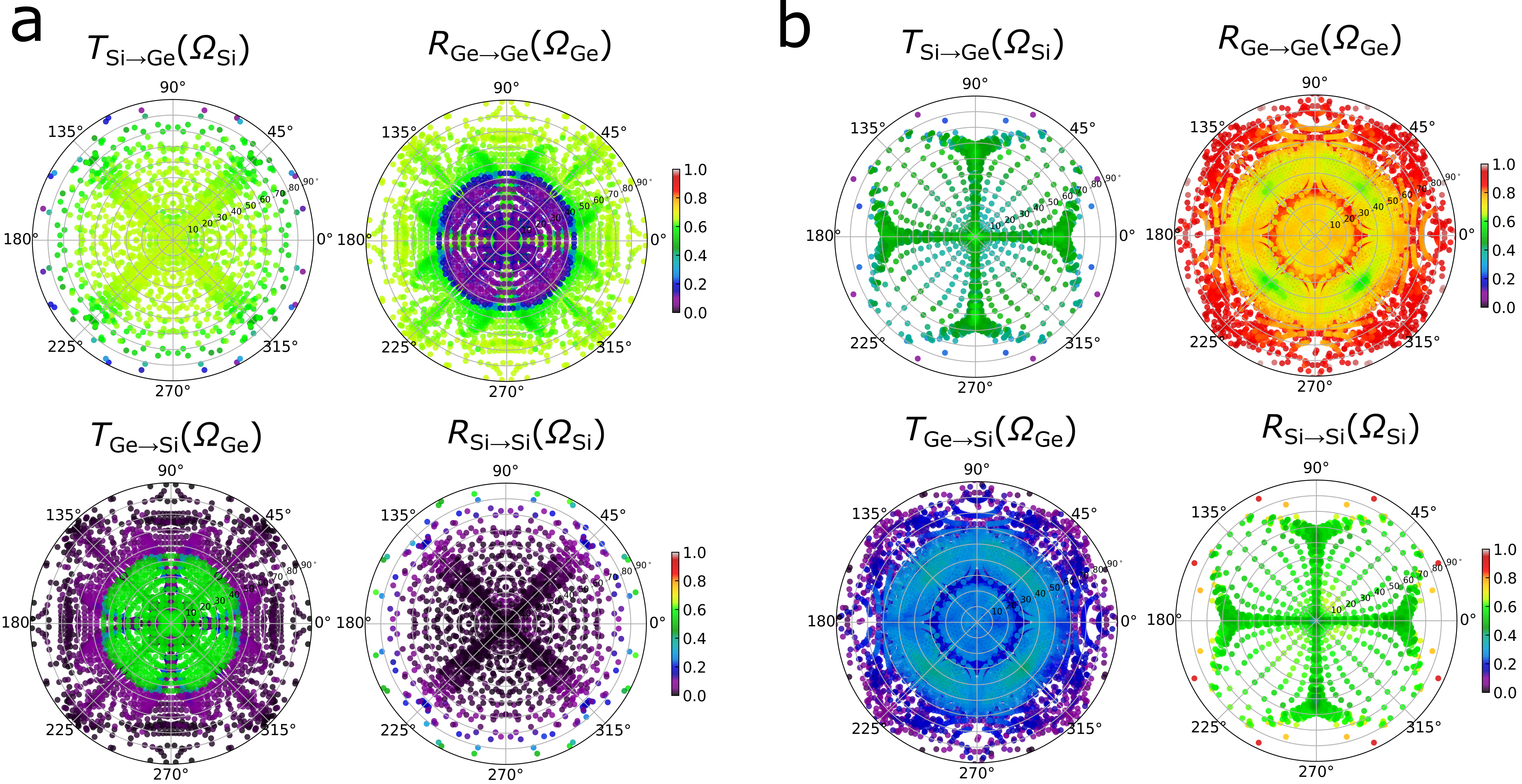}
\caption{The directional total transmittance and total reflectance for 8 ml structures at (a) $\omega$ = 3 THz and (b) $\omega$ = 10 THz.}
\label{low_high_tot}
\end{figure}

\begin{figure}
    \includegraphics[width=0.9\textwidth]{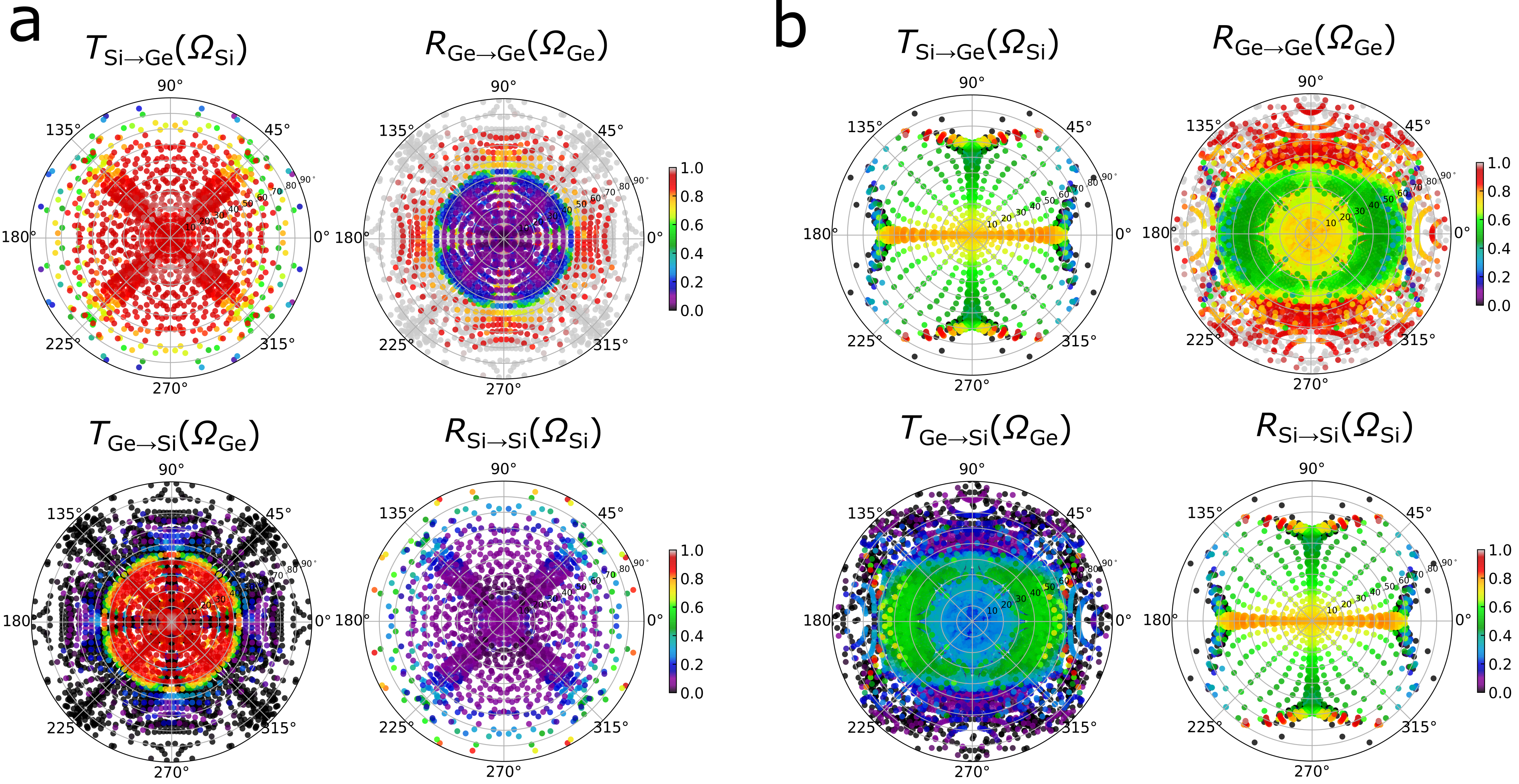}
\caption{The directional total transmittance and total reflectance for an ideal Si/Ge interface at (a) $\omega$ = 3 THz and (b) $\omega$ = 10 THz.}
\label{low_high_perfecvt}
\end{figure}

We have presented the directional diffuse transmittance and reflectance for a rough interface with $\omega = 5$ THz in  Fig. 2 of the main text. We are also interested in the directional diffuse transmittance and reflectance at other frequencies, as shown in Fig.~\ref{low_high}. The total transmittance and reflectance is shown in Fig.~\ref{low_high_tot}. We find that the diffuse scattering probabilities generally have a weaker angular dependence compared with the total scattering probabilities. In Fig.~\ref{low_high_perfecvt}, we examine the total transmittance and reflectance for an ideal interface. The scattering probabilities for the ideal interface have a stronger angular dependence compared with the rough interfaces. 
%What's more, the patterns of the total scattering probabilities for an ideal Si/Ge interface clearly reveal pmm symmetry (no 90-degree rotation symmetry), which is also the symmetry for the ideal [001] Si/Ge interface structure. In contrast, the ensemble-averaged transmittance and reflectance seems to 

\bibliographystyle{apsrev4-2}
\bibliography{sref}